\documentclass[12pt]{article}
\usepackage[centertags]{amsmath}
\usepackage{amsmath, amssymb,graphicx}

\usepackage{epsfig}
\usepackage{ulem}
\usepackage{yfonts}
\usepackage[english]{babel}
\usepackage{float}
\usepackage{caption}
\usepackage{subcaption}
\usepackage{array}
\usepackage{esint}
\usepackage{amsthm}
\usepackage{latexsym}
\usepackage{mathrsfs}
\usepackage{amsbsy}
\usepackage[mathcal]{euscript}
\pdfoutput=1
\usepackage{epsfig}
 \usepackage{jheppub}
 
 \usepackage{ulem}

\usepackage{jheppub}

\usepackage{mathdots}
\usepackage{MnSymbol}
\usepackage{multirow}

\newcommand{\nn}{\nonumber}
\newcommand{\be}{\begin{equation}}
\newcommand{\ee}{\end{equation}}
\newcommand{\bea}{\begin{eqnarray}}
\newcommand{\eea}{\end{eqnarray}}

\def\ben{\begin{equation}}
\def\een{\end{equation}}
\def\bea{\begin{eqnarray}}
\def\eea{\end{eqnarray}}

\input amssym.def
\input amssym.tex

\definecolor{darkraspberry}{rgb}{0.53, 0.15, 0.34}

\definecolor{darkblue}{rgb}{0., 0, 1}
\newcommand{\IA}{\textcolor{darkblue}}

\definecolor{dgreen}{rgb}{0.,0.6,0.}

\def\cO{{\cal O}}

\def\cQ{{\cal Q}}

\title{Complete evaporation of black holes and\\
Page curves
 }
\author{Irina Aref'eva and Igor Volovich}

\affiliation{Steklov Mathematical Institute, Russian Academy of Sciences,\\Gubkina str. 8, 119991, Moscow, Russia}

\emailAdd{arefeva@mi-ras.ru, volovich@mi-ras.ru}

\abstract{
The problem of complete evaporation of Schwarzschild black holes raised by Hawking is that one has an explosion of the temperature $T=1/8\pi M$ for vanishing black hole mass $M$. 
We consider the
Reissner-Nordstrom  black hole and note that if  mass $M$ and charge $Q<M$ satisfy the bound $Q>M-C M^3$, $C>0$ for small $M$ then the complete evaporation of  black holes  without blow-up of temperature is possible. We describe  curves on the surface of state equations such that the motion along them provides   complete evaporation.
The radiation entropy  follows the Page curve  and vanishes at the end of evaporation.
 Similar results  for rotating  Kerr, Schwarzschild-de-Sitter  and  Reissner-Nordstrom-(Anti)-de-Sitter black holes are discussed. 

}
\begin{document}

\maketitle

\newpage

\section{Introduction}
 It was suggested by Hawking that the Schwarzschild  black holes produce radiation like black bodies with temperature 
of $
T = 1/8 \pi M $, where $ M $ is the mass of the black hole \cite{Haw1}. 
During this process,  the mass of the black hole is lowered and the temperature is increased. It is an open question whether this process continues until the entire mass of the black hole has been converted to radiation or whether it stops when the mass is close to the Planck mass \cite{Haw1,Haw2,Page:2013dx,Susskind,FN}.   In this paper, we study under which conditions in classical gravity the complete evaporation of black holes is possible.
\\

According to the Stefan-Boltzmann law, the energy density $E$ of black body radiation with temperature $T$ is given by $
E=\pi ^2 T^4/15. $
Therefore from Hawking's formula for the temperature, it follows that the energy density of the radiation, emitted by a black hole behaves at small $ M $ as $ M ^ {- 4} $.
If the mass of the black hole disappears during evaporation, then the black hole releases an infinite amount of energy, which is clearly unphysical. Hawking drew attention to this problem in his first paper on the evaporation of black holes with the title "Black hole explosions?" \cite{Hawk3}. 
 \\

 The  information loss problem \cite{Haw2,Page:1993wv} is closely related to this unphysical  behaviour, since the radiation entropy $S_{R}$ diverges for small $ M $ as $ M ^ {- 3} $.
 One can say that because of this infinity, complete evaporation never occurs in nature, and we cannot obtain complete evaporation, which ends only by thermal radiation being obviously a mixed state.
  Just the evolution of  the initially pure quantum state  to  this mixed state  breaks unitarity and leads to information paradox.
\\

In this paper,  a possible mechanism for the complete evaporation without temperature explosion  of black holes in the framework of classical gravity is discussed. This mechanism also ensures  that the entropy of radiation vanishes in the limit when the mass of the black hole tends to zero, which is consistent with unitary evolution. 
We note that the complete evaporation without explosion of  temperature and energy is possible if the black hole possess addition parameters -- a charge $Q$ or an angular momentum $a$, or we deal with non-zero cosmological constant and we are in near-extremal regimes. \\

\newpage

In the case of the Reissner-Nordstrom black hole  if the mass $M$ and charge $Q<M$  satisfy also the bound
\be
\label{Q>M}Q>M-C M^3,\quad C>0,\ee
for small mass $M$, then the Hawking temperature for the charged black hole 
\be
\label{TQM}
T=
\frac{1}{2\pi}\frac{\sqrt{M^2-Q^2}}
{(M+\sqrt{M^2-Q^2})^2}
\ee
tends to 0 when $M\to 0$.
\\

We can describe special curves in the  domain \eqref{Q>M}.
The expression \eqref{TQM}  defines the surface $\Sigma$ of the state equation $T=T(M,Q)$ in the 3-dimensional space with coordinates $(M,Q,T)$. Evaporation process is described by a curve $\sigma$ on $\Sigma$: $M=M(t)$, $Q=Q(t)$, $T=T(M(t),Q(t))$, where $t$ is a time variable (see Fig.\ref{fig:3D-RN}.A in Sect.\ref{sect:RN}). It is convenient to make the change of variables $(M,Q)\to (M,\lambda)$, $\lambda=\sqrt{M^2-Q^2}$.
We describe an evaporation curve $\sigma$ by a function $\lambda=\lambda(M)$
and
 the charge belong this curve is
 \be
\label{Q-M}
Q^2=M^2-\lambda(M)^2,
\ee 
where $0<\lambda(M)< M$. Obviously, the Reissner-Nordstrom solution of the Einstein equations with parameters $M$ and $Q$ will still be a solution if the charge $Q$ is taken to depend on the mass $M$.
The Hawking temperature $T$  becomes 
\be\label{T-lambda2}
T=\frac{\lambda(M)}{2\pi(M+\lambda(M))^2}.
\ee
If we take $\lambda (M)$ such that for small $M$ it obeys   
$\lambda (M)=o(M^2)$,
then  $T$ tends to $0$  as $M\to 0$  and we get the complete evaporation of the black hole. Therefore, for small $M$, the evaporating black hole must be in a state close to the extreme one.
\\

In this paper we consider two particular examples of near extremal Reissner-Nordstrom black holes. \begin{itemize}
\item We take  
 \be\label{lambdaMgamma}\lambda (M)=CM^{\gamma},\qquad C>0,\quad \gamma >2.
 \ee
 This means that for small $M$ we are dealing
with almost extreme regime.
An interesting case is when $\gamma=2$, i.e. $\lambda (M)=CM^2$. In this case, the  limit of temperature when $M\to 0$ is not equal to zero, but is equal to $C/2\pi$, although the mass and charge are vanishing. In the cases with $\gamma \geq 2$ the mass dependence of charge has a deformed bell-shaped form (see Fig.\ref{fig:RN}.A below).
\item
We also consider the case then  the function $T=T(M)$ is given. In this case  one can solve the quadratic equation \eqref{T-lambda2} and find the function $\lambda (M)$.
We take as an example the temperature of the form
\bea\label{T-M-M0}
T(M)&=&C\sqrt{M(M_0-M)},\eea
where $C$ and $M_0$ are positive  constants and $0\leq M\leq M_0$. The radiation entropy $S_{rad}(M)$ is proportional to $T^3$
so in this case one has $S_{R}(M_0)=S_{R}(0)=0$.
\end{itemize}
$\,$\\

As mentioned above, the problem of  temperature explosion during evaporation of the black hole is closely related with the information paradox. In the context of studying the information paradox,  Page \cite{Page:1993wv,Page:2013dx}
 proposed that the Schwarzschild  black holes evaporate completely and  the radiation entropy of evaporating
black holes first increases but then decreases and tends to zero when the black hole mass vanishes. This hypothetical behavior is known as the Page curve. Recent works devoted to the information paradox are aimed to obtain the Page curve
\cite{Penington:2019npb,Almheiri:2019psf,Almheiri:2019hni} for the entanglement entropy  of   radiation. In this work we deal with the usual thermodynamic entropy for radiation that is proportional to $T^3$. So if temperature decreases with vanishing of mass then the entropy of radiation decreases too. To get the time dependence of this evolution we consider charge and mass change during the black hole evaporation.
\\

The loss of the mass and charge during evaporation of
the Reissner-Nordstrom  black hole
is a subject of numerous considerations. Changes in mass and charge during the evaporation of a  RN black hole satisfy a system of two coupled equations (see below eqs. \eqref{Mdot} and \eqref{Qdot}).  Assuming that the relation between mass and charge is fixed, we are left with a single non-linear differential equation
\be\label{seq}
\frac{dM}{dt}=-f(M),
\ee
where an explicit  expression for the function $f(M)$ is given in Sect.\ref{sect:TE}. For small $M$ and $\lambda(M)=CM^{\gamma},\,\,\gamma  > 2$ we get mass evolution in the form 
\be
M(t)=\frac{M_0}{(1+Bt)^{1/(3\gamma-6)}},\,\,\,\, \,\,t\geq 0,
\ee
where $M_0$ and $B$ are positive constants. This form of time dependence of mass of evaporation  black hole together with radiation entropy dependence as $T^3$ provide the Page form of time evolution of the radiation entropy, Fig.\ref{fig:Page-curve-mass-time}.A - Fig.\ref{fig:Page-curve-mass-time}.C.   If $\gamma =2$ then $M(t)=M_0 e^{-Bt}.$ In this case the leaving time of black hole is infinite and the the evolution of the radiation has the form presented in Fig.\ref{fig:Page-curve-mass-time}.C
and Fig.\ref{fig:Page-curve-mass-time}.E.
$\,$\\

For Kerr, Schwarzschild de Sitter
and Reissner-Nordström-(Anti)-de Sitter black holes, we also indicate the curves on the equation of the state surfaces along which the complete evaporation of black holes occurs without thermal explosions.
\\

The paper is organized as follows. In Sect.\ref{sect:RN}  models of  complete evaporation of the  Reissner-Nordstrom black hole
accompanied by the temperature  goes to $0$ are considered. In Sect.\ref{sect:Kerr} the  models of
complete evaporation of the Kerr black hole are investigated and in Sect.\ref{sect:KN} these results are generalized to the Kerr-Newman black hole. Complete evaporation of the Schwarzschild-de-Sitter black hole   is considered in Sect.5.
In Sect.\ref{sect:CERNds} we discuss complete evaporation of Reissner-Nordstrom-(Anti)-de-Sitter black holes.
Sect.\ref{sect:SD} contains summary and discussions.

\section{Complete evaporation of the  Reissner-Nordstrom black hole}\label{sect:RN}
We consider a model of complete evaporation of a Reissner-Nordstrom (RN) black hole with the following metric
\be
\label{ds-RN}
ds^2=-f(r)dt^2+f(r)^{-1}dr^2+r^2d\Omega^2,
\ee
where
\be
\label{f-RN}
f(r)=1-\frac{2M}{r}+\frac{Q^2}{r^2},
\ee
here $M$ and $Q$ are mass and charge of the black hole. It is assumed that $M^2 \geq Q^2 $.
The blackening function \eqref{f-RN} can be presented as 
\be\label{f-rpm-RN}
f=(r-r_+)(r-r_-)/r^2,
\ee
where
\be
r_\pm =M\pm\sqrt{M^2-Q^2}.
\ee
The temperature of the  Reissner-Nordstrom black hole is
\be\label{RN-TMQ}
T=
\frac{1}{2\pi}\frac{\sqrt{M^2-Q^2}}
{(M+\sqrt{M^2-Q^2})^2}
\ee
\\

\subsection{Evaporation curves  and bell-shaped temperature}

We  take $Q$ to be a function on $M$ of the form \eqref{Q-M}. The Hawking temperature $T$ under this constraint   becomes equal to
\be\label{T-lambda}
T=\frac{\lambda(M)}{2\pi(M+\lambda(M))^2}.
\ee
Depending on the behavior of the function $\lambda(M)$ we have:
\begin{itemize}
\item 
if $\lambda(M)$  satisfies for small $M$ the bounds
\be
\label{lambda-bound}
0<\lambda (M)\leq C M^\gamma,\,\,\,\,\,\,C>0,\,\,\gamma>2
\ee
then 
$T\to 0$ as $M\to 0$ and one gets the complete evaporation of black hole;
\item  if the function $\lambda(M)$ satisfies the bounds
\be
\label{restr}
0<\lambda (M)\leq C\frac{M^\gamma}{A+M^{\gamma+1}},\,\,\,\,\,C>0,\,A>0,\,\gamma>2
\ee
then the temperature   $T\to 0$ also for $M\to\infty$;  
\item
if the function $\lambda(M)$ is
\be
\label{restr}
\lambda (M)= \frac{M^\gamma}{A+M^{\gamma-1}},\,\,\,\,\,\,A>0,\,\gamma>3,
\ee
then the temperature   $T\to 0$  and $Q\to 0$ also for $M\to\infty$. Note that the asymptotic of $T$ at $M\to \infty$
coincides with the Schwarzschild case.
\end{itemize}
$\,$\\

 The entropy and the free energy under constraint
 \eqref{Q-M}  are  equal
\bea\label{S-RN}
S_{RN}&= & \pi r_+ ^2=
 \pi (M+\lambda (M))^2\\
\label{F-RN}
G_{RN}&= &M-TS
=
 M-\frac{1}{2} \lambda(M) \eea

Note that the entropy $S_{RN}$ and  the free energy $G_{RN}$  go to $0$ as $M\to 0$ for $\lambda$
satisfying \eqref{lambda-bound}.
\\

  Let us suppose that the temperature  is given by a function $T=T(M)$. We solve the quadratic equation \eqref{T-lambda} and find two functions
\bea
\label{lam-pm}
\lambda_{\pm}(M)&=&\frac{1-4 \pi  M T\pm\sqrt{1-8 \pi  M T}}{4 \pi 
   T}\eea
   Both solutions are real  for 
   \be
   \label{real}
   1-8 \pi  M T(M)\geq 0.
   \ee

\subsection{Examples}
\subsubsection{Deformed  bell-shaped dependence of charge on mass $M$}
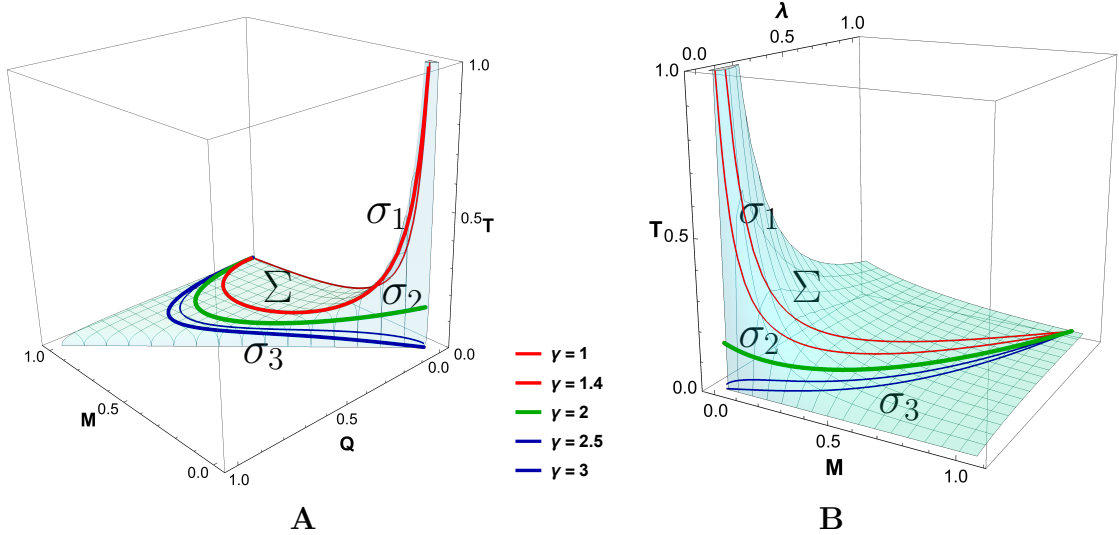
\begin{figure}[h]
  \centering
   \begin{picture}(0,0)
\put(100,70){
{\Large$\Sigma$}}
\put(92.5,47){
{\Large$\sigma_3$}}
\put(139,100){
{\Large$\sigma_1$}}
\put(145,69){
{\Large$\sigma_2$}}
\put(300,70){
{\Large$\Sigma$}}
\put(333,28){
{\Large$\sigma_3$}}
\put(280,100){
{\Large$\sigma_1$}}
\put(280,53){
{\Large$\sigma_2$}}
\end{picture}
   \includegraphics[scale=0.24]{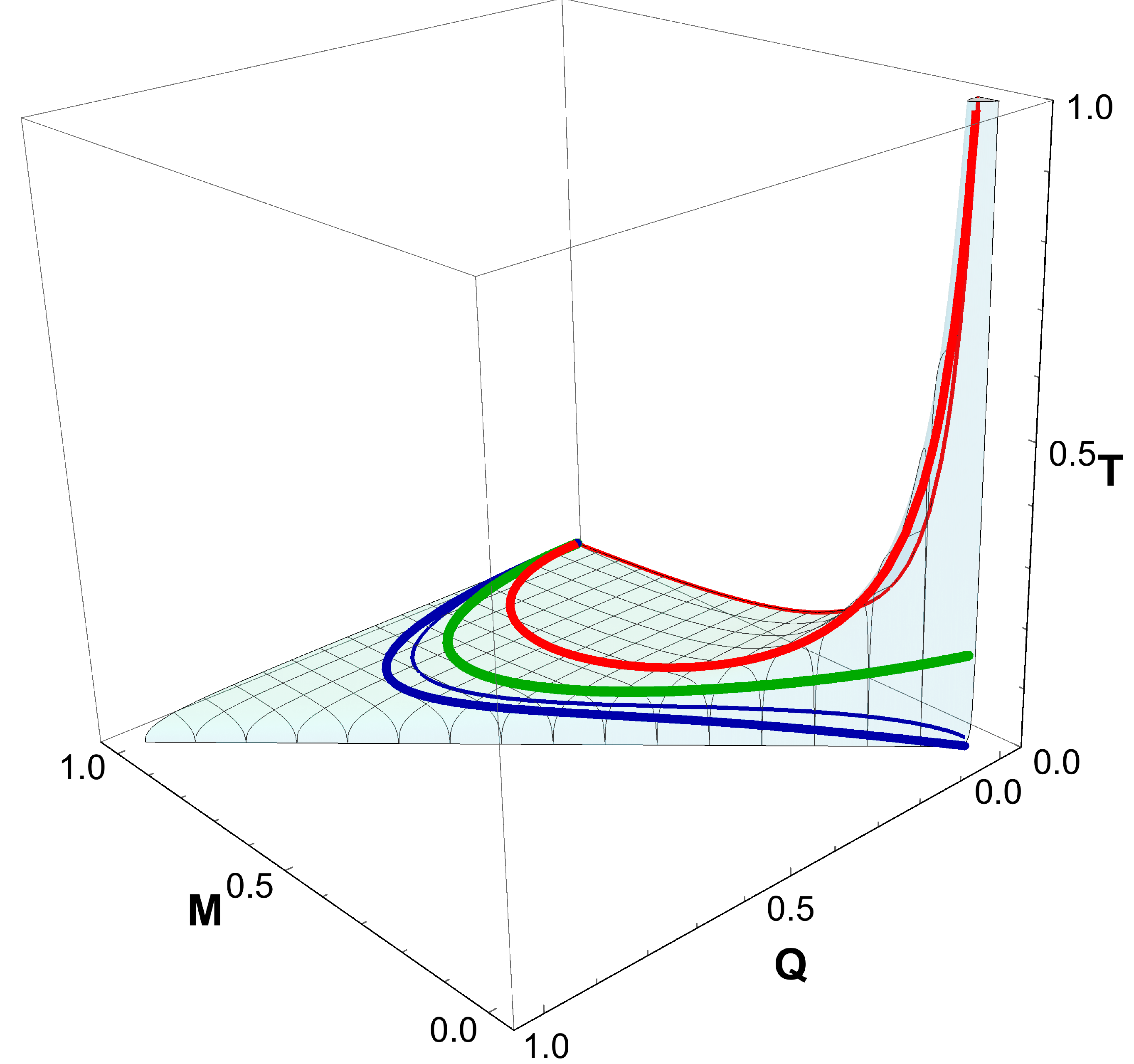} \includegraphics[scale=0.25]{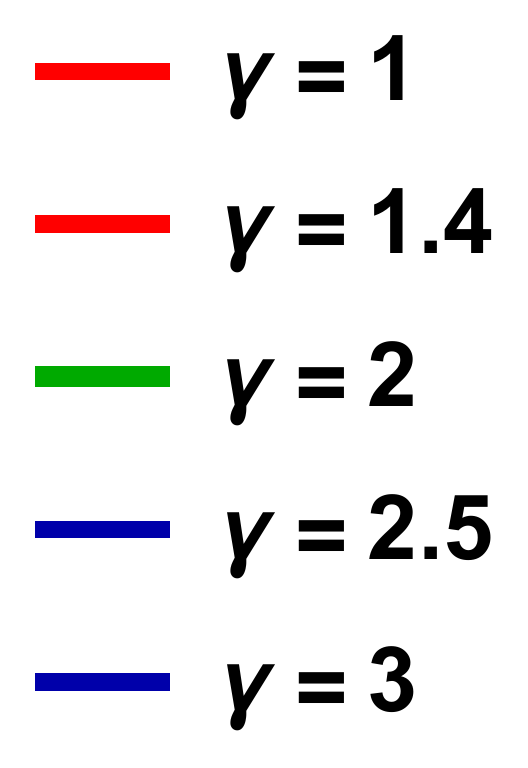}
 \includegraphics[scale=0.28]{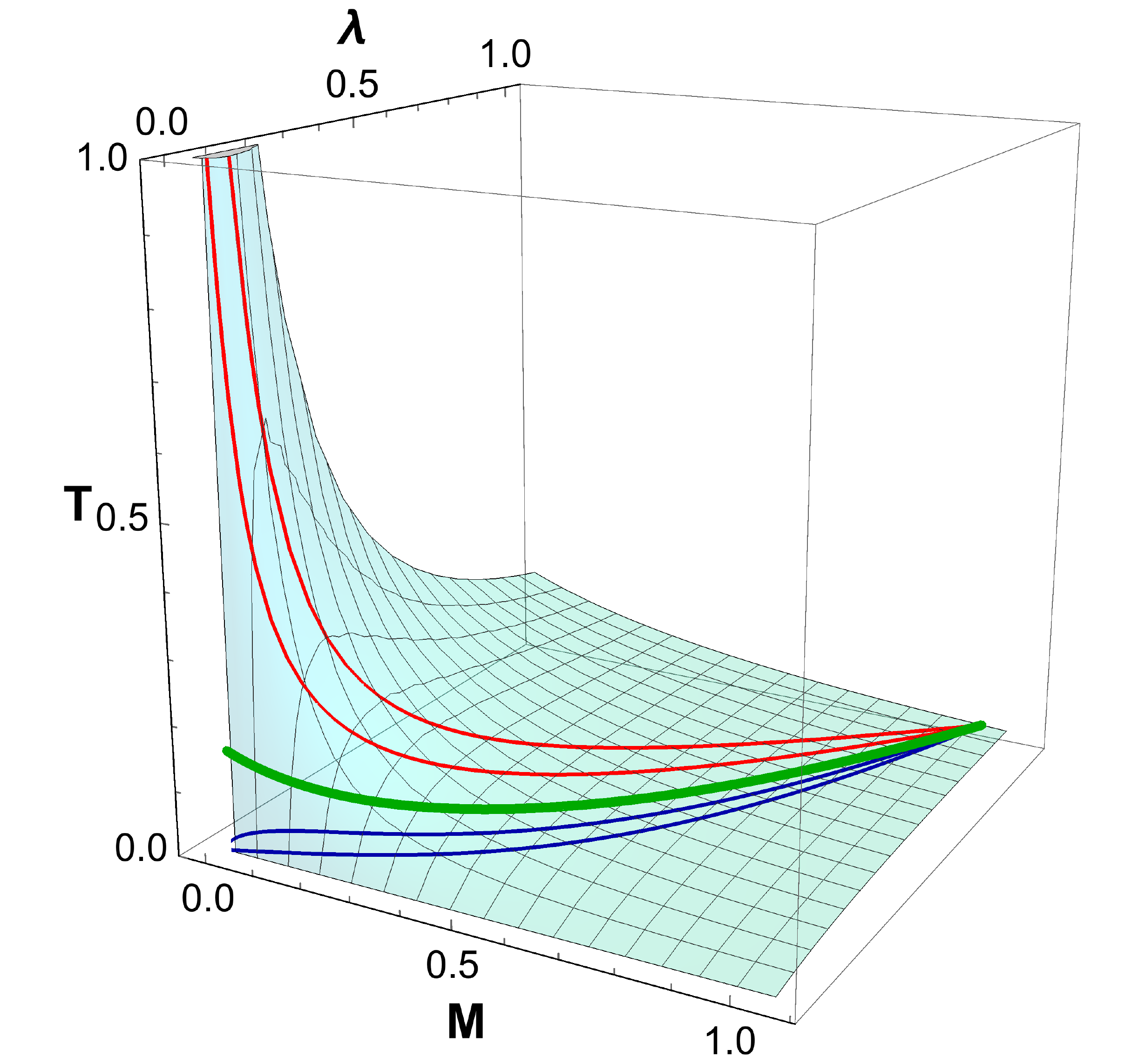}  \\ {\bf A}\hskip190pt{\bf B}
\caption{A) The 3D plot shows  the dependencies of the temperature on the mass $M$ and the charge $Q$. 3D curves
$\sigma_\gamma$ show the dependence of  the temperature on the mass along the constrains \eqref{Q-M} with $\lambda(M)=M^\gamma$,  $\gamma=1,1.4,2,2.5,3$.  B) 
The 3D plot shows  the dependencies of the temperature on the mass $M$ and  $\lambda$. 3D curves $\sigma_\gamma$ show the dependence of  the temperature on the mass along the constrain \eqref{Q-M} with $\lambda(M)=M^\gamma$,  $\gamma=1,1.4,2,2.5,3$. 
}
  \label{fig:3D-RN}
\end{figure}

Equation \eqref{RN-TMQ} defines the surface $\Sigma$ of the state equation for Reissner-Nordstrom black hole. This surface is shown in Fig.\ref{fig:3D-RN}.A. The same surface is shown in  Fig.\ref{fig:3D-RN}.B in $(M,\lambda,T)$ coordinates, eq. \eqref{T-lambda}. 
The 3D curves in Fig.\ref{fig:3D-RN}.A and Fig.\ref{fig:3D-RN}.B show the dependence of temperature on mass along the curves
\be
\label{lambda-gamma}
\lambda(M)=\left(\frac{M}{m_0 }\right)^{\gamma }.
\ee
 with different  $\gamma=1,1.4,2,2.5,3$.
 We see that on some curves the temperature tends to infinity as $M\to 0$ (for these curves $\gamma<2$), while on curves with $\gamma>2$ the temperature tends to zero as
   $M\to 0$.
$$\,$$

 \begin{figure}[h]
  \centering
   \includegraphics[scale=0.23]{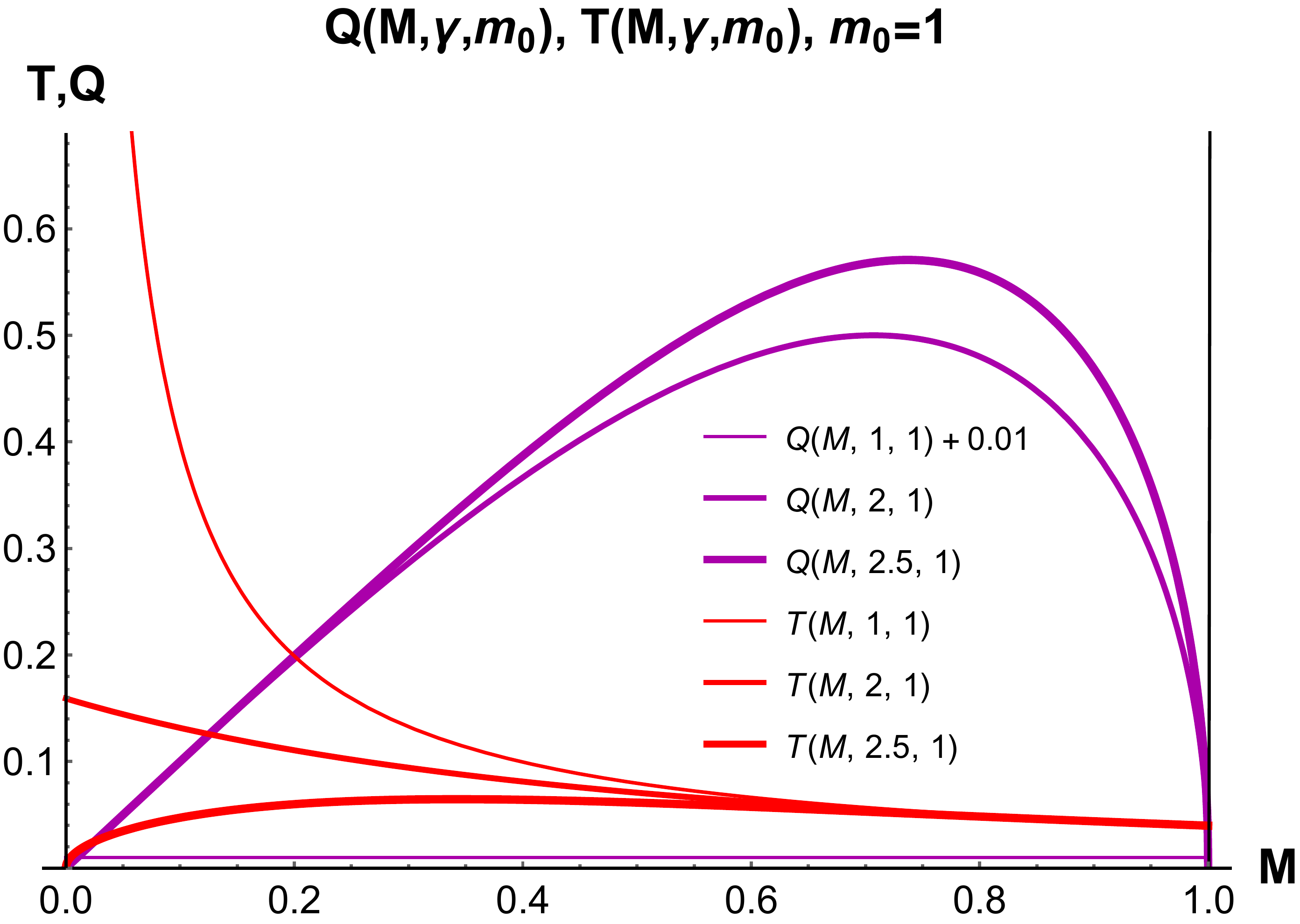} \quad
    \includegraphics[scale=0.17]{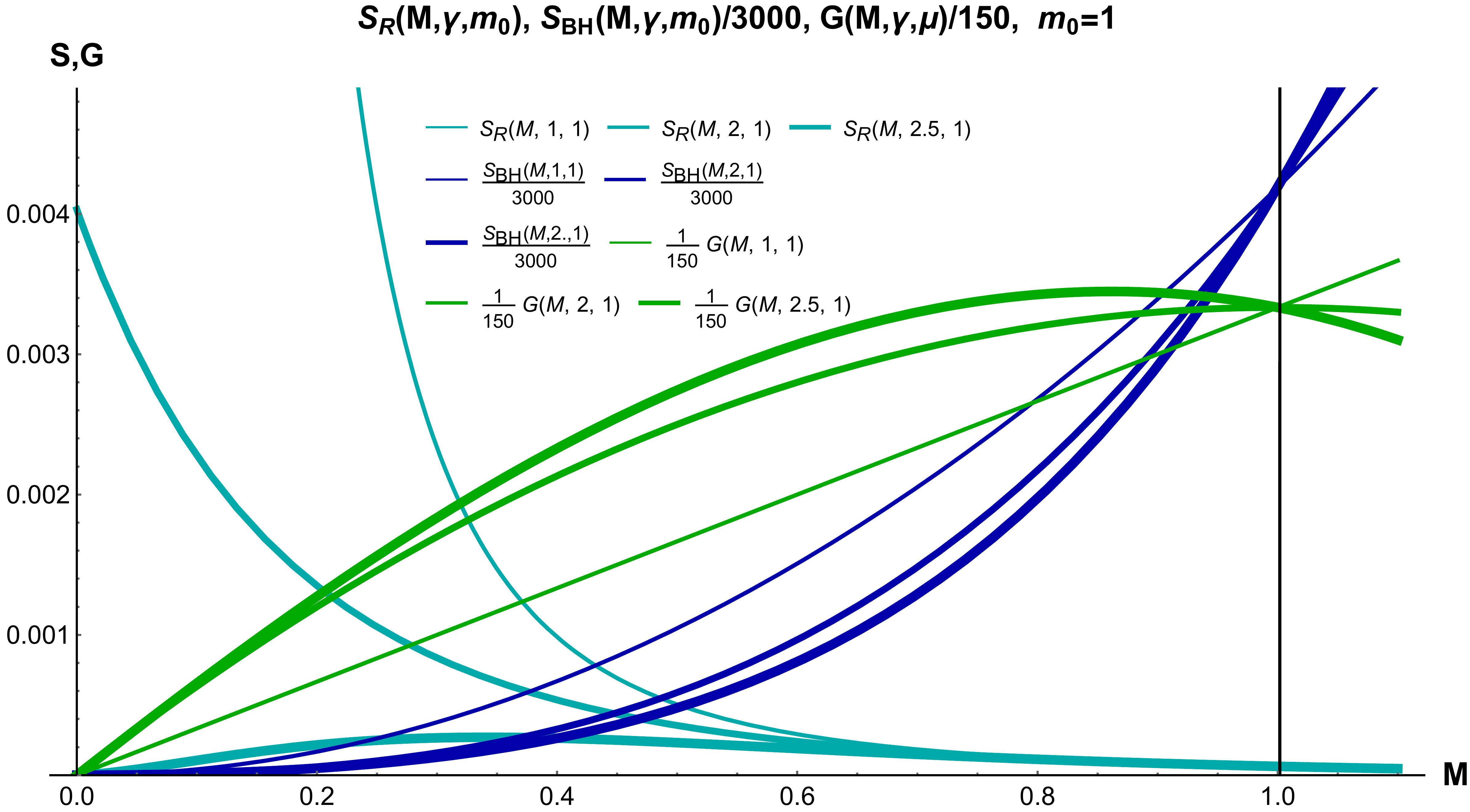} 
\\ {\bf A}\hskip190pt{\bf B}
\caption{A) The graph shows charge $Q$ (magenta) and temperature $T$ (red) versus $M$ for different values of the scaling parameter $\gamma$, $\gamma=1,2,2.5$. B) The graph shows dependencies of free energy $G$ (green), black hole entropy $S$ (blue) and radiation entropy (dark cyan) on $M$ for scale parameter $\gamma=1.2, 2.5$. 
 The black lines show the boundaries of the allowed regions for $M$. 
 }
  \label{fig:RN}
\end{figure}
 Mass dependence of charge $Q$, temperature $T$, entropy and free energy at
     \eqref{lambda-gamma} with different $\gamma$ parameters and the same $\mu=1$ parameter are shown in fig.\ref{fig:RN}.
We see that for all $\gamma> 1$ there is a restriction on
$M$, $M\leq 1$.  The temperature and entropy of the radiation tend to zero at $M\to 0$ for $\gamma>2$, to a nonzero constant for $\gamma=2$, and to infinity for $\gamma<2$. In this case, the temperature and radiation entropy $S_R$ at $\gamma>2$, starting from the initial value at $M=1$, increase to a certain maximum value, then decrease to zero, i.e. the mass dependencies $T $ and $S_R$ have deformed bell shapes (the thickest red and dark cyan lines in Fig.\ref{fig:RN}.A and B, respectively). In the case of a slow dependence of mass on time during black hole evaporation, the dependence of radiation entropy on mass, represented in Fig.\ref{fig:RN}.B by the dark cyan line, leads to the Page form of the time evolution of radiation entropy, see  Fig.\ref{fig:Page-curve-mass-time}.
The entropy of a black hole and the free energy tend to zero at $M\to 0$ for all values of $\gamma$. We also see, Fig.\ref{fig:RN}.A, that the shape of Q versus $M$ is a deformed bell (except in the case of $\gamma=1$ and $Q=0$, which corresponds to the Schwarzschild case).
Note that in fig.\ref{fig:RN}.B one can see that the free energy increases as the black hole mass decreases. This corresponds to the region $M$, where the charge increases with decreasing mass $M$.

In our recent paper \cite{Arefeva:2022cam} we found restrictions on $\gamma $
in \eqref{lambda-gamma} under which the entanglement entropy calculated with the island formula  has no  explosion at $M\to 0$.

 \subsubsection{Semi-circle  dependence of temperature on mass $M$}  
One can take the semi-circle form of $T$ dependence on $M$ \eqref{T-M-M0} with $C=1$.
The condition \eqref{real}  gives a restriction on possible values on admissible $M_0$.
Indeed, substituting \eqref{T-M-M0} to the condition \eqref{real} we get
\bea
   \label{rest-exp}
   1-8 \pi  M \sqrt{M(M_0-M)}\geq 0, \eea
   that should be satisfied 
 for all $M<M_0$.  This can be realized for $M_0\leq M_{0,cr}=2^{1/2}/\pi^{1/2} 3^{3/4}\approx 0.35$.
 \\
 
  \begin{figure}[h!]
  \centering
  \begin{picture}(0,0)
\put(92.5,24){\color{blue}
{\Huge$\mathbb *$}
\put(220,-19){\color{blue}
{\Huge$\mathbb *$}}}
\end{picture}
\includegraphics[scale=0.21]{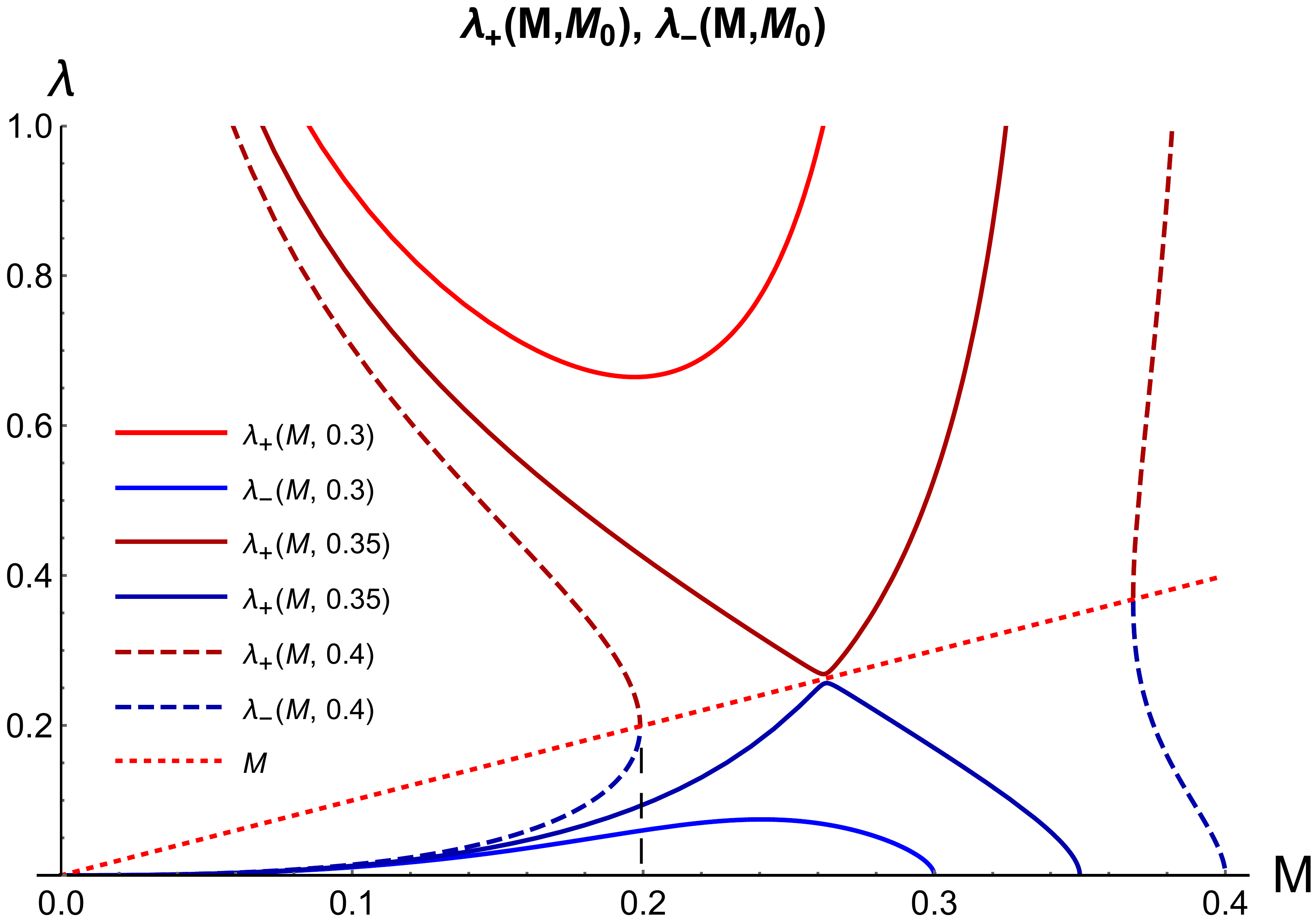}
\qquad
\includegraphics[scale=0.28]{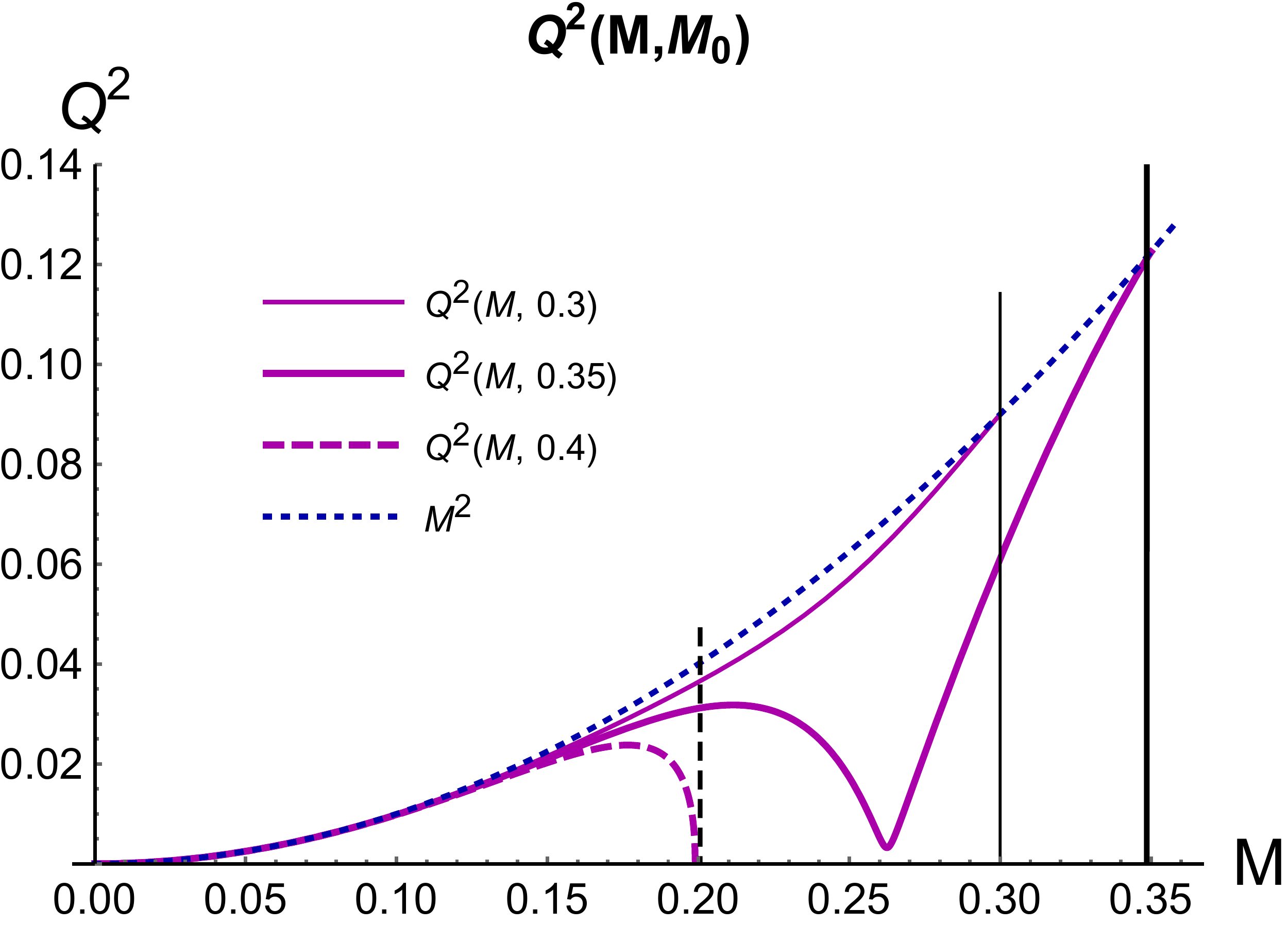}
\\ {\bf A}\hskip190pt{\bf B}
 \caption{
A) $\lambda$ as a function of $M$ for two branches $\lambda _\pm =\lambda _\pm (M)$ given by \eqref{lam-pm}, where $T$ is defined by \eqref{T-M-M0}
 The dashed red line corresponds to $\lambda=M$ and we see that the $\lambda _+$ is not a physical solution, since $\lambda _+>M$. We also see that real solutions exist for all $M\leq M_0$  if $M_0<M_{0,cr}\approx 0.35$
 (blue solid lines) and for $M<M_{cr}(M_0)$ for $M_0>M_{0,cr}$ (dashed blue line). We do not consider the right part of $\lambda _-$, since it does not admit the $M\to 0$ limit. B) $Q^2$ vs $M$  for $\lambda _-(M)$ as in Fig.A. Solid magenta lines correspond to  $M_0<M_{0,cr}$ and the dashed magenta  line to $M_0>M_{0,cr}$. The points indicated by {\large$\mathbb *$} correspond to the Schwarzschild case.
  }
  \label{fig:rest}
\end{figure}
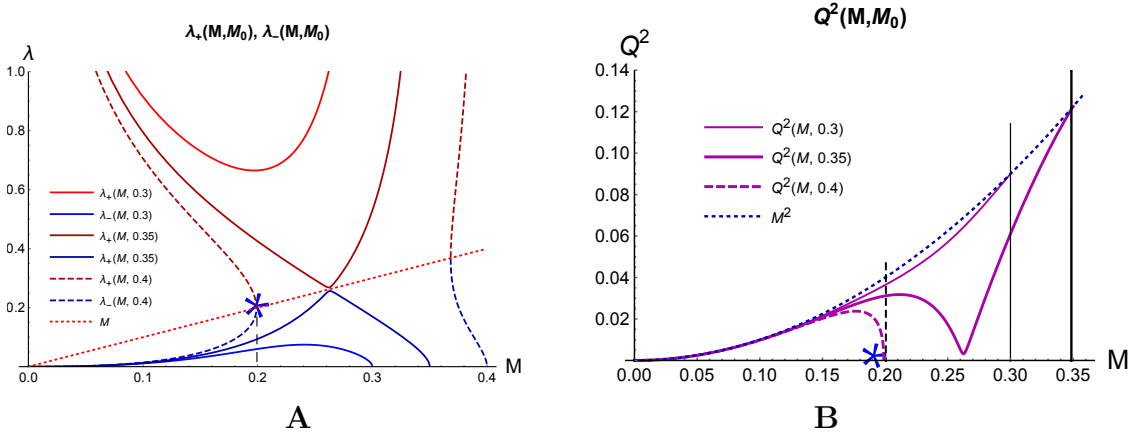
 
 We plot in  Fig.\ref{fig:rest}.A
 $\lambda$ as a function of $M$ for two branches $\lambda _\pm =\lambda _\pm (M)$ given by \eqref{lam-pm}, where $T$ is defined by \eqref{T-M-M0}. 
 We see that for small $M_0 \lessapprox 0.35$ both $\lambda _\pm$ are real. 
For $M_0  \gtrapprox
0.35$ two solutions $\lambda_{\pm}$ are real only on parts of the interval $[0,M_0]$ (dashed blue lines in Fig.\ref{fig:rest}.A). For small $M$ and $M_0 
\lessapprox  0.35$ we have
\bea
\lambda_-(M, M_0)&=&2 \pi  M^{5/2} \sqrt{M_0}+\cO\left(M^{7/2}\right),\\
 \lambda_+(M, M_0)&=&\frac{1}{2 \pi 
   \sqrt{M} \sqrt{M_0}}+\cO\left(M^{1/2}\right),  
   \eea
   and in accordance with \eqref{real} we consider only the $\lambda_-$-branch.
 \\

We plot in  Fig.\ref{fig:rest}.B  the mass dependence of the charge $Q$, that is defined by \eqref {Q-M} with $\lambda=\lambda_-(M)$.

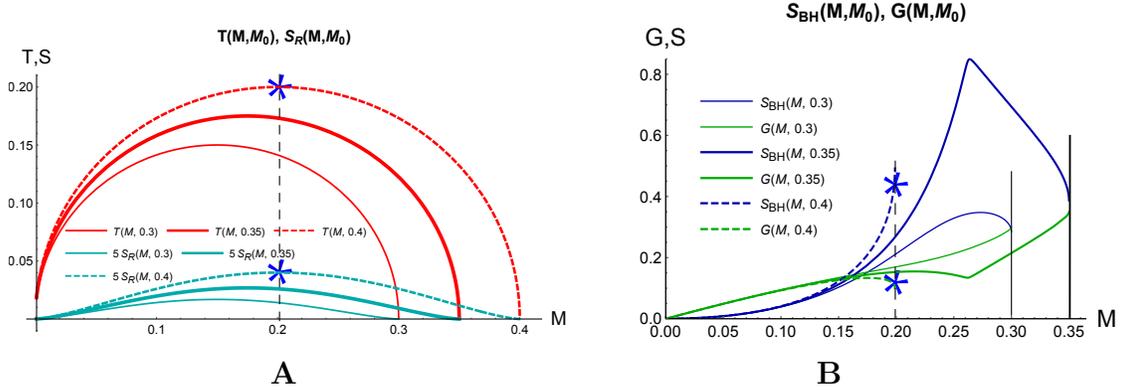
\begin{figure}[h!]
  \centering
    \begin{picture}(0,0)
\put(101,20){\color{blue}
{\Huge$\mathbb *$}}
\put(101,90){\color{blue}
{\Huge$\mathbb *$}}
\put(334,16){\color{blue}
{\Huge$\mathbb *$}}
\put(334,53){\color{blue}
{\Huge$\mathbb *$}}
\end{picture}
  \includegraphics[scale=0.2]{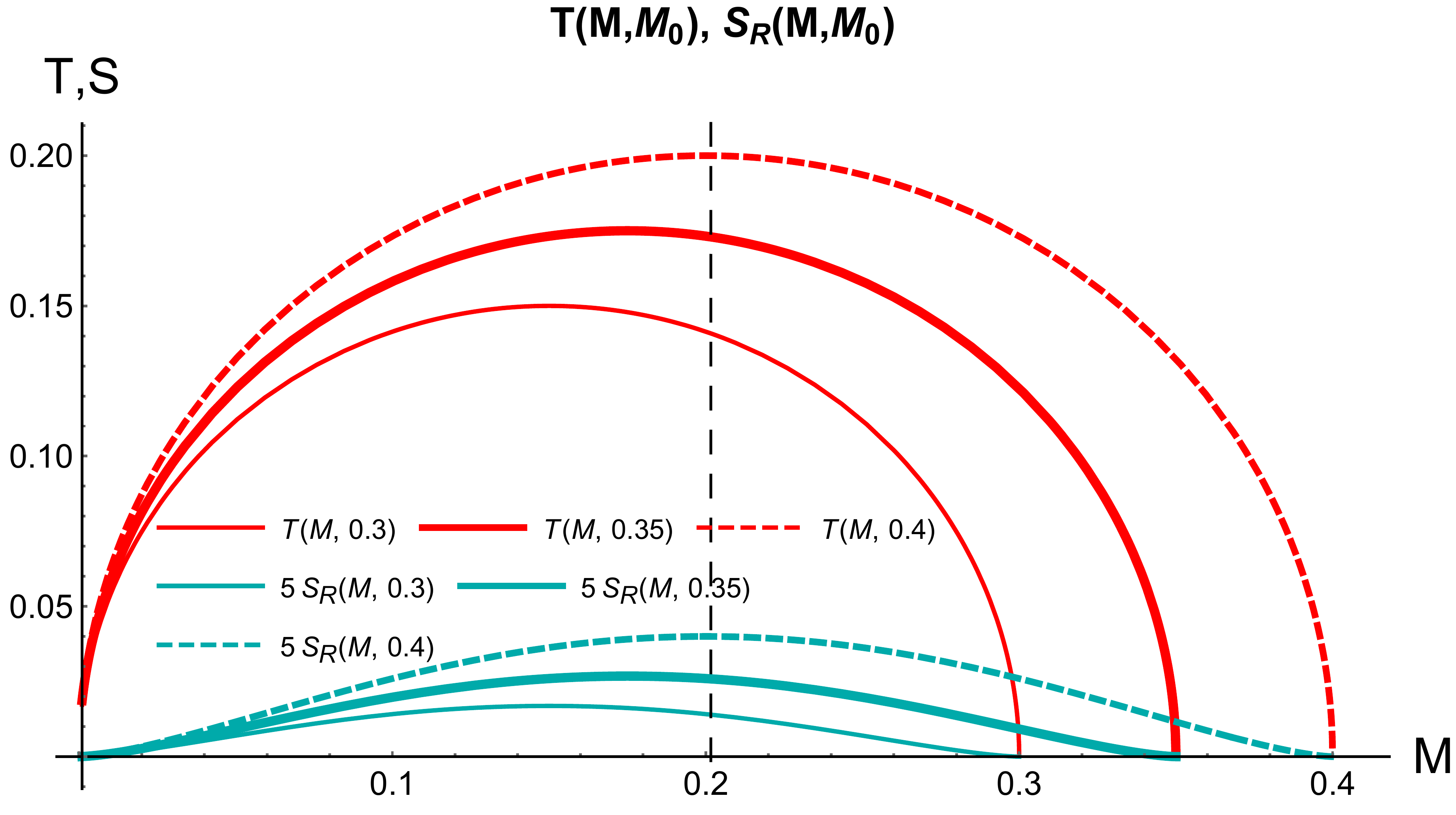} \qquad
 \includegraphics[scale=0.25]{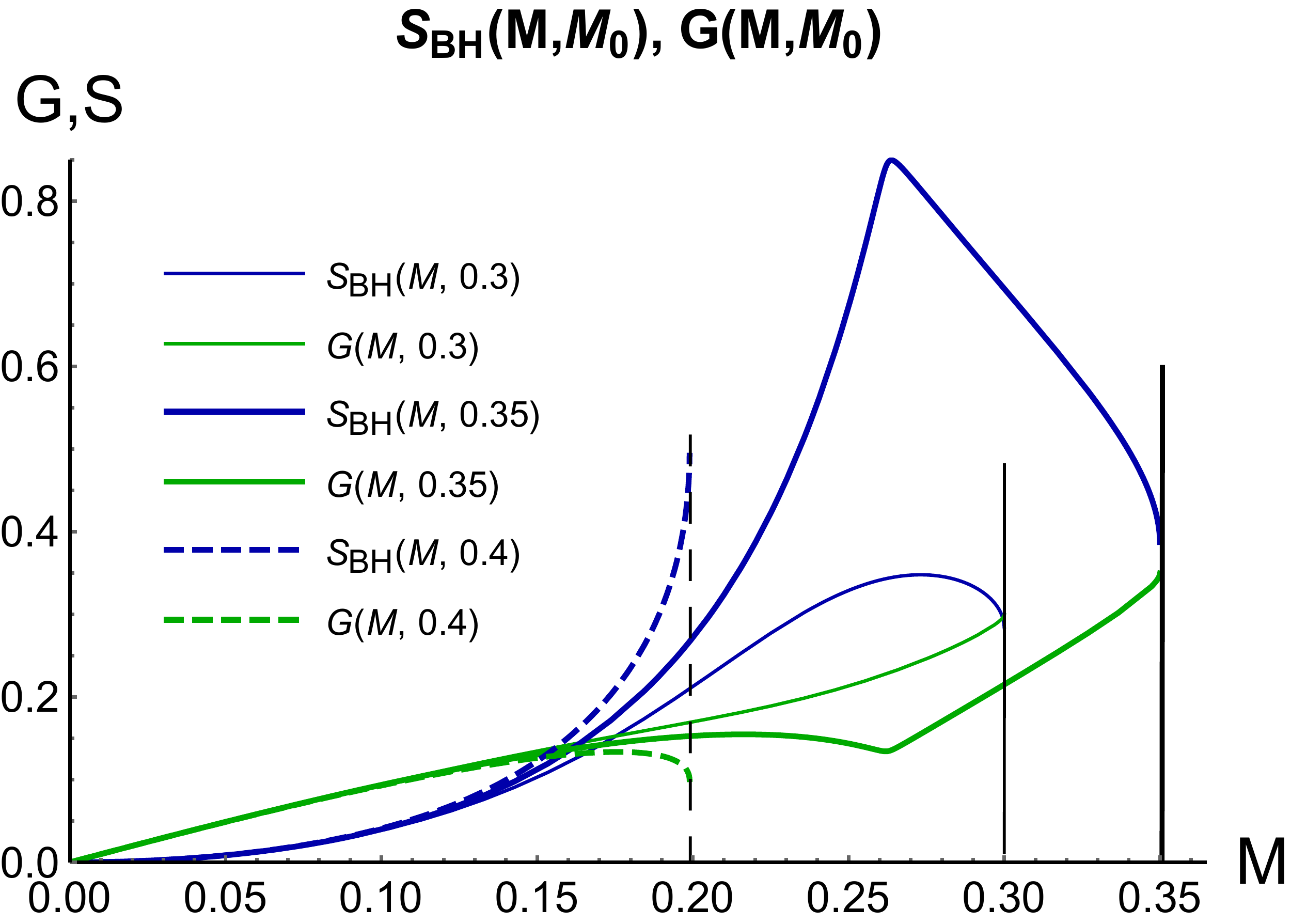}   
  \\
                  {\bf A }
\qquad\qquad\qquad\qquad\qquad \qquad  \qquad \qquad{\bf B }\\
\caption{A)
The plot shows mass dependences of the temperature (red) and the entropy of radiation (cyan)  for three different $M_0$: $M_0=0.3$, $M_0=0.35$ and $M_0=0.4$. In the first two cases (solid lines) $M\leq M_0$, and in the third one (dashed line) $M<M_{cr}(M_0)$, $M_0>M_{0,cr}=0.3501$.
B) 
 The plot shows  mass dependences of the entropy (blue) and free energy (green) of the Reissner-Nordstrom  black hole  for the same $M_0$ as in A).  The vertical solid black lines indicate  the value of $M_0$ and  $M\leq M_0$ and the dashed vertical line shows  
 $M_{cr}(0.4)$. 
 }
  \label{fig:all}
\end{figure}

We have the following asymptotic behaviour of $T(M, M_0)$, $S_{BH}(M, M_0)$, $G(M, M_0)$ and $S_{R}(M, M_0)$
for small $M$
\bea
 T(M, M_0)&=&\sqrt{M} \sqrt{M_0}+\cO\left(M^{1/2}\right),\quad
S_{BH}(M, M_0)= \pi  M^2+\cO\left(M^{7/2}\right)\\
 G(M, M_0)&=& M+\cO\left(M^{5/2}\right), \qquad\quad\quad
S_{R}(M, M_0)=M^{3/2} M_0^{3/2}+\cO\left(M^{5/2}\right).
\eea
 \\
 
 The plot in Fig.\ref{fig:all}.A shows the dependences of the temperature and the entropy of radiation on $M\leq M_0$ for two different $M_0$,  
$M_0<M_{0,cr}$ and $M_0=M_{0,cr}\approx 0.35$. As has been noted above, there is no a real solution for $M_0>M_{0,cr}$.
The plot  in Fig.\ref{fig:all}.B  shows dependences of the entropy (blue) and free energy (green) of the Reissner-Nordstrom  black hole   on $M$ calculated  on the branch $\lambda_-$ for different $M_0\leq M_{0,cr}$.  The vertical black lines indicate  the value of $M_0$, $M\leq M_0$. 
 It is interesting to note that both entropies, the black hole entropy and the radiation entropy have the form suitable to realize the Page curves - they first increase from some initial values, in particular from zero values for the case of the radiation entropy, up to some maximal value, then start to decrease  to zero  at $M\to 0.$ 


\subsection{Time evolution}\label{sect:TE}
The loss of the mass and charge during evaporation of RN black hole
is a subject of numerous consideration \cite{Gibbons-1875,Zaumen,Carter,Damour,
Page-1976,
Hiscock-1990,Gabriel-2000,Sorkin:2001hf,
Ong:2019vnv} and refs therein.
\\

In the case of fixed relations \eqref{Q-M}, we  consider the following system of equations\bea
\label{Mdot}
\frac{dM}{dt} &=& -A\sigma T^4 + \frac{Q}{r_+}\frac{dQ}{dt},\\
\label{Qdot}
\frac{dQ}{dt} &=&
\frac{M-\lambda\lambda^{\prime}}{\sqrt{M^2-\lambda^2}}
\frac{dM}{dt},\eea
where $A$ is a positive constant and the cross-section $\sigma$ is proportional to $M^2$ for small $M$. 
The first equation in the system of equations \eqref{Mdot} and \eqref{Qdot} coincides with the equation considered in \cite{Hiscock-1990}, and the second is obtained by simply differentiating the relation \eqref{Q-M}.
From \eqref{Mdot} and \eqref{Qdot} we get
\bea\label{ddM}
\frac{dM}{dt} &=& - \frac{A\sigma T^4(M+\lambda)}{\lambda (1+\lambda^{'})}.
\eea
For $\lambda(M)=CM^{\gamma}$ and small  $M$  one gets 
 \be
\frac{dM}{dt} =-C_1 M^{3\gamma-5}.\ee This equation  for  $\gamma  > 2$ and $M(0)=M_0$ has a solution
\bea
M(t)&=& \frac{M_0}{  \left(1+B\,
   t\right)^{\frac{1}{3(
   \gamma -2)}}}, \quad B=\frac{3(\gamma-2)C_1}{M_0^{6-3
   \gamma }},  \quad \gamma > 2,
   \eea
where $M_0$ and $B$ are positive constants. 
For $\gamma=2$  we have
\be
M(t)=M_0 e^{-C_1 t}.\ee

One gets an infinite large time of the complete evaporation of charged black hole under constraint
\eqref{Q-M}.
\\

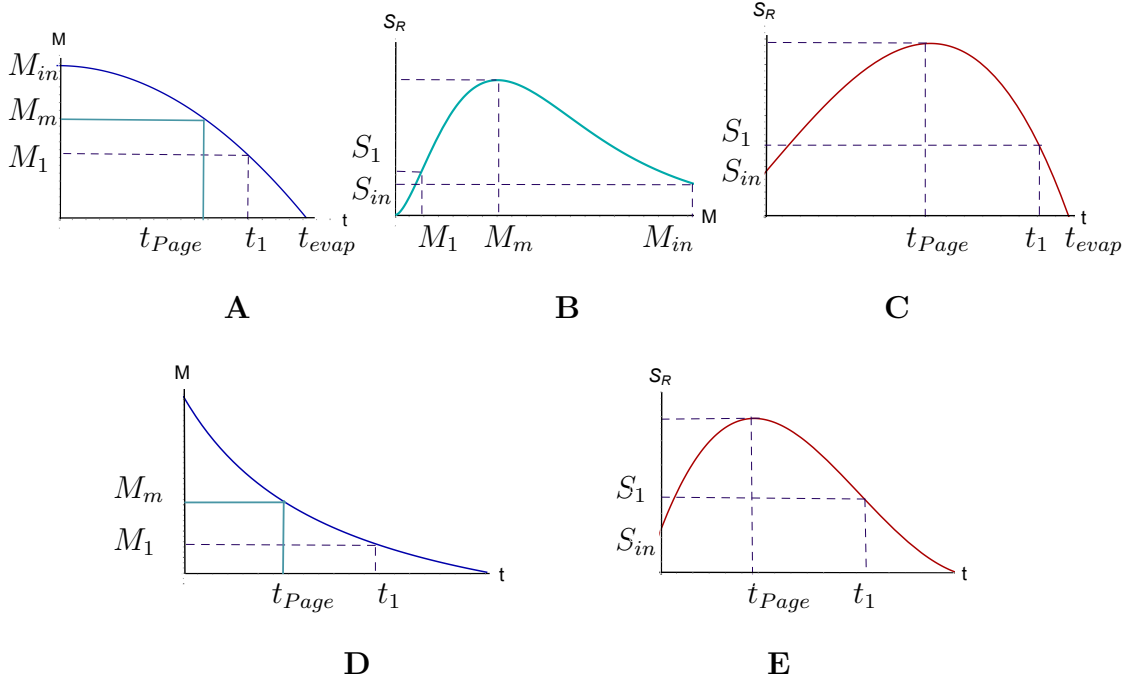
\begin{figure}[t]
 \centering
 \begin{picture}(0,0)
 \put(35,-5){$\Large{t_{Page}}$}
 \put(75,-5){$\Large{t_{1}}$}
 \put(95,-5){$\Large{t_{evap}}$}
 \put(140,-5){$\Large{M_{1}}$}
 \put(165,-5){$\Large{M_{m}}$}
 \put(225,-5){$\Large{M_{in}}$}
 \put(-15,60){$\Large{M_{in}}$}
 \put(-15,25){$\Large{M_{1}}$}
 \put(-15,43){$\Large{M_{m}}$}
 \put(370,-5){$\Large{t_{1}}$}
 \put(255,35){$\Large{S_1}$}
  \put(255,20){$\Large{S_{in}}$}
 \put(115,29){$\Large{S_1}$}
  \put(115,13){$\Large{S_{in}}$}
 \put(385,-5){$\Large{t_{evap}}$}
 \put(325,-5){$\Large{t_{Page}}$}
 \put(85,-140){$\Large{t_{Page}}$}
 \put(125,-140){$\Large{t_{1}}$}
 \put(265,-140){$\Large{t_{Page}}$}
 \put(305,-140){$\Large{t_{1}}$}
 \put(25,-100){$\Large{M_{m}}$}
 \put(25,-120){$\Large{M_{1}}$}
  \put(215,-120){$\Large{S_{in}}$}
 \put(215,-100){$\Large{S_{1}}$}
 \end{picture}\includegraphics[width=40mm]{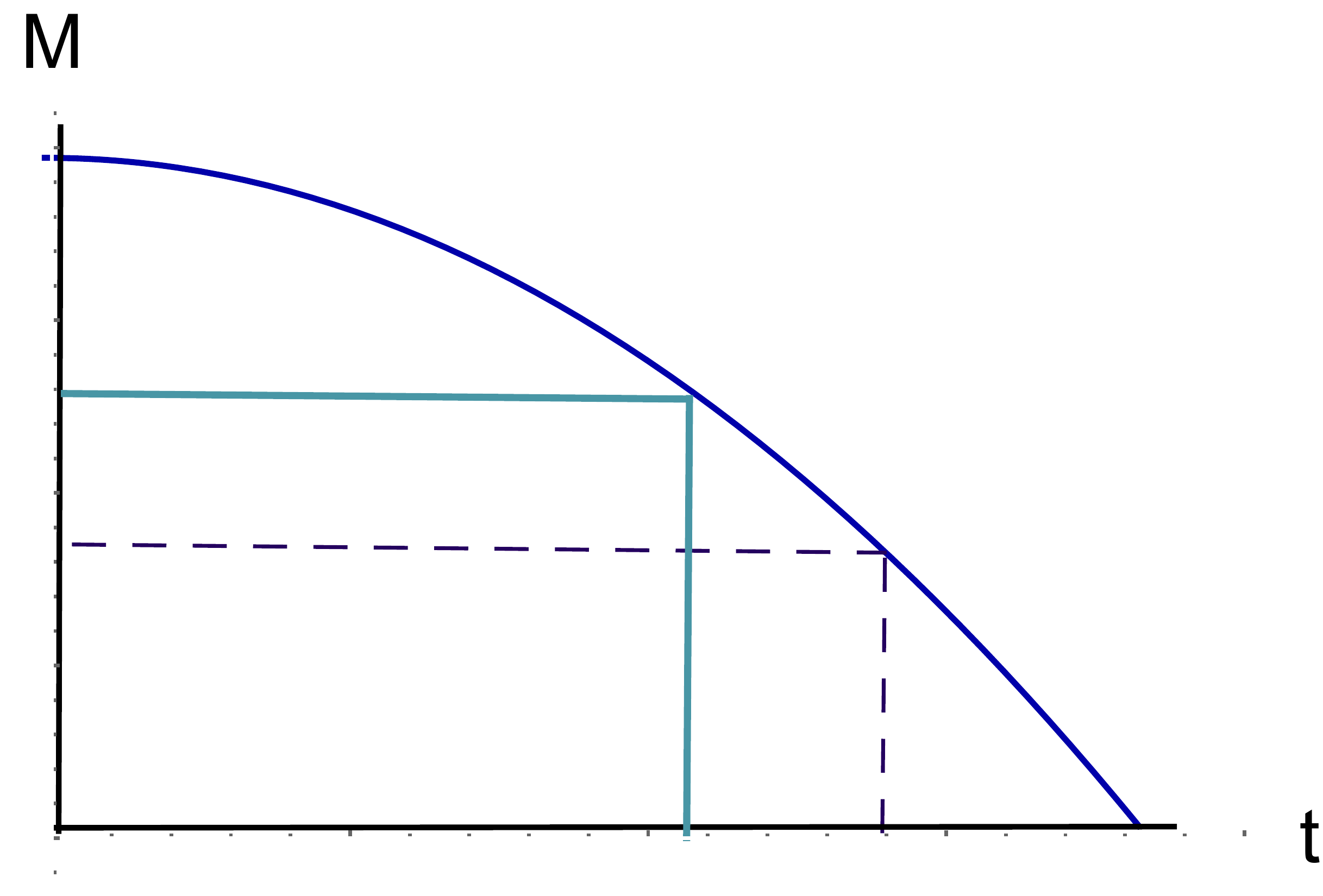}\quad\includegraphics[width=45mm]{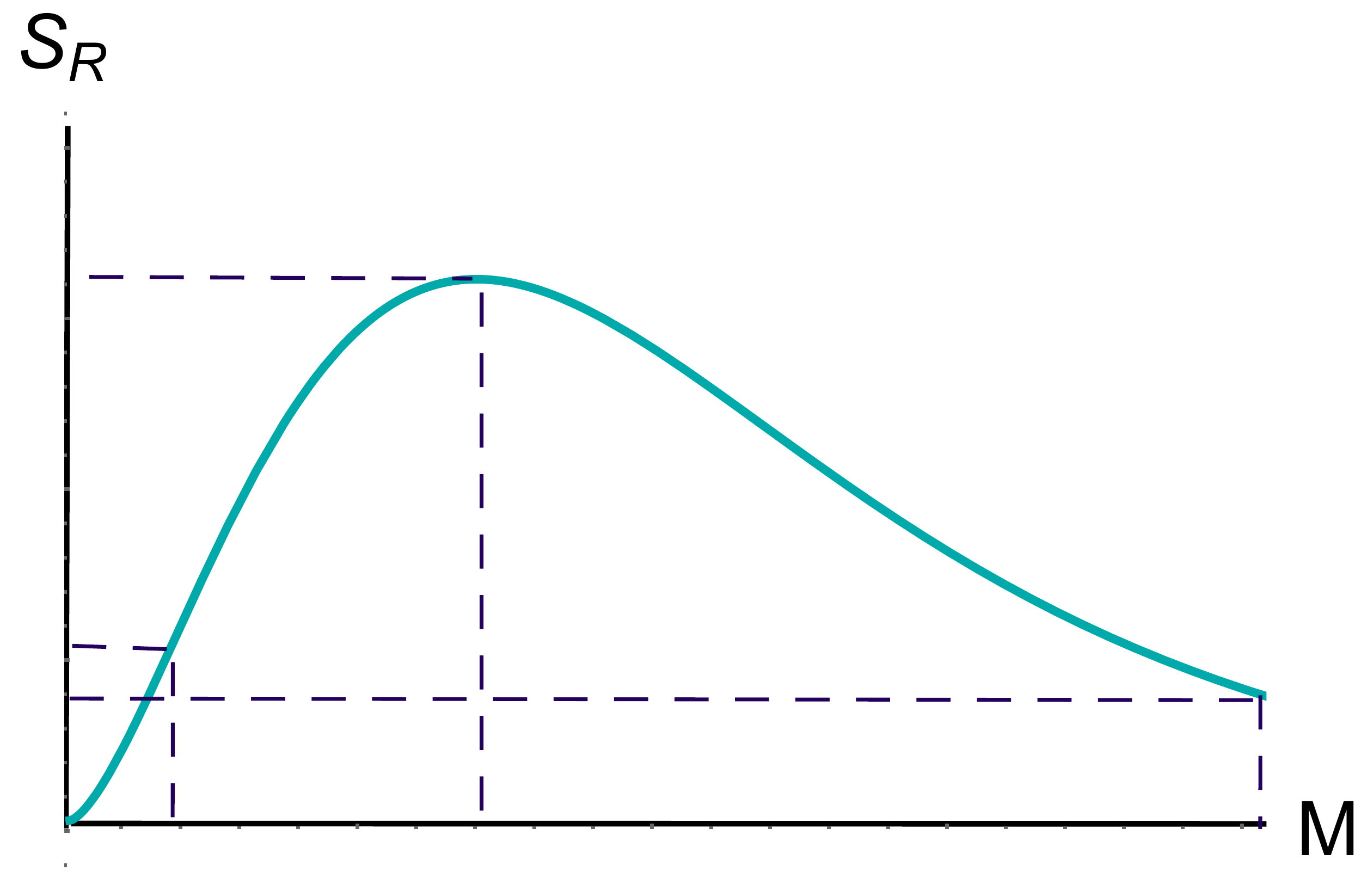}\quad\includegraphics[width=45mm]{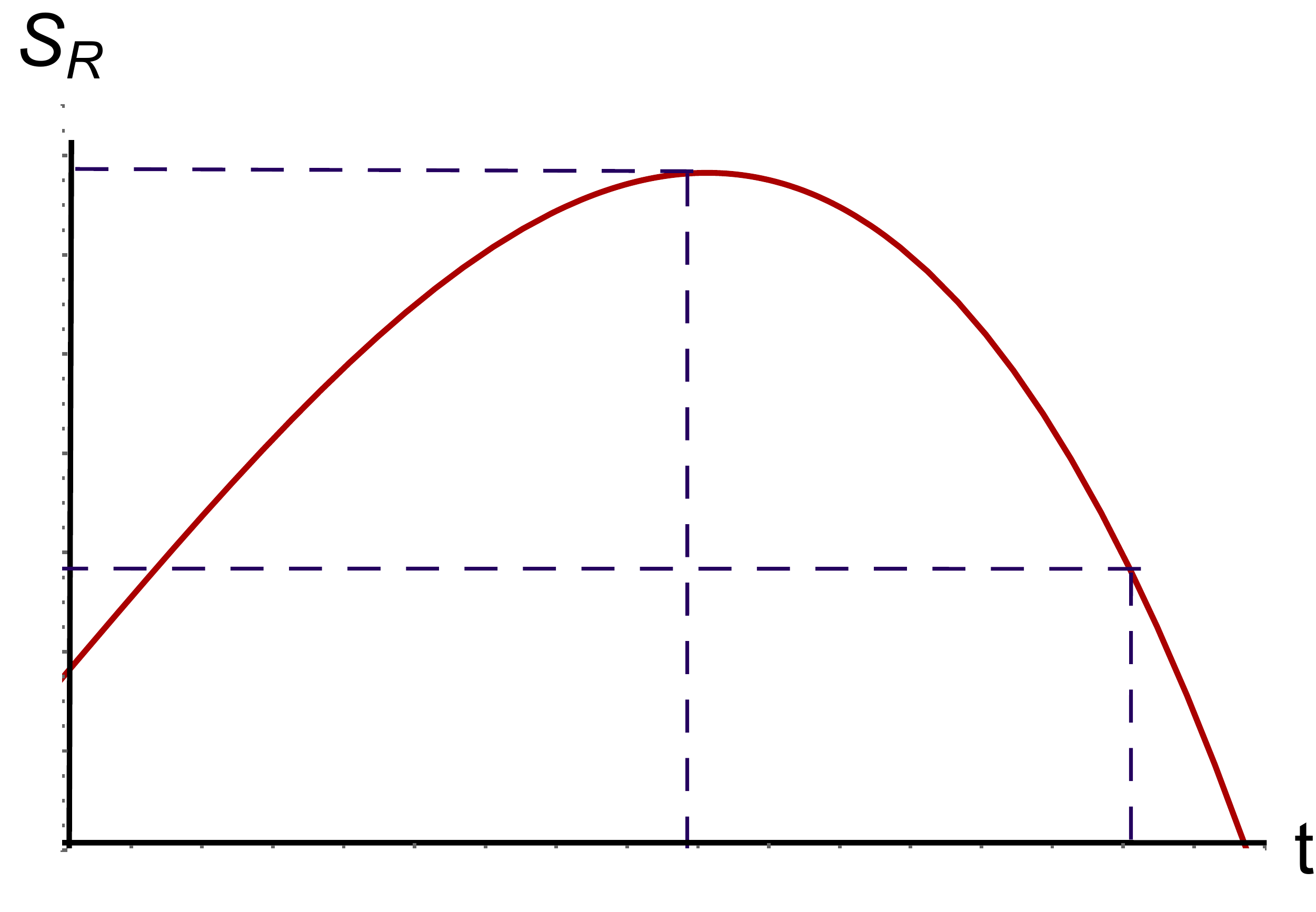}\\$\,$\\{\bf A}\hskip115pt {\bf B}\hskip115pt {\bf C}	\\$\,$\\\includegraphics[width=45mm]{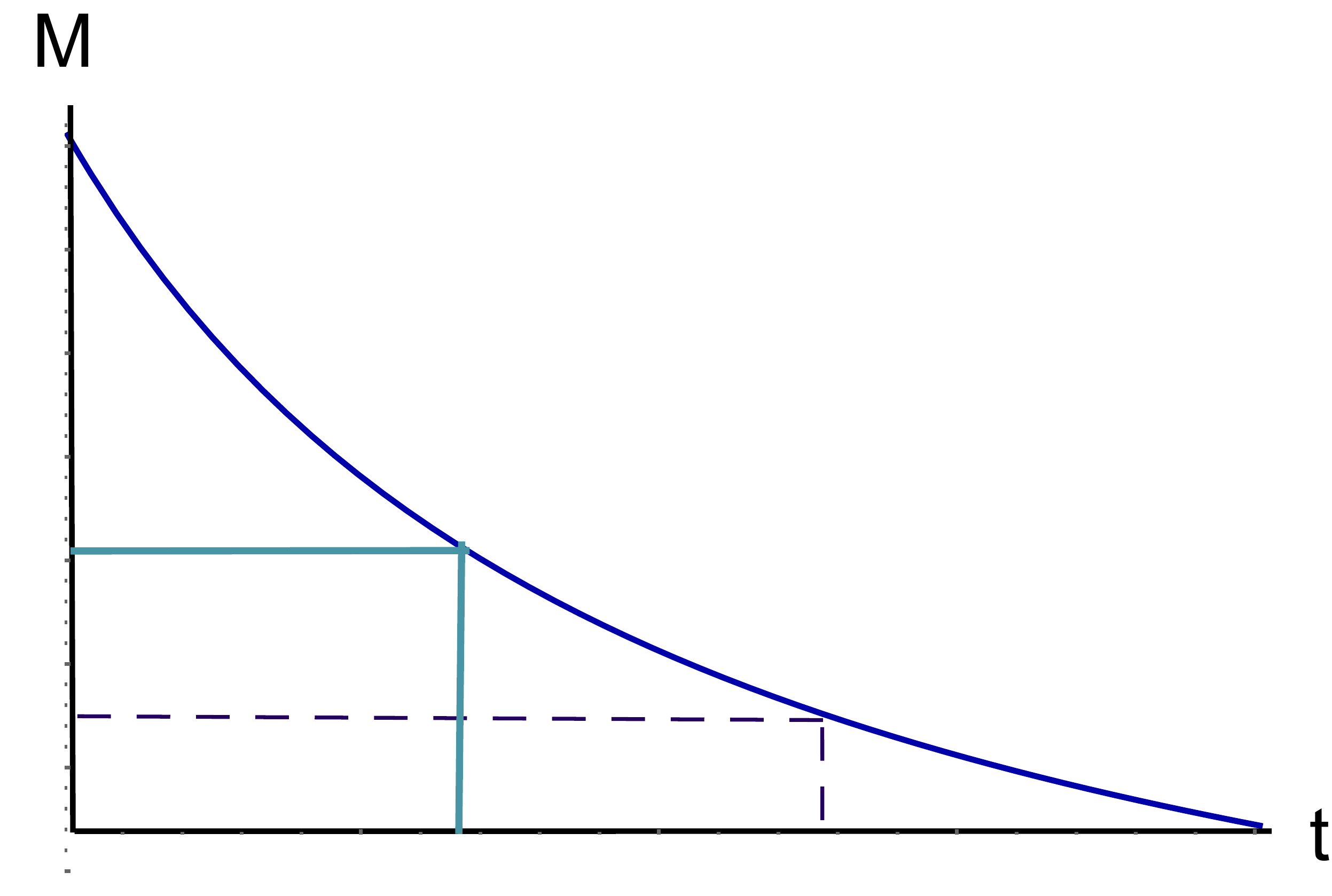}\qquad\qquad\includegraphics[width=45mm]{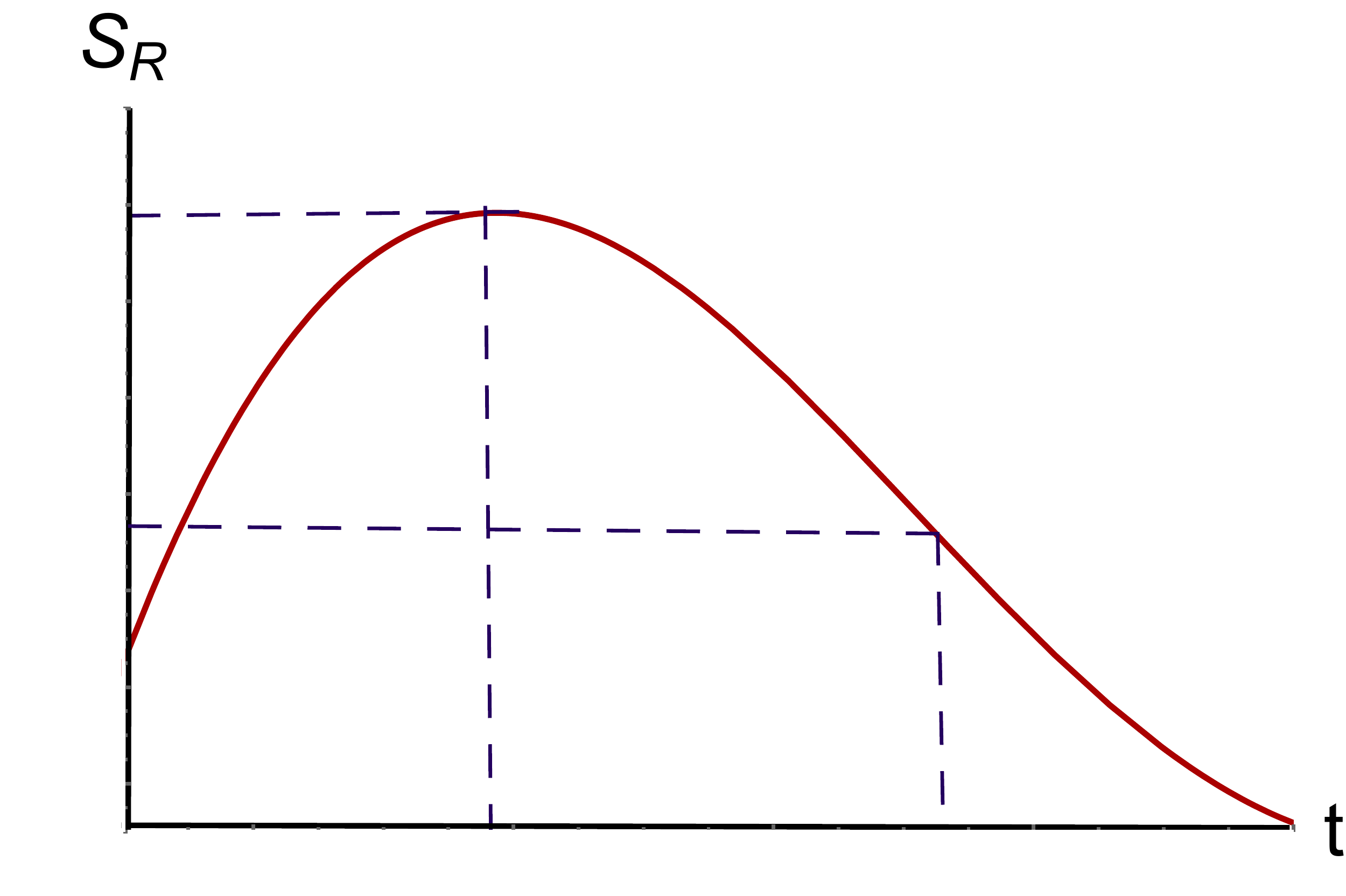}\\$\,$\\{\bf D}\hskip150pt {\bf E}	\\$\,$\\
 \caption{Page  curves for  radiation entropy and mass dependencies of the radiation entropy. 
         A) Time dependence of the mass of a black hole that evaporates in a finite time. 
        B) Dependence of the radiation entropy of on the mass of a black hole. 
       C) Time dependence of the radiation entropy of a black hole evaporating in a finite time.
      D) Time dependence of the black hole mass,
which evaporates infinitely. 
     E) Time dependence of the radiation entropy of a black hole which evaporates indefinitely. 
    	}
   	\label{fig:Page-curve-mass-time}
  \end{figure}

In Fig.\ref{fig:Page-curve-mass-time} we present two different regimes  of mass versus time, 
Fig.\ref{fig:Page-curve-mass-time}.A and Fig.\ref{fig:Page-curve-mass-time}.D, the former with a finite decay time and the latter with an infinite decay time. Because of the dependence of radiation entropy on mass shown in fig.\ref{fig:Page-curve-mass-time}.B, these two different types of time dependence on mass result in two different entropy versus time dependences, Fig.\ref{fig:Page-curve-mass-time}.C and
Fig.\ref{fig:Page-curve-mass-time}.E. But these two graphs do not differ significantly if we consider them on the interval $(0,t_1)$, here $t_{Page}<t_1$.
In both cases, entropy first increases with time, then decreases after Page time $t_{Page}$.
Note that if $M_m>M_1>M_{Plack}$, we can ignore the effects of quantum gravity in the above consideration.

\newpage
\section{Complete evaporation of the Kerr black hole }\label{sect:Kerr}

The  Kerr metric in Boyer-Linquist coordinates reads
\begin{multline}
 ds^2 = -\frac{\Delta-a^2 \sin^2 \theta}{\Sigma}dt^2 - 2a\sin^2 \theta
  \frac{r^2+a^2-\Delta}{\Sigma}dtd\phi\\
  + \frac{(r^2+a^2)^2-\Delta a^2
  \sin^2 \theta}{\Sigma}\sin^2 \theta d\phi^2 +
  \frac{\Sigma}{\Delta}dr^2 + \Sigma d\theta^2 
\end{multline}
where 
\begin{equation}
 \begin{split}
  \Sigma &= r^2 + a^2 \cos^2 \theta \ , \\
  \Delta &= r^2 - 2Mr + a^2 = (r-r_+)(r-r_-)\ .
 \end{split}
\end{equation}
The outer and inner horizon are located at $r=r_+,r_-$ 
respectively and
\be
r_\pm =M\pm\sqrt{M^2-a^2}
\ee
The temperature of the Kerr black hole is
\be
\label{T-kerr}
T_{Kerr}=\frac{1}{4\pi}\frac{\sqrt{M^2-a^2}}{M(M+\sqrt{M^2-a^2})}
\ee
When $M$  equal to $a$  one gets an extremal black hole with $T=0$.
\\

If the  angular momentum  $a<M$ also satisfies  the bound
\be
\label{a>M}
a>M-C M^3,\quad C>0,\ee
for small mass $M$, then the Hawking temperature 
$T_{Kerr}$
tends to 0 when $M\to 0$. 
\\

If we take $a$ to be a function of $M$ of the form
\be
\label{a-M-lambda}
a^2=M^2-\lambda(M)^2,
\ee 
where the function $0<\lambda(M)\leq M$, 
then the temperature    $T$ becomes equal to
\be
\label{T-Kerr-l}
T_{Kerr}=\frac{\lambda(M)}{4\pi M(M+\lambda(M))},
\ee
and the entropy and the free energy are
\bea\label{SS-Kerr}
S_{Kerr}&= & \pi r_+ ^2=
 \pi (M+\lambda (M))^2,\\
\label{F-Kerr}
G_{Kerr}&= &M-T_{Kerr}S_{Kerr}
=
M- \frac{\lambda(M)(M+\lambda (M))}{4 M}.
\eea
If $\lambda(M)=o(M^2)$ as $M\to 0$, then $T_{Kerr}\to 0$.
\\

Similar to the RN case, if one takes
 the function $\lambda(M)$ as
\be
\label{restr-Kerr}
\lambda (M)= \frac{M^\gamma}{A+M^{\gamma-1}},\,\,\,\,\,\,A>0,\,\gamma>3,
\ee
then the temperature   $T\to 0$  and $a\to 0$ also for $M\to\infty$.

\subsection{Examples}
\subsubsection{Deformed bell-shaped evaporation curves}
\begin{figure}[t!]
  \centering
 \includegraphics[scale=0.23]{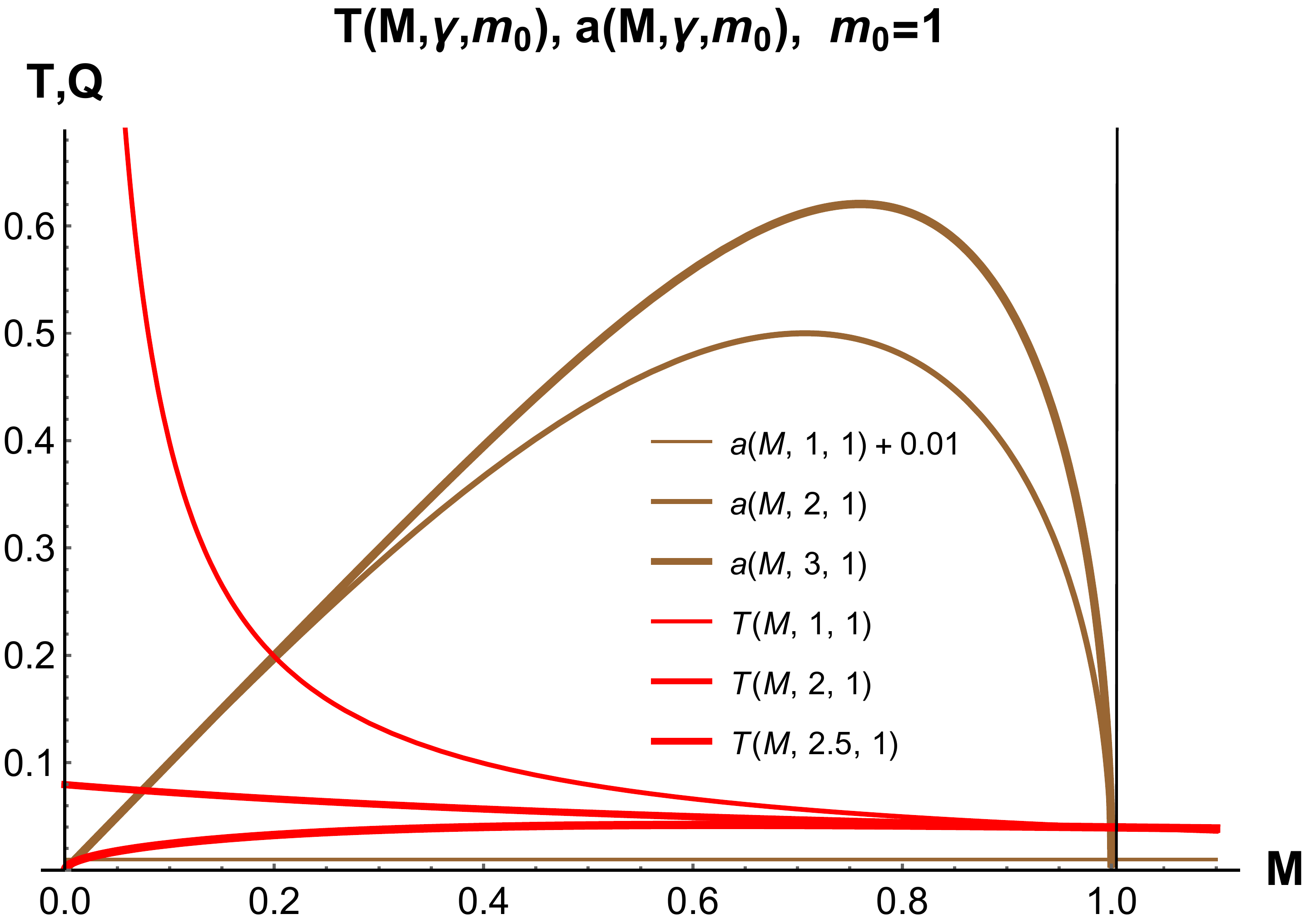} \qquad
  \includegraphics[scale=0.15]
  {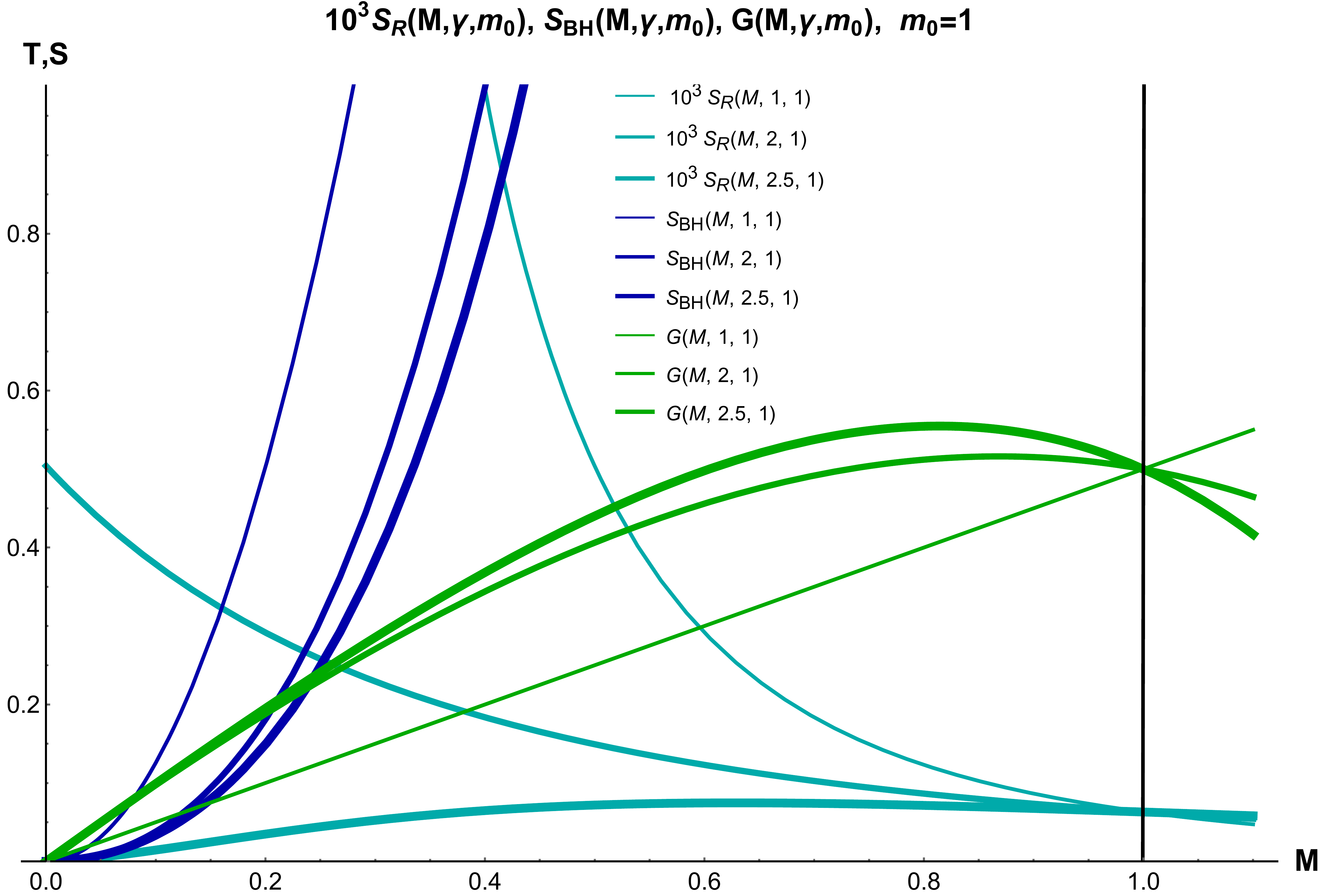} \\
  {\bf A}\hskip190pt{\bf B}
   \caption{The plot shows the dependence of the temperature $T$ (red), the angular momentum $a$(brown), the free energy $G$(green), 
   the black hole entropy $S$ (blue) and  the radiation entropy (cyan)  for  the curves \eqref{a-M-lambda} with   function $\lambda$ given by \eqref{lambda-gamma} with different scaling parameter $\gamma=1,2,2.5,3$ and $\mu=1$. 
   }
  \label{fig:Kerr}
\end{figure}

As in the previous Sect.\ref{sect:RN}, we first consider the case \eqref{lambda-gamma}. The  behaviour of  the charge $a$,  the temperature $T$, the entropy and the free energy as functions of  $M$ for the curves
 \eqref{a-M-lambda}  with different scaling parameter $\gamma$ and the same parameter $\mu=1$ are presented in Fig.\ref{fig:Kerr}.
 These plots are very similar to the plots presented in Fig.\ref{fig:RN} for the RN case. 
  For all $\gamma> 1$ there is a restriction on 
$M$,  $M\leq 1$ and  the temperature and the radiation entropy go to zero as $M\to 0$ for $\gamma>2$, to a non-zero constant for $\gamma=2$ and to the infinity for $\gamma<2$. Moreover, the radiation entropy at $\gamma>2$, starting from the initial value at $M=1$, first increases to a certain maximum value, then decreases to zero, i.e. has a form that, in the case of a slow decrease in mass during evaporation, provides the form of the Page dependence of the radiation entropy  on time, see Fig.\ref{fig:Page-curve-mass-time}.  The black hole entropy (blue lines, Fig.\ref{fig:RN}.B) and the free energy (green lines, Fig.\ref{fig:RN}.B) go  to zero when
 $M\to 0$ for $\gamma \geq 1$.   We also see, Fig.\ref{fig:Kerr}.A,   that 
 the form of dependence of $a$ on $M$ is a deformed bell,
 compare with Fig.\ref{fig:RN}.A.
\\

 \newpage
 \subsubsection{Semi-circle  dependence of temperature on mass $M$}  
 If the function $T(M)$ is given, then from equation \eqref{T-Kerr-l} we get
\be
\label{lambda-M-K}
\lambda(M)=\frac{4 \pi  M^2 T(M)}{1-4 \pi  M T(M)}.\ee
Unlike  \eqref{T-lambda}, now the equation relating $T$ and $\lambda$, i.e. equation \eqref{T-Kerr-l}, is linear on $\lambda$ 
which gives us  
\eqref{lambda-M-K}.
Assuming the temperature dependence on $M$ is the same as for the RN case,  \eqref{T-M-M0},
we get
\be\label{Kerr-lambda-m}
\lambda(M,M_0)=\frac{4 \pi  M^2 \sqrt{M(M_0-M)}}{1-4 \pi  M \sqrt{M(M_0-M)}}.\ee
Note that the explicit forms of $\lambda(M)$ in the RN and Kerr cases are different, compare \eqref{lam-pm} with \eqref{lambda-M-K}.
\\

\begin{figure}[t]
  \centering
    \begin{picture}(0,0)
\put(68.5,4){\color{blue}
{\Huge$\mathbb *$}}
\put(309,6){\color{blue}
{\Huge$\mathbb *$}}
\put(68,-82){\color{blue}
{\Huge$\mathbb *$}}
\end{picture}
   \includegraphics[scale=0.25]{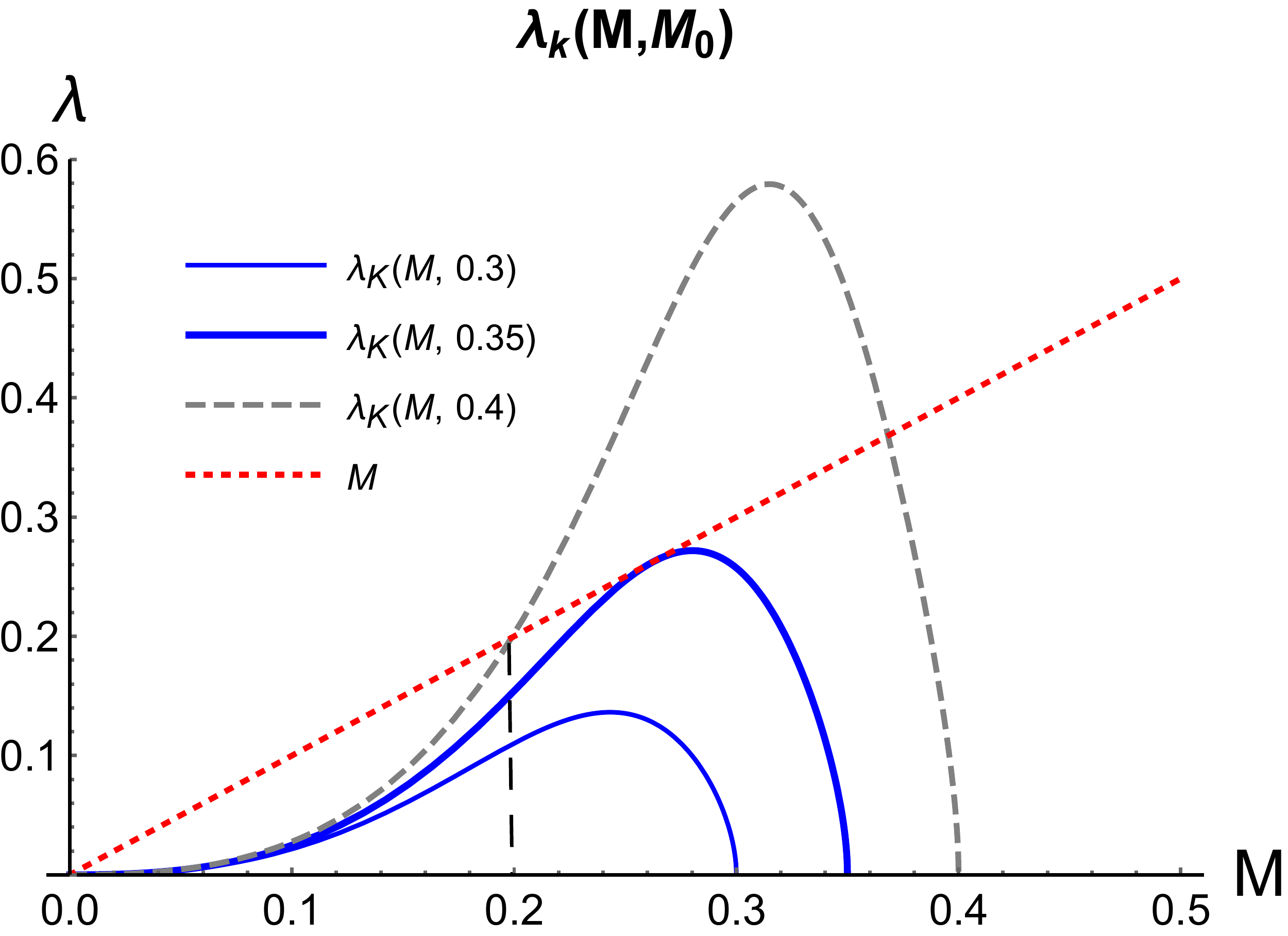}\qquad
   \includegraphics[scale=0.27]{plots/EOS-RN.pdf} \\
   {\bf A }\qquad\qquad \qquad  \qquad\qquad  \qquad{\bf B }\\$\,$\\
\includegraphics[scale=0.18]{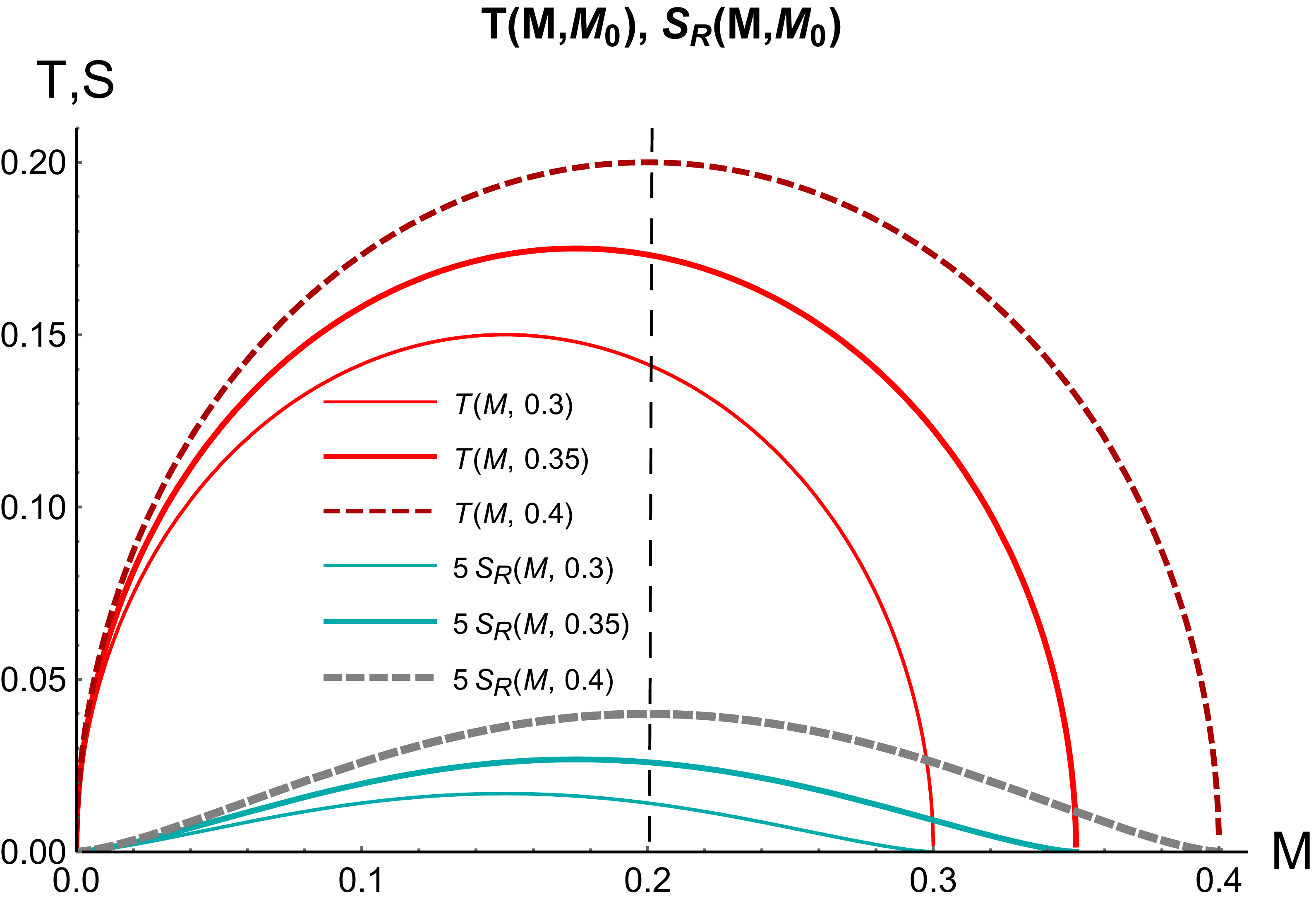}\qquad\qquad
\includegraphics[scale=0.23]{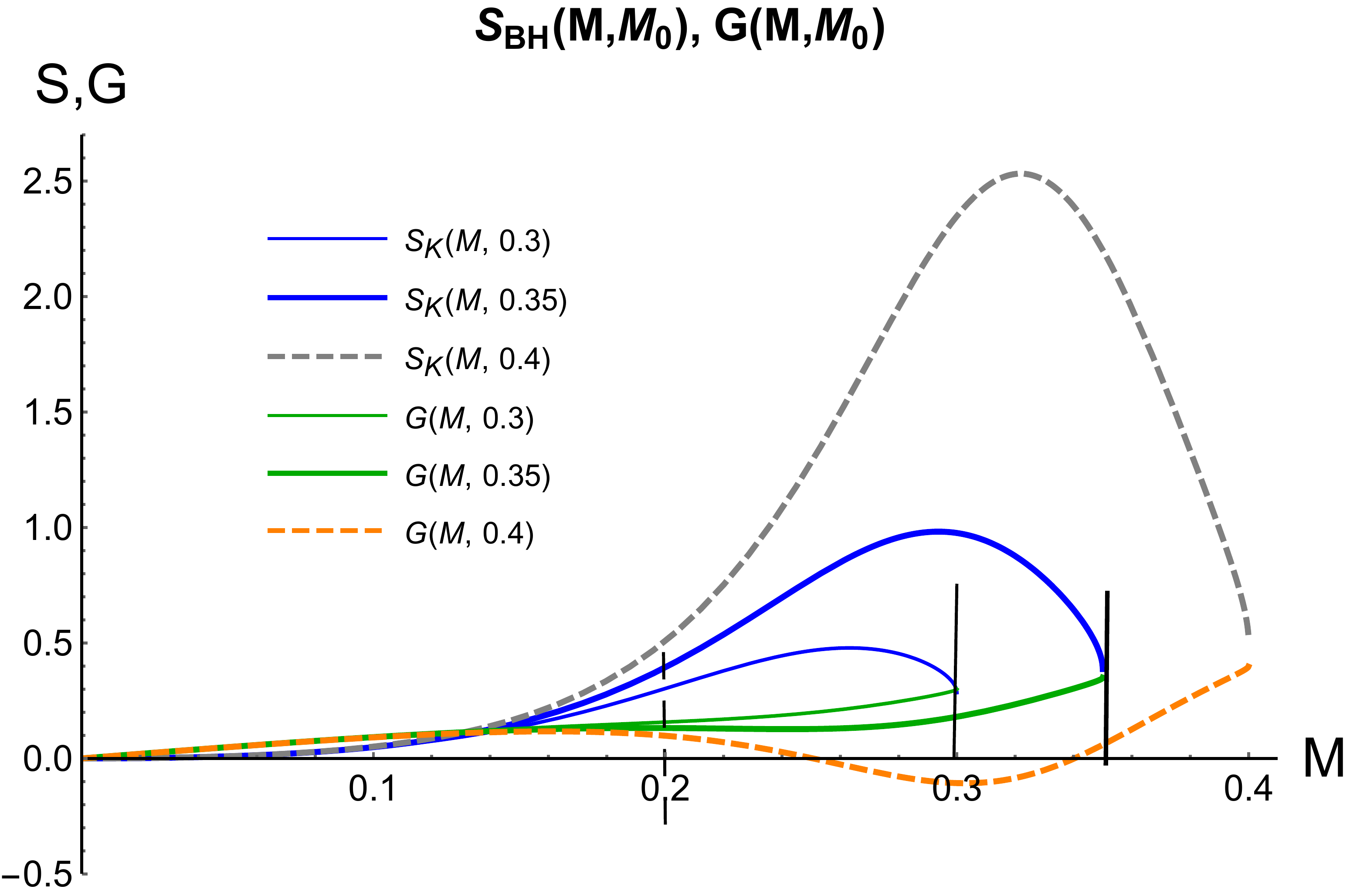}
 \\
{\bf C } \qquad\qquad \qquad  \qquad\qquad  \qquad{\bf D }\\

   \caption{A) The plots show the dependence of $\lambda$ on $M$ given by \eqref{Kerr-lambda-m} for $M_0=0.3,0.35,0.4$.
   B) The plots show curves corresponding to the curves  presented in A).  Plots in C) show the mass dependencies of 
   the  temperature and the radiation entropy for the curves  presented in B), and  the mass dependencies of the  free energy  and the black hole entropy 
 for these curves  are presented in D).
 }
  \label{fig:semi-circle-Kerr}
\end{figure}

For small $M_{0}$ from \eqref{Kerr-lambda-m} follows that $\lambda(M,M_0)<M$ and the constraint \eqref{a-M-lambda} corresponds to positive values of $a^2$. At the critical value of $M_{0,cr}$ the following equation 
\be
\label{cr-Kerr}
M^2-\frac{16 \pi ^2 M^5 (M_0-M)}{\left(1-4 \pi  M
   \sqrt{M (M_0-M)}\right)^2}=0,
   \ee
has one real solution, $M_{0,cr}\approx 0.35003$, and $M_1\approx0.261$. For $M_0>M_{0,cr}$ equation \eqref{cr-Kerr}
has two real solutions, for example, for $M_0=0.4$ shown in Fig.\ref{fig:semi-circle-Kerr}, there are two real solutions $M_{1}=0.199$,
$M_{2}=0.368$. Indeed, 
in the plots in Fig.\ref{fig:semi-circle-Kerr}.A  we see that $\lambda(M,M_0)<M$ for $M_0<0,35$ and for $M_0>0,4$ there are two intersections of the gray dashed line represented $\lambda(M,M_0)$  with the red dotted line. The case of one real root of equation \eqref{cr-Kerr} is shown by thick blue line in  Fig.\ref{fig:semi-circle-Kerr}.A and thick blue line in  Fig.\ref{fig:semi-circle-Kerr}.B.
\\

The plots in Fig.\ref{fig:semi-circle-Kerr}.C  show mass dependences   of the temperature $T(M,M_0)$ and the entropy of radiation, $S(M,M_0)$ and the plot in Fig.\ref{fig:semi-circle-Kerr}.D show mass dependences of the black hole entropy and the free energy on $M$ for different $M_0$. 

 \newpage
 \newpage
\section{Complete evaporation of the Kerr-Newman black hole}\label{sect:KN}

The Kerr-Newman (KN) metric is 
\bea
\label{ds-KN}
ds^2 &=& - \frac{1}{\rho ^2}\left( {\Delta _r -  a^2\sin
^2\theta } \right)dt^2 + \frac{\rho ^2}{\Delta _r }dr^2 + \rho
^2d\theta ^2
\\
&+& \frac{1}{\rho ^2}\left[ { (r^2 + a^2) - \Delta
_r a^2\sin ^2\theta } \right]\sin ^2\theta d\varphi ^2
 - \frac{2a}{\rho ^2}\left[ { (r^2 + a^2) - \Delta _r }
\right]\sin ^2\theta dtd\varphi ,\nn
\eea
where
$\rho ^2= r^2 + a^2\cos ^2\theta ,
\quad
 \Delta _r = (r^2 + a^2) - 2Mr + q^2$. 
Here the parameters $M$, $a$ and $q$ are the
mass, the angular momentum, and the charge  of the black hole,
respectively.
The electromagnetic potential is 
\begin{equation}
\label{A-KN} A_\mu = - \frac{qr}{\rho ^2}\left( {1,0,0, - a\sin
^2\theta } \right).
\end{equation}

There are two horizons $r_\pm$ and 
\bea
\label{r+KN}  r_ +=M+\sqrt{M^2-(a^2+q^2)}.
\eea
The Hawking temperature of the black hole horizon is given by 
\bea
\label{KN-T} T_ {KN} &=&
\frac{ r_ +   -  M
}{2\pi  (r_ + ^2 + a^2)}.
\eea

\begin{figure}[t!]
  \centering
\includegraphics[scale=0.27]{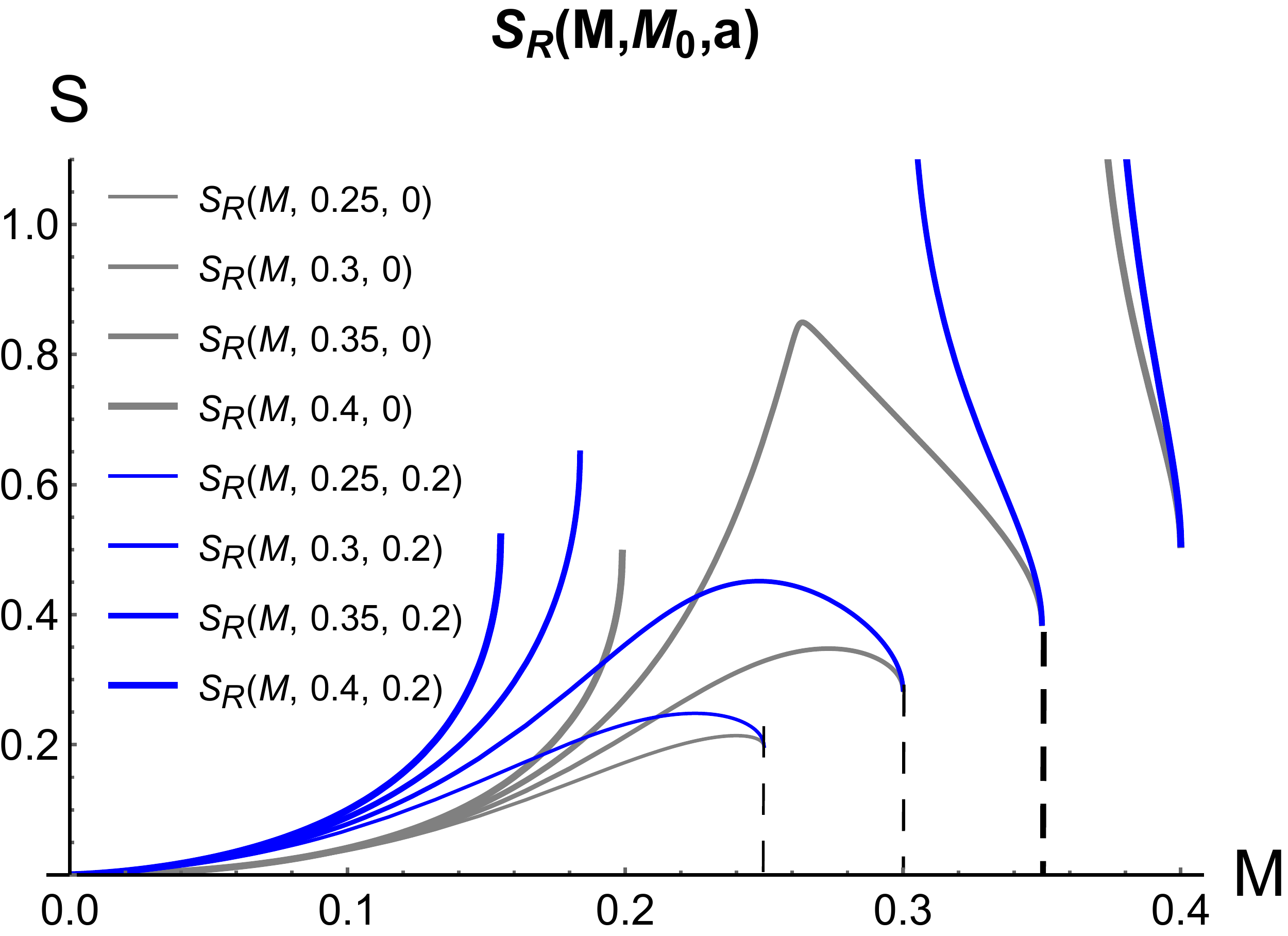}\qquad\qquad
\includegraphics[scale=0.27]{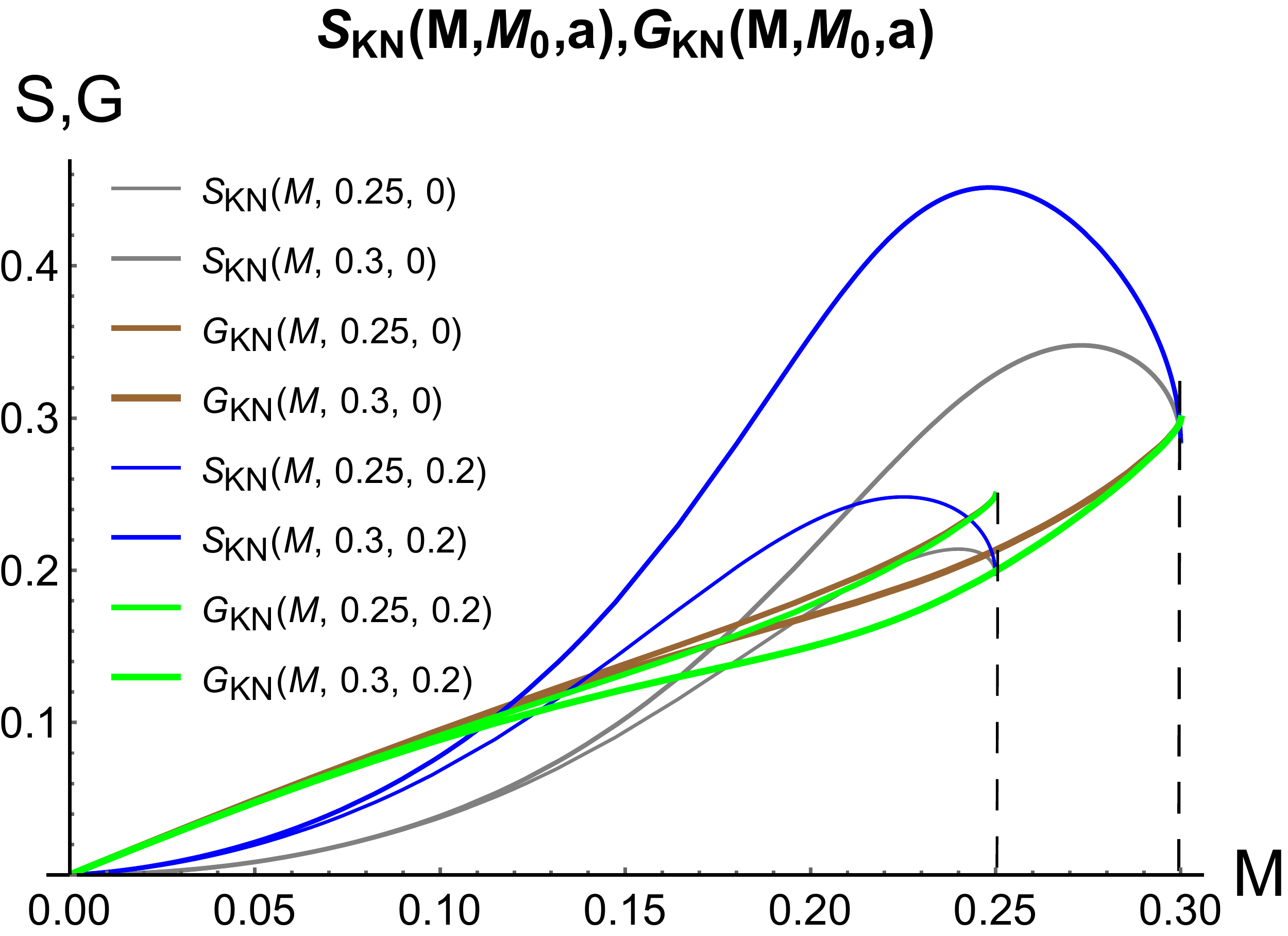}\\ {\bf A}\hskip220pt{\bf B}
  \caption{A) The KN black hole entropy as function of $M$ for special constraint providing the semi-circle dependence of temperature on $M$ for zero $a$ (gray lines)  and $a=0.2$ (blue lines). B) The KN black hole entropy and free energy as functions of $M$ for zero $a$ (gray lines for entropy and brown lines for free energy)  and $a=0.2$ (blue lines
  lines for the entropy and green lines for the free energy) for two choices of $M_0$ in eq.\eqref{T-sgrt}, $M_0=0.25,0.3$. To different choices of $M_0$ correspond the lines of different thickness. 
  }
  \label{fig:semi-circle-Newmann}
\end{figure}
We can put the constraint
\be
\label{KN-lambda}
a^2+q^2=M^2-\lambda ^2(M),\ee
where $\lambda (M)$ is a function of $M$. Substituting \eqref{KN-lambda} into \eqref{KN-T} we get
\bea
\label{KN-T-lambda}
 T_{KN}=
   \frac{\lambda(M)}{2 \pi 
   \left(2M^2+2\lambda M-q^2\right)}.
   \eea
    From the first equality in \eqref{KN-T-lambda} for $q=0$ we get
  \eqref{T-Kerr-l} and from the second  for $a=0$ we get \eqref{T-Kerr-l}.
   \\
   
Assuming that $\lambda (M)<C M^\gamma$, $\gamma>1$, from \eqref{KN-lambda} we get that $a^2+q^2=\cO(M^2)$.  The requirement that $T\to 0$ for $M\to 0$ forces us to assume  $\gamma>2$. 
  \\
  
Solving  equation \eqref{KN-T-lambda} in respect to $\lambda$
we get
\be
\label{KN-sol-lambda}
\lambda _{\pm}=\frac{\pm\sqrt{-16 \pi ^2 a^2 T^2-8 \pi  M
   T+1}-4 \pi  M T+1}{4 \pi  T}
\ee
 Note, that $q$ does not enter to \eqref{KN-sol-lambda}. This equation \eqref{KN-sol-lambda} is analogous to  equation \eqref{lam-pm}, which does not contain  $Q$. Equation \eqref{lam-pm} follows from \eqref{KN-sol-lambda} for $a=0$.
  The equivalent representation is 
  \be
\lambda(M) = \frac{2 \pi  T \left(2 M^2-q^2\right)}{1-4 \pi  M T},\ee
from which we get \eqref{lambda-M-K} for $q=0$.
 \\
 
Assuming that 
\be
\label{T-sgrt} T=\sqrt{M(M_0-M)}\ee
and substituting this expression in \eqref{KN-sol-lambda}, we get 
\be
\label{KN-lambda-semi-circle}
\lambda _{\pm}=\lambda_{\pm}(M,M_0,a).
\ee
$\lambda _{\pm}$ have different asymptotic for small $M$
\bea
\lambda _{+}
&=&\frac{1}{2 \pi  \sqrt{M_0 M}
   }+\cO\left(M^{1/2}\right),\\
 \lambda _{-}&
 =
 &2 \pi  a^2 \sqrt{M_0 M}+\cO\left(M^{3/2}\right),
\eea
and we take the second branch $ \lambda _{-}$. On this branch we find the form of the KN black hole entropy
and the free energy
\bea
S_{KN,-}&=&\pi  (M+\lambda_-)^2\\
G_{KN,-}&=&M-   \frac{\lambda_- (M+\lambda_-)^2}{2  
   \left(\left(\lambda_-+M\right)^2+a^2\right)}
\eea
The dependence of the entropy of radiation of KN black hole  on $M$ for the constraint  corresponding to   $ \lambda _{-}$, is presented in 
Fig.\ref{fig:semi-circle-Newmann}.A. The dependence of the KN black hole entropy on $M$ for the constraint  corresponding to   $ \lambda _{-}$, is presented in Fig.\ref{fig:semi-circle-Newmann}.B. Here these dependences are shown by  gray lines for  $a=0$ 
and  blue lines for $a=0.2$. 
In Fig.\ref{fig:semi-circle-Newmann}.B we show also  the free energy as functions of $M$ for zero $a$ (brown lines for free energy)  and $a=0.2$ (green lines for free energy) for two choices of $M_0$ in eq.\eqref{T-sgrt}, $M_0=0.25,0.3$. To different choices of $M_0$ correspond the lines of different thickness.
\section{Complete evaporation of the Schwarzschild-de Sitter \\ black hole}\label{sect:SdS}
The line element has the form
\be
ds^2=-f(r)dt^2+f(r)^{-1}dr^2+r^2d\Omega^2,
\ee
where
\be
f(r)=1-\frac{2M}{r}-\frac{\Lambda}{3}r^2,
\ee
$M$ is the mass of the black-hole and $\Lambda>0$ is  the positive cosmological constant.
For  
\be\label{M-L}
0<3M\sqrt{\Lambda}  <1 
\ee 
there are 3 horizons, the black hole horizon 
\bea
\label{rpSdS}
r_+&=&\frac{2}{\sqrt{\Lambda}}\sin\Big(\frac{1}{3}\arcsin 3M\sqrt{\Lambda}\Big),
\eea
the cosmological horizon
\bea
r_c&=&\frac{2}{\sqrt{\Lambda}}\sin \left(\frac{1}{6} \left(2 \arccos
   \left(3 \sqrt{\Lambda }
   M\right)+\pi \right)\right)\eea
   and the negative non-physical one. 
\\

 The Hawking temperature of Schwarzschild-de Sitter is
\be
T_{SdS}=\frac{1}{4 \pi r_{+}}\left(1-\Lambda r_{+}^{2}\right), 
\ee
where $r_+$ is given by \eqref{rpSdS}.
We can represent the temperature as 
\be
T_{SdS}(M,\Lambda)=\frac{1-4
   \sin ^2\left(\frac{1}{3} \arcsin
   \left(3 \sqrt{\Lambda }
   M\right)\right) }{8 \pi \sin
   \left(\frac{1}{3} \arcsin
   \left(3 \sqrt{\Lambda }
   M\right)\right)}\sqrt{\Lambda } \ee
   We see that for fixed $\Lambda$ the temperature becomes infinite when $M\to 0$.
   We note that the nominator is equal to zero at $\Lambda=1/9M^2$ and this value of $\Lambda$ realizes the bounded value  of $\Lambda$ admissible by inequality \eqref{M-L}.
   \\

   By analogy with the cases of RN and Kerr considered in the previous sections, we can consider $\Lambda$ to be dependent on $M$ and parametrize this dependence by a function $\lambda=\lambda(M)>0$, i.e.
   \be
   \Lambda=\frac{1-\lambda(M)^2}{9M^2}.
   \ee
   We assume that $\lambda=\lambda(M)$ satisfies the bounds $0<\lambda\leq 1$. One can check that if
   \be
   \label{lambda-dS}
   \lambda (M)=o(M),\,\,\,\,M\to 0,
   \ee
   then $T\to 0$ as $M\to 0$. 
   \\
   
   Indeed, since now the temperature is
   \be
T_{SdS}=\frac{\sqrt{1-\lambda^2}}{3M} \,\frac{1-4
   \sin ^2\left(\frac{\pi}{6} -\frac{1}{3}\arcsin
   \lambda\right) }{8 \pi \sin
   \left(\frac{\pi}{6} -\frac{1}{3}\arcsin
   \lambda\right)} \ee   
   and assuming \eqref{lambda-dS} we get the asymptotic expansion  for small $M$ as
   \bea 
   T_{SdS}&=&\frac{\lambda
   }{6 \sqrt{3} \pi  M}+\frac{\lambda ^2}{27 \pi  M}+{\it O}(\lambda^3).
   \eea 
We see that if $\lambda\sim M^{\gamma}$, $\gamma>1$, then $T\to 0$ when $M\to 0$.
\\
   
\begin{figure}[t!]
  \centering
\includegraphics[scale=0.7]
 {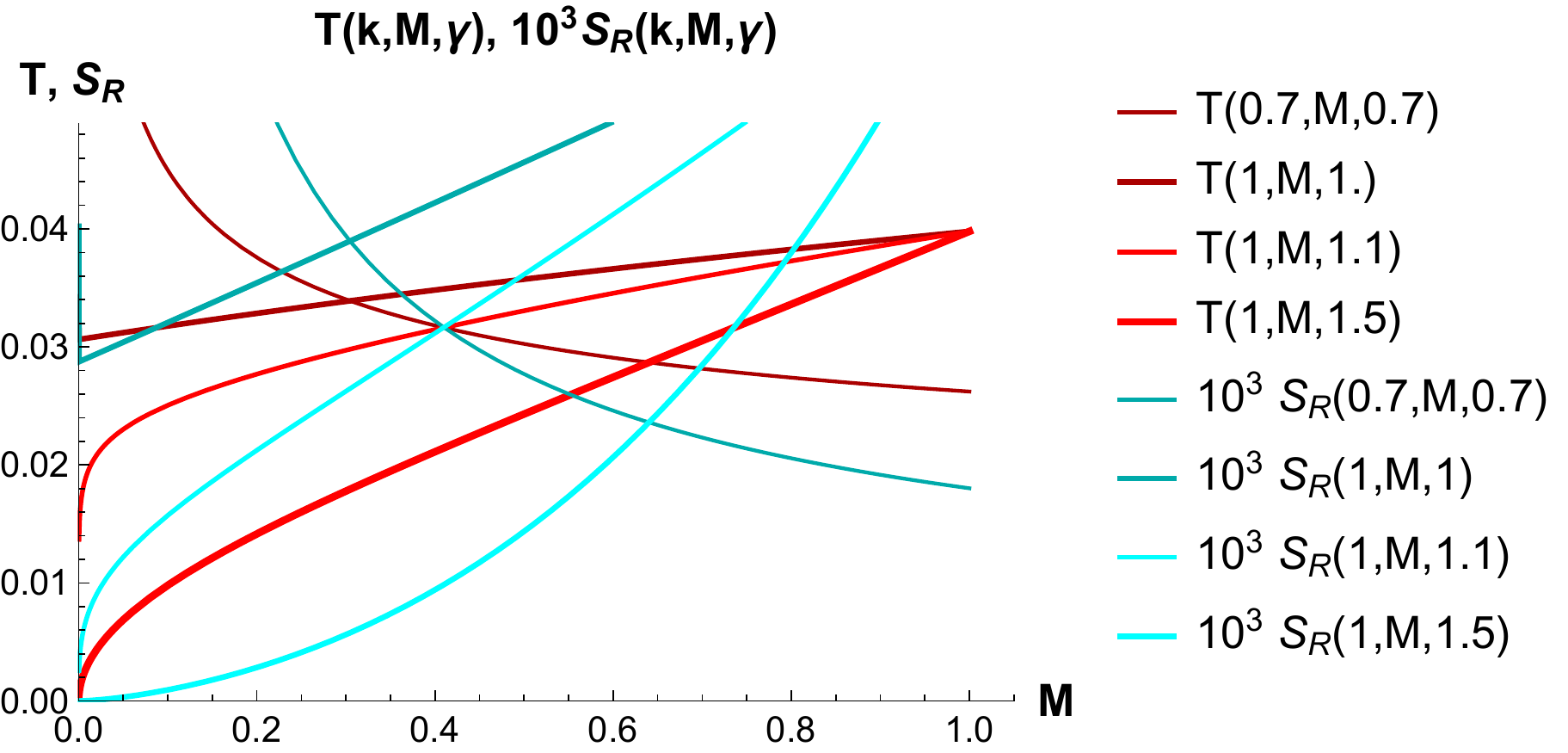}\\$\,$\\
 \includegraphics[scale=0.7]
 {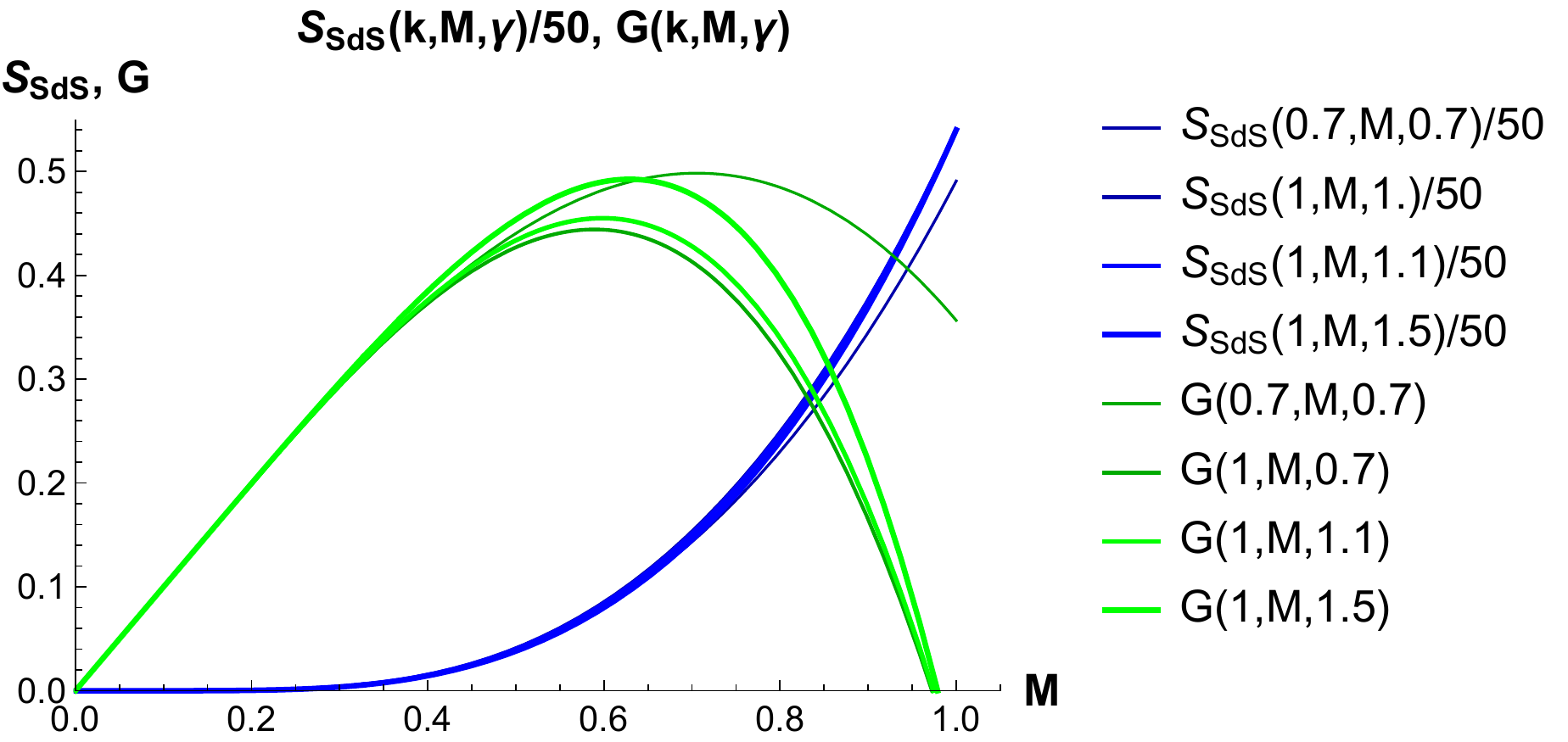}
   \caption{The plot shows the dependence of the temperature $T$ (red), the entropy $S_{SdS}$ (blue), the free energy $G$(green) 
   and the radiation entropy $S$ (cyan) on $M$ for functions $\lambda(M)=k M^\gamma$   with different scaling parameter $k$ and $\gamma$. 
   Plots in light tones correspond to parameters describing dependencies with $T\to 0$
   at $M\to 0$. Darker tones correspond to increasing temperature at $M\to 0$, or $T(0)\neq 0$ (thick darker red line, $\gamma=1$).  
   }
  \label{fig:SdS}
\end{figure}

   In Fig.\ref{fig:SdS} we present dependence of the temperature (red lines), the black hole entropy (blue lines),  
   the free energy $G$(green) 
   and the radiation entropy $S$ (cyan) on $M$ for different forms of $\lambda(M)$, $\lambda(M) =k M ^{\gamma}$.
  Here $\gamma=0.7,1,1.1$ and $1.5$. 
   Plots in darker  tones correspond to $\gamma \leq 1$ and light tones to $\gamma > 1$. We see that  for the later cases  $T\to 0$
   at $M\to 0$.   To guaranty $ \Lambda >0$ it is assumed that $ \lambda(M)<1$. 
%
%
\newpage
$$\,$$
\section{Complete evaporation of RNdS/AdS black holes}\label{sect:CERNds}
The blackening functions for the Reissner-Nordstrom de Sitter/Anti de Sitter (RNdS/AdS)  black holes are 
\bea
f&=&1  - \frac{2M}{r} + \frac{q^2}{r^2}+ \varepsilon\,\frac{r^2}{\ell^2}, \qquad \frac{1}{3}\Lambda=\frac1{\ell^2},\eea
where
$\varepsilon=+1$ for AdS and $\varepsilon=-1$ for dS. 

For the dS case there is the domain of parameters ($0\leq q\leq M\leq M_{cr}(q,\ell)$), where the equation $f(x)=0$ has 4 real roots 
$r_{--} \leq r_- \leq r_+\leq r_c$ and 3 last roots ( $r_-$ horizon,
event horizon $r_+$ and cosmological horizon $r_c$ ) are positive and the first $ r_{--}$ is negative. 
For the AdS case there is also the domain of parameters where two roots (event horizon $r_+$ and cosmological horizon $r_c$, we keep for them the same notations) are positive. The last two roots are complex. In both cases, dS and AdS, in
 the domain of existence of the event horizon,  we get the expressions for the black hole mass and the temperature
 in term of $x\equiv r_+$ (see for example \cite{Li:2016zca})
\bea
\label{M-RNAdSmm} M&=&\frac12 x+\frac{q^2}{2x}+\varepsilon
 \frac{x^3}{2\ell^2},\\
\label{T-RNAdSmm}
T &=&
\frac{1}{4\pi}(\frac{1}{x}-  \frac{q^2}{ x^3  }+\varepsilon \frac {3 x}{\ell^2} ).  \eea
We consider the following curve on the equation of state surface 
\be\label{q-x}
q^2=x^2-\mu^2 (x),\ee
where the function $\lambda (x)$  satisfies the bounds 
\be\label{bound}
0<\mu (x)\leq x.\ee
Eq.\eqref{q-x} gives the following expressions for $M$ and $T$ in terms of $\lambda$
\bea
\label{M-RNAdSlambdammm} M&=& x-\frac{\mu^2}{2x}+\varepsilon\frac{x^3}{2\ell^2},\\
\label{T-RNAdSlambdammm}
T &=&
 \frac{1}{4\pi  x}\left(  \frac{\mu^2}{x ^2  }+ \varepsilon\frac{3x ^2}{\ell^2} \right).  \eea

We denote by $m(x)$ the expression 
\be
m(x)=x-\frac{\mu^2}{2x}+\varepsilon\frac{x^3}{2\ell^2}\ee
The function $\lambda(x)$ should be  such that
equation 
 \eqref{M-RNAdSlambdammm}
\be\label{Mm}
M=m(x),\ee
has a positive solution  $x=x(M)$ for  sufficiently small $M$.
We suppose that $m'(x)>0$ for small $x\geq0$, $m(0)=0$, then $m(x)$ is an increasing function and equation \eqref{Mm} has an unique positive solution for sufficiently small $M$. Hence the function $\lambda(x)$ should satisfy the following relation for small $x$
\bea
\left(\frac{\mu(x)^2}
{2x}\right)^\prime<1+
3\varepsilon\frac{x^2}{2\ell^2}\eea
Furthermore, to obtain $M$
and $T$ vanishing in the limit $x\to 0$ we assume
the bound 
\be\mu(x)=o(x^{3/2})\ee
In the next subsections we demonstrate numerically behaviour of temperature along the curves \eqref{q-x}

\subsection{RNdS}
\begin{figure}[h!]
  \centering
 \includegraphics[scale=0.28]
 {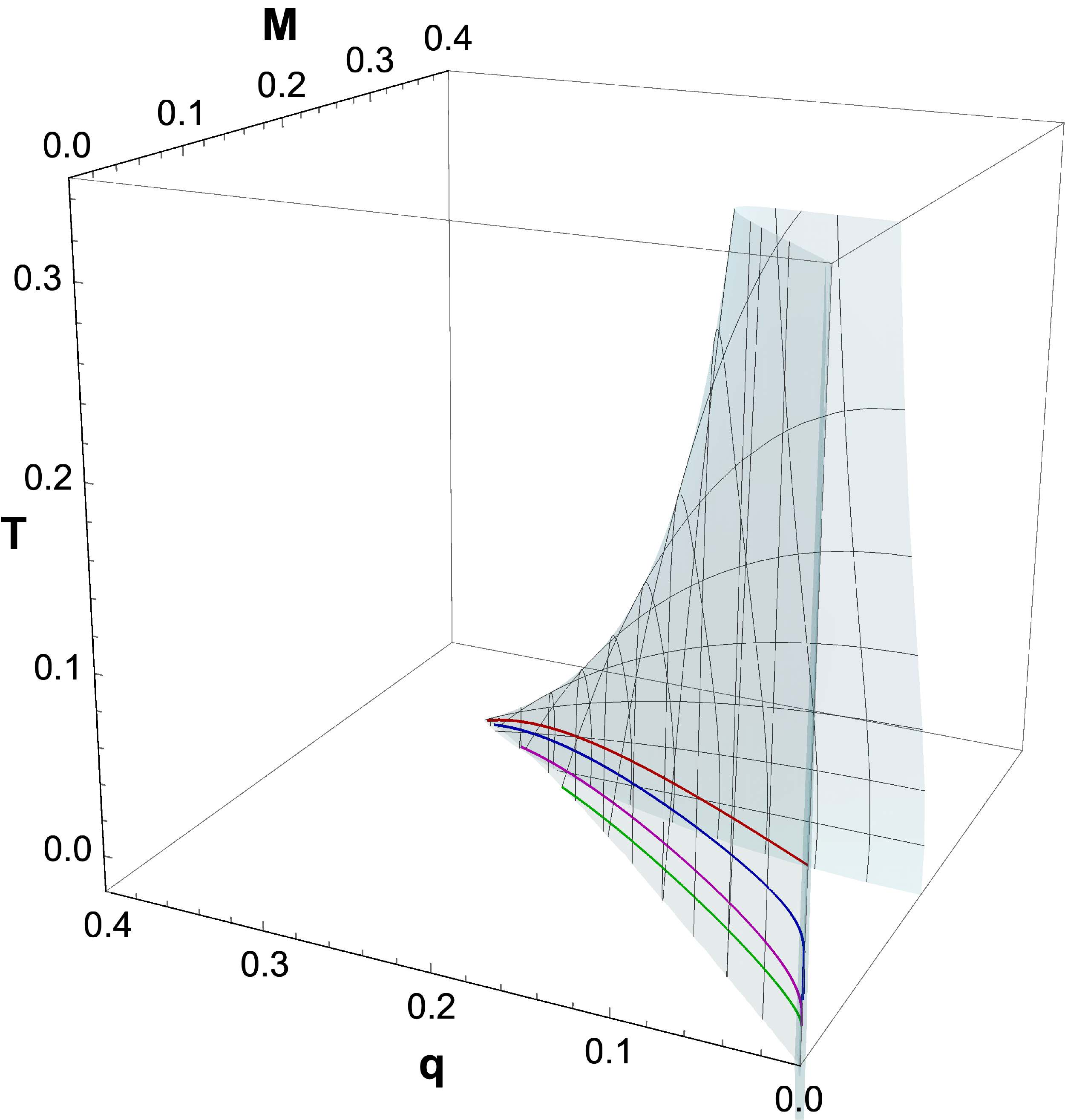}\qquad
 \includegraphics[scale=0.27]
 {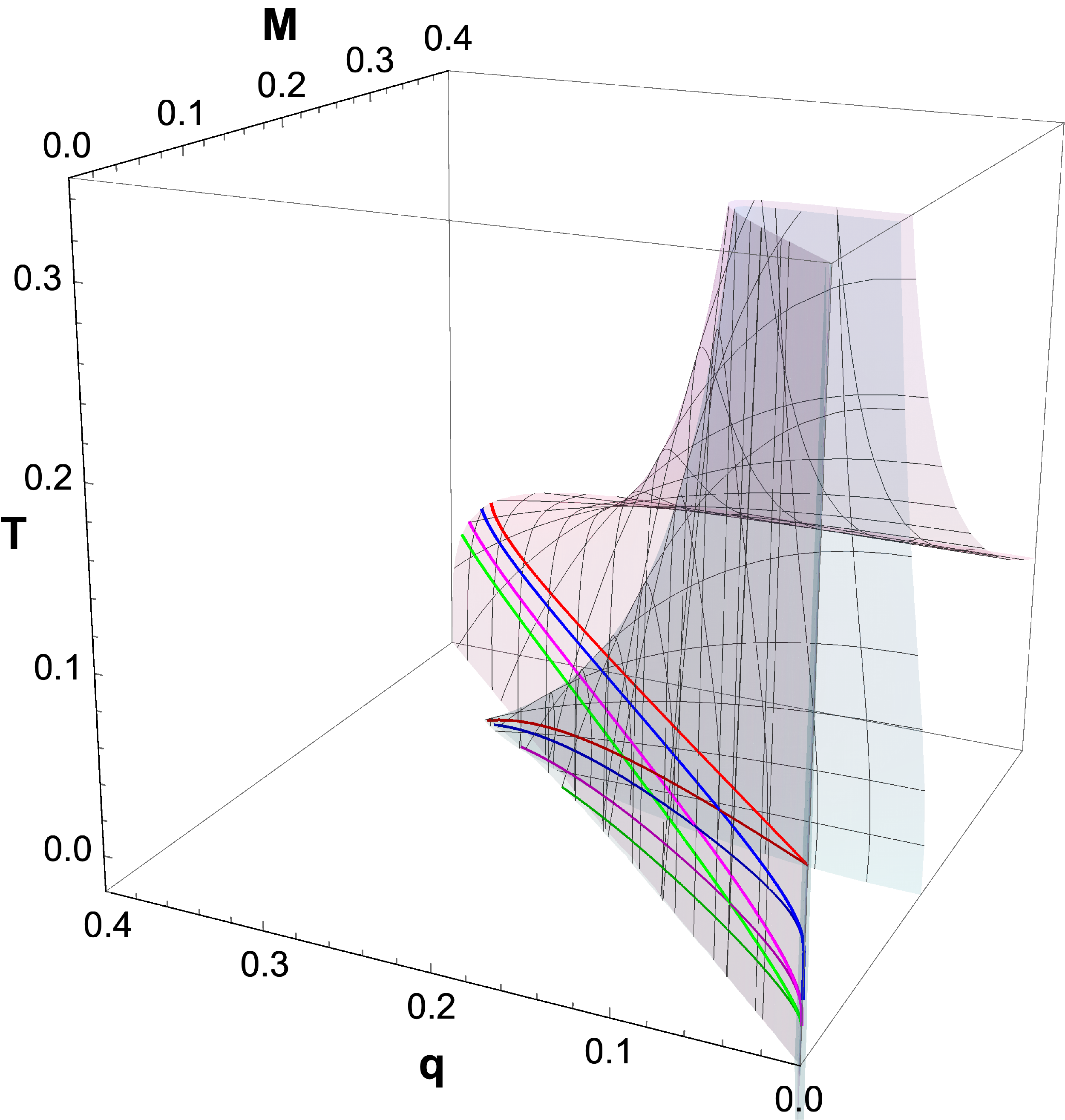}
 \\
 \includegraphics[scale=0.19]{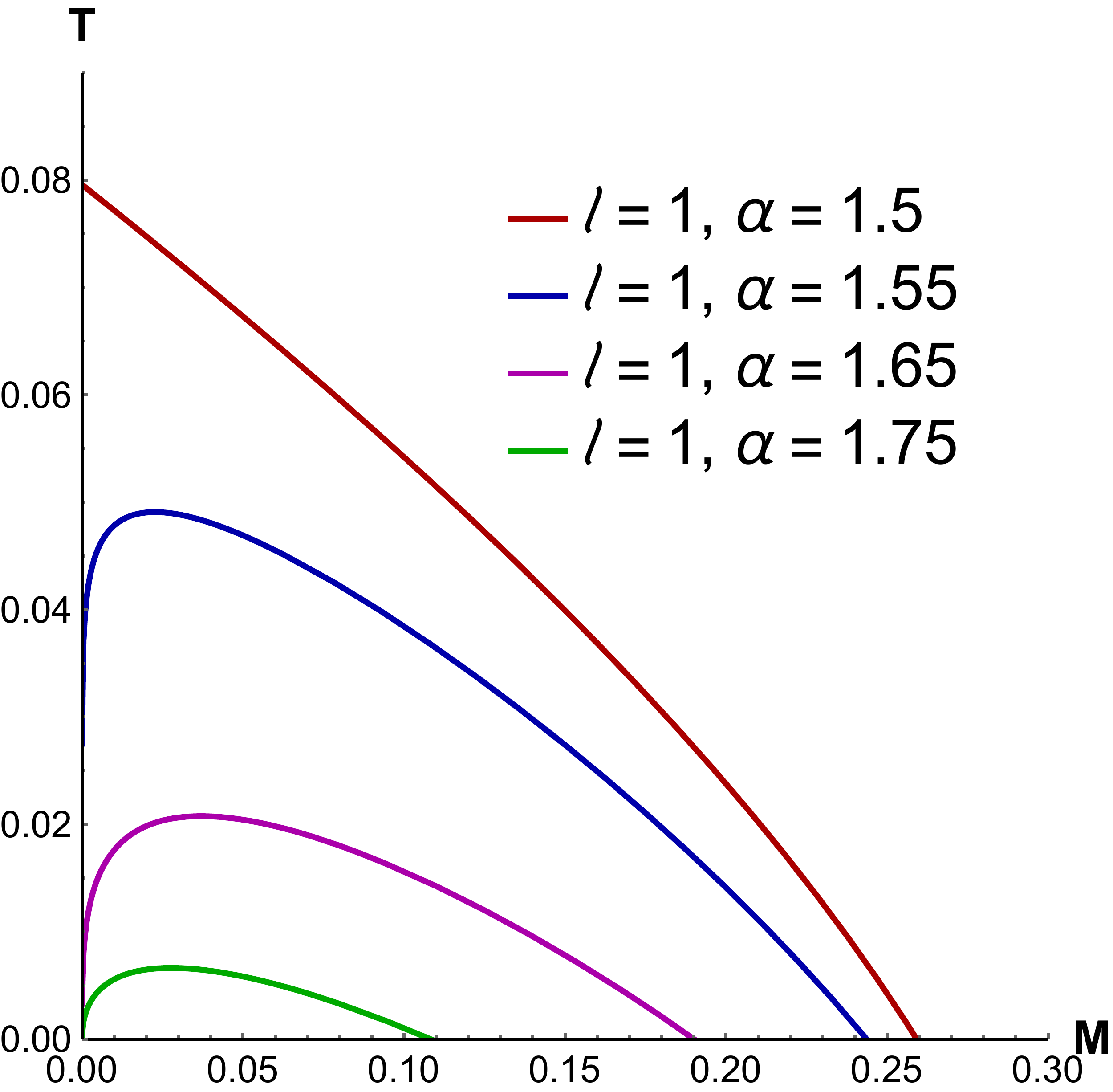}
  \qquad  \qquad \qquad
  \includegraphics[scale=0.19]
 {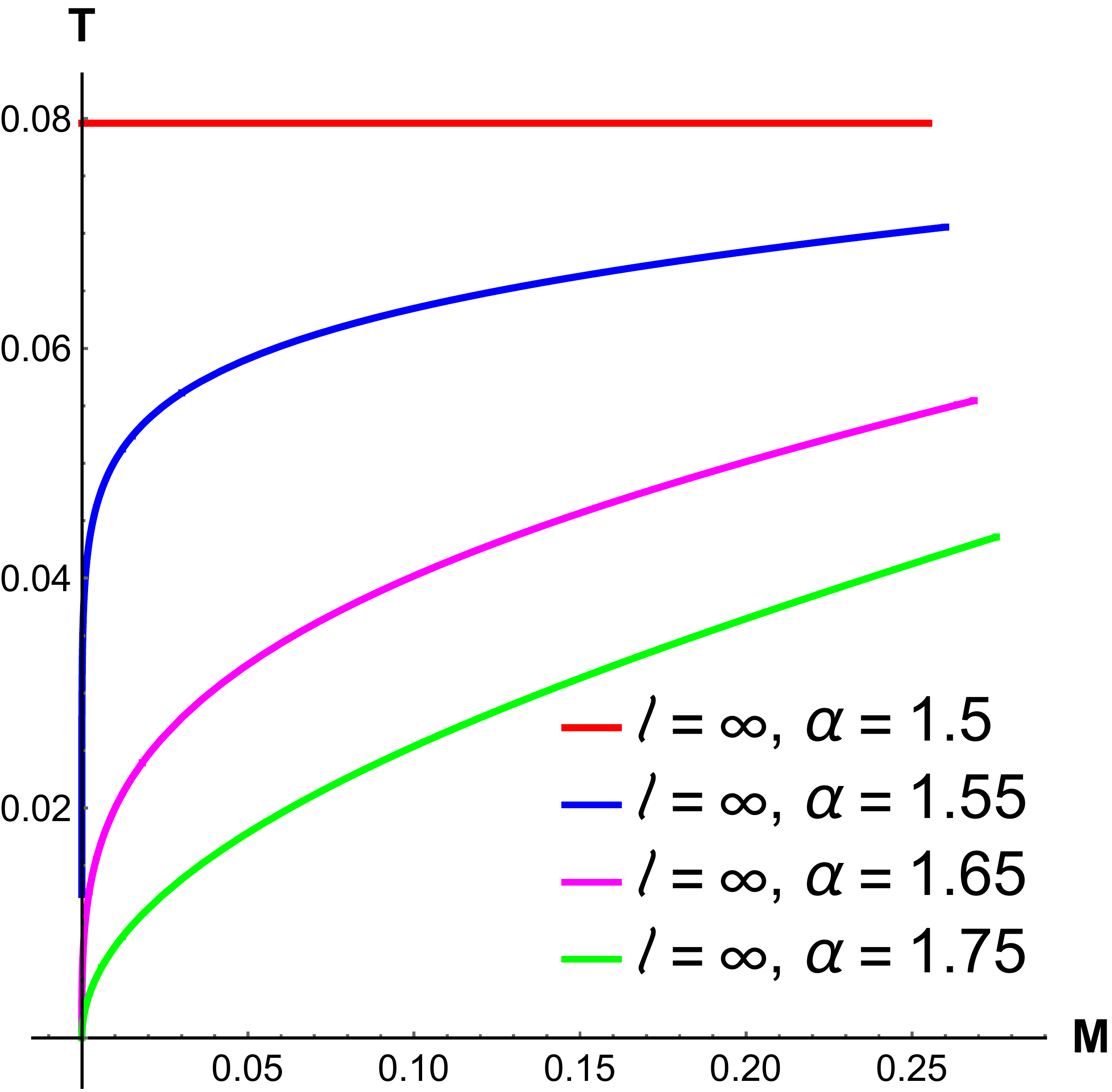}\\
  {\bf A}\hspace{150pt} {\bf B}
  \caption{{\bf A}.  Top:  The cyan surface shows the dependence of temperature on mass $M$ and $q$. The color curves show dependencies of temperature along curves $q=q(x)$ given by equation \eqref{q-x} and \eqref{lambdac} with different $\gamma$. Here $\ell=1$. Bottom: dependence of the temperature along the curves given by equations 
  \eqref{q-x} and  \eqref{lambdac}  with  different $\gamma $ indicated on the legends to this plot.
 {\bf B}. Top: The pink surface  show  the dependence of temperature on mass $M$ and $q$
for flat case ($\ell=\infty$). 
 The light color curves show dependencies of temperature along curves $q=q(x)$ given by equations \eqref{q-x} and \eqref{lambdac} with different $\gamma$ for flat case. The cyan surface and darker curves are the same as on {\bf A}.
  }
   \label{fig:M-T-ds}
\end{figure}

\begin{figure}[t!]
  \centering
   \includegraphics[scale=0.32]{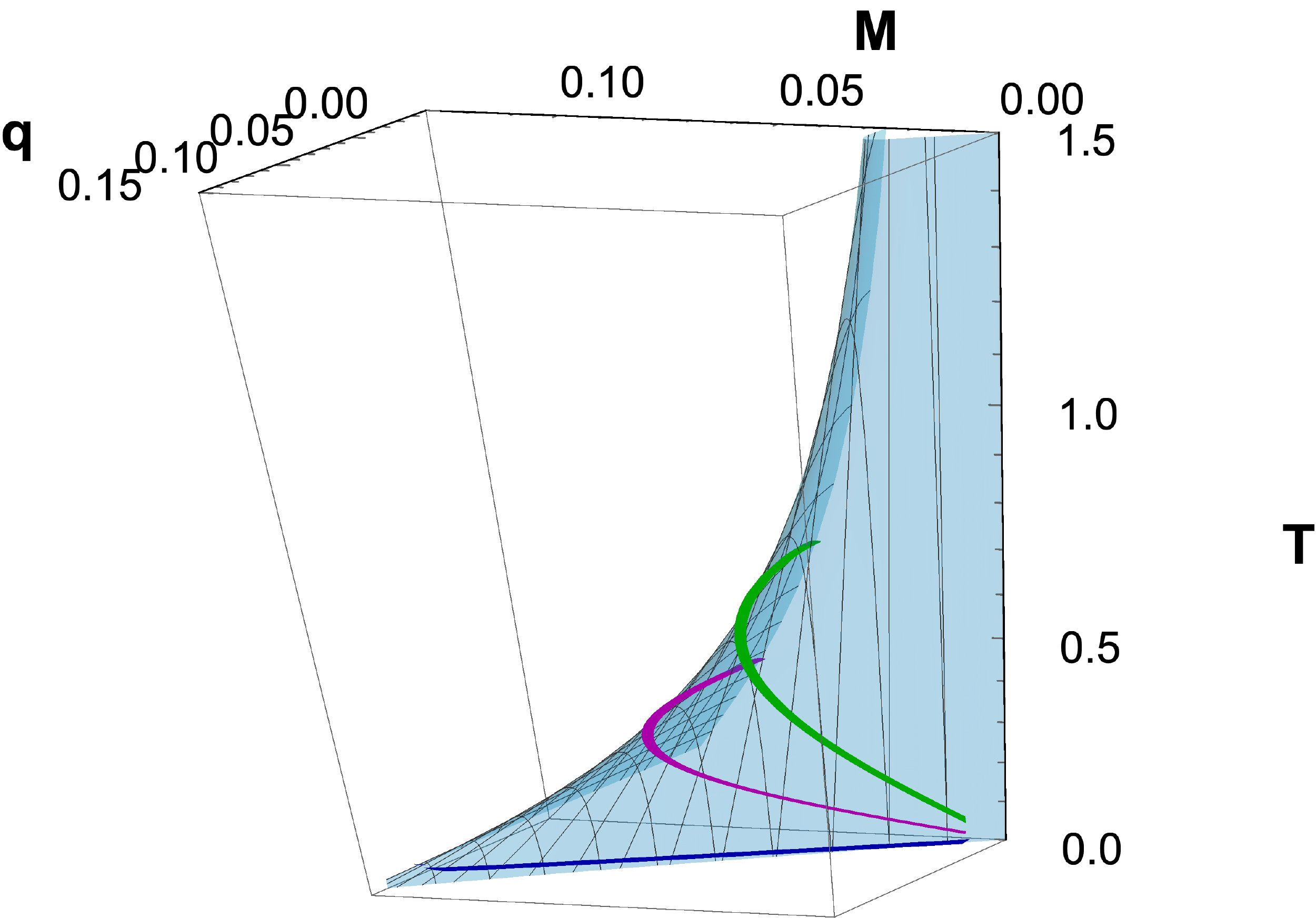}
   \includegraphics[scale=0.19]{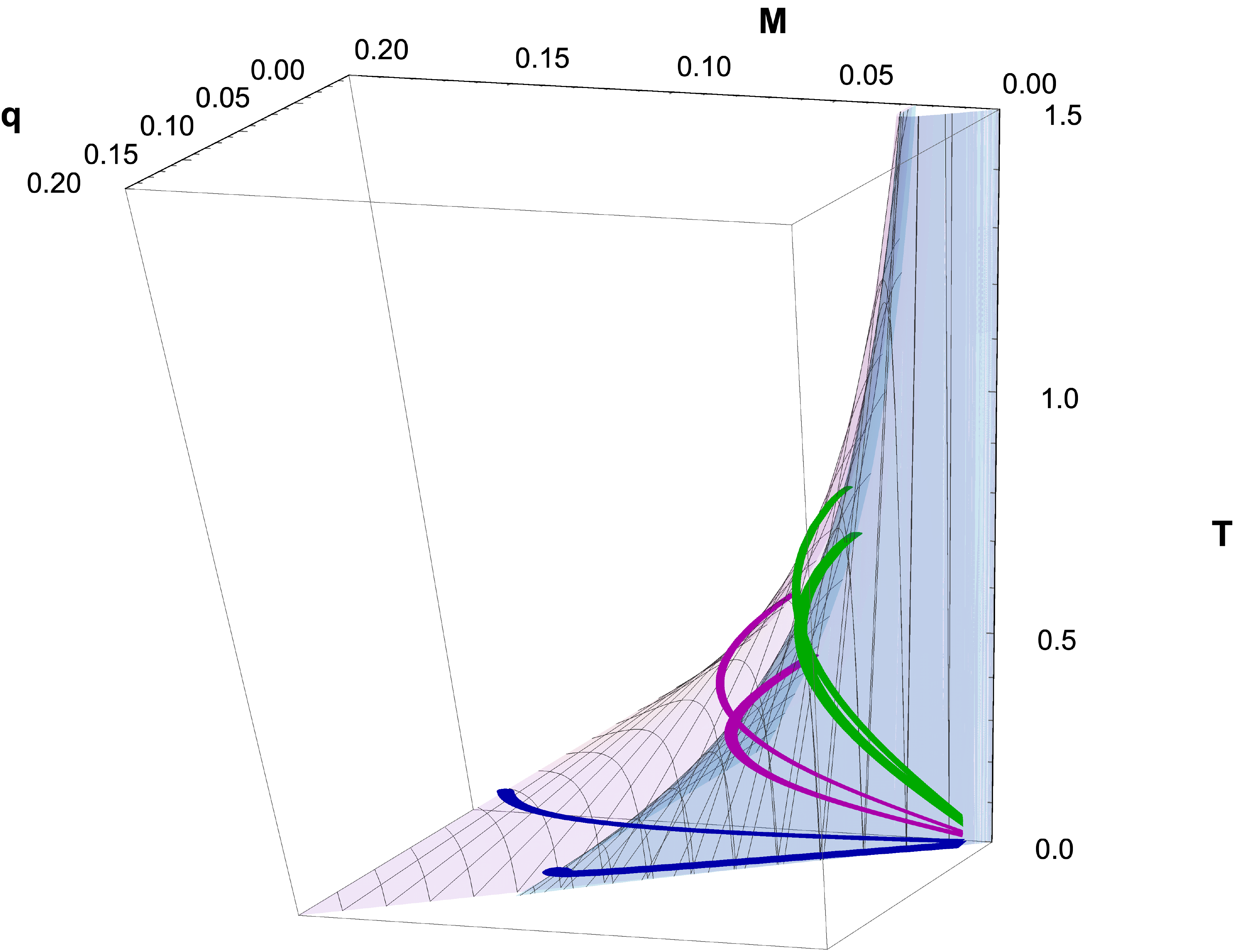}\\
   \begin{subfigure}{.5\textwidth} 
   \qquad\qquad\includegraphics[scale=0.19]{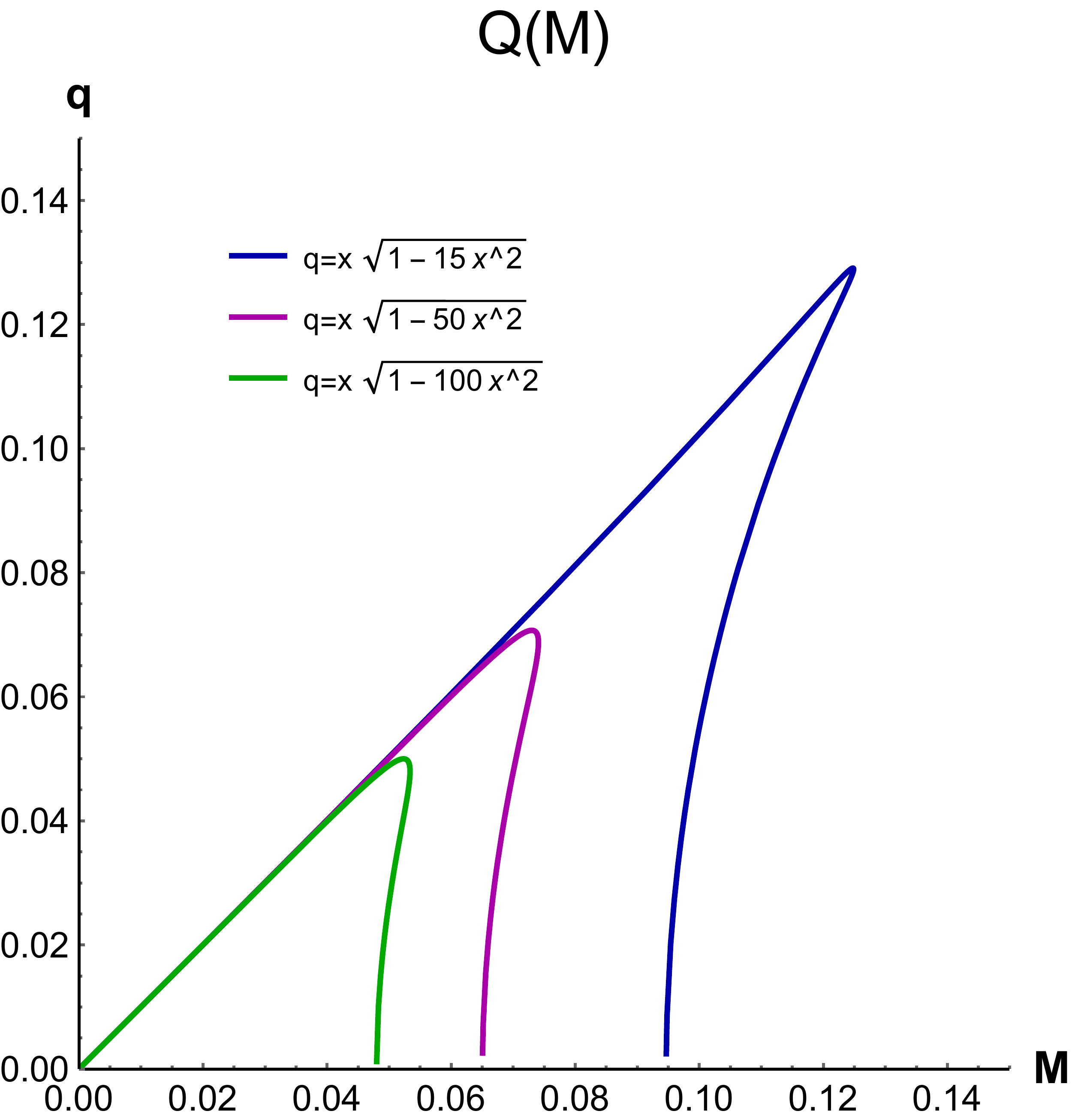}
 	\end{subfigure}
 	\qquad \begin{subfigure}{.4\textwidth}	\qquad\includegraphics[scale=0.3]
 {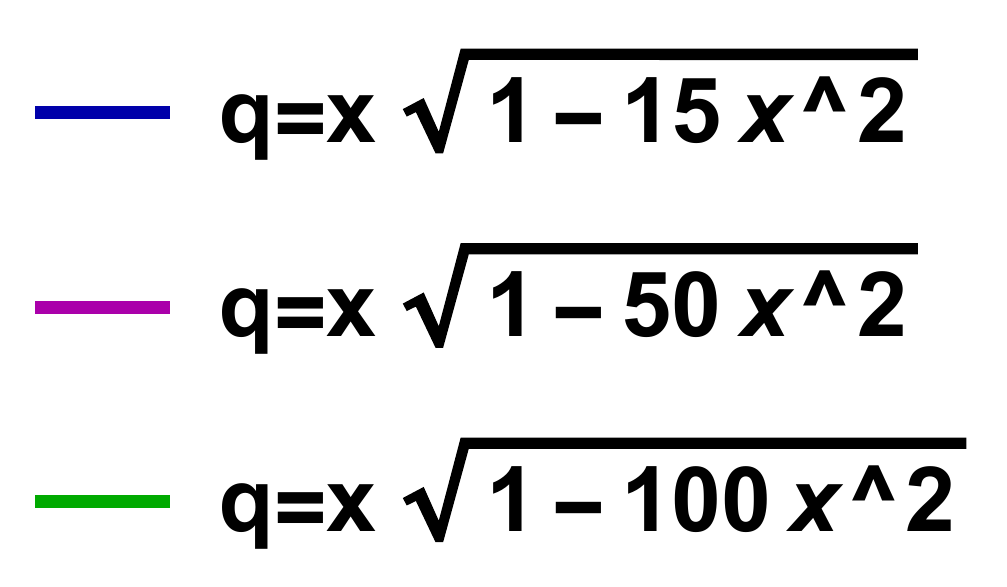}
  	\end{subfigure}
 	\\$\,$\\
  {\bf A}\hspace{250pt} {\bf B}
  \caption{ 
The cyan surface in the top of the coulomb {\bf A} shows the temperature dependence of the black hole temperature on the black hole mass $M$ and the charge $q$ for  $\ell=0.5$. 
The colored curves show dependence of the temperature along spacial curves $q=x\sqrt{1-Cx^2}$. A top view on the curves depicted in the 3D plot shown in the top line of {\bf A} is shown in the bottom of {\bf A}.  The pink surface in the top of the coulomb {\bf B} shows the temperature dependence of the black hole temperature on the black hole mass $M$ and the charge $q$ for  $\ell=\infty$. The cyan surface here is the same as in  {\bf A}. The colored curves here correspond to $q=x\sqrt{1-Cx^2}$ on both surfaces.
  }
   \label{fig:M-T-ds-gamma2}
  \end{figure}

Let us first consider the RNdS case,
$\varepsilon=-1$, in more details. 
For instance,  we can take 
\be
\label{lambdac}
\mu(x)=C x^\alpha.
\ee
\begin{itemize}
\item  
All the requirements (including the  positivity of the temperature) are satisfied if 
\be\label{rest}
\frac{3}{2}<\alpha<2,\ee
see Fig.\ref{fig:M-T-ds}.
In the top of 
Fig.\ref{fig:M-T-ds}.{\bf A} we present  the dependence of temperature on mass $M$ and $q$ (the cyan surface). The color curves show dependencies of temperature along curves $q=q(x)$ given by equation \eqref{q-x} and \eqref{lambdac} with different $\gamma$ under restriction \eqref{rest} and $C=1$, here $\ell=1$. In the right plot we also show this dependence at $\ell=\infty$ (the pink surface).  In both cases the curves are very closed to the critical line $M=q$.

\item $\gamma=2$ corresponds to
\be
q=x\sqrt{1-Cx^2}\label{q-gamma2}\ee
The behaviour of temperature  along the curves \eqref{q-gamma2}
for $x\to 0$ is presented on Fig.\ref{fig:M-T-ds-gamma2}.  As in 
Fig.\ref{fig:M-T-ds} cyan and pink
 surfaces  show  the dependence of temperature on mass $M$ and $q$ for
$\ell=1$ and  $\ell=\infty$, respectively.
The colored curves show dependence of the temperature along  curves
\eqref{q-gamma2} for different $C$.
As in the previous case the curves are very closed to the critical line $M=q$
\end{itemize}

\newpage
$$\,$$
\newpage
$$\,$$
From Fig.\ref{fig:M-T-ds} we see that all curves with $\alpha>\alpha_{cr}=1.5$ that show vanishing of the temperature for $M\to 0$. This concerns the finite $\ell$, Fig.\ref{fig:M-T-ds}.A, as well as $\ell \to \infty$, Fig.\ref{fig:M-T-ds}.B. Note that    vanishing of the temperature at $M\to 0$ show all curves  with $\gamma>\gamma_{cr}=2$ presented in Fig.\ref{fig:RN}.A. Let us explain relation between these critical 
$\alpha_{cr}=1.5$ and $\gamma_{cr}=2$.
To this purpose let compare  the parametrization used near $M\to 0$ in Sect. and parametrization near $M\to 0$ used here,
\bea
\label{qM}
q^2&=&M^2-\lambda ^2(M)\\
\label{qxp}
q^2&=&x_+^2-\mu ^2(x_+)\eea
Taking into account relation
\eqref{M-RNAdSlambdammm} at $\ell=\infty$, i.e.
\bea
\label{M-RNAdSlambdammmm} M&=& x_+-\frac{\mu ^2}{2x_+},
\eea
and \eqref{qM} and 
\eqref{qM} we get
\bea
\mu^2(x_+)=x^2_+-(x_+-\frac{\mu^2}{2x_+})^2+\lambda ^2(x_+-\frac{\mu^2}{2x_+})\eea
Supposing that 
\be\mu(x)=x^{\alpha},\qquad \lambda (M)=\left(\frac{ M}{m_0}\right)^\gamma
\ee
we get equation 
   \be
  F(x,\alpha,\gamma)= \left(\frac{x}{m_0}\right)^{2 \gamma } \left(1-\frac{1}{2} x^{2 (\alpha -1)}\right)^{2
   \gamma }-\frac{1}{4} x^{4 \alpha -2}\ee
   If $\alpha >0$ for small $x$ in the leading order we get identity if
   \be
   m_0^\gamma=2,\qquad \gamma= 2\alpha -1.\ee
   In particular, we see that $\alpha=3/2$ correspond to $\gamma=2$. This is in agreement with plots presented in Fig.\ref{fig:RN}.A and Fig.\ref{fig:M-T-ds}.B, since in both case the temperature go to constant values for $\alpha=3/2$ and  $\gamma=2$, respectively. 
   \\
   
   One can use another parametrization of mass, charge and cosmological constant and obtain results similar to results obtained above. In this parametrization
 mass and cosmological constant are written in terms of ratio of event horizons to cosmological horizon, $x=r_+/r_c$ and the electric charge $q$ \cite{Zhang:2022aqg}

\bea\label{M}
M&=&\frac{(x+1) \left(x^2 r_c^2+q^2 x^2+q^2\right)}{2 x \left(x^2+x+1\right) r_c},\qquad x=r_+/r_c\\
\label{La}
\Lambda &=& \frac{3 \left(x r_c^2-q^2\right)}{x \left(x^2+x+1\right) r_c^4} 
\eea
The expressions for temperatures for black hole horizon 
reads
\bea\label{Tp}
T_+&=&\frac{(x-1) \left(q^2 (x (3 x+2)+1)-r_c^2 x^2 (2 x+1)\right)}{4 \pi  x^3 \left(x^2+x+1\right) r_c^3}
\\\label{Tc}
  T_c&=&\frac{(x-1) \left(x (x+2) r_c^2-q^2 (x
   (x+2)+3)\right)}{4 \pi  x \left(x^2+x+1\right)
 r_c^3}
\eea

We set 
\be\label{Qrho}
q^2=x^2 r_c^2 -\rho^2(x)\ee
One can see that if $\rho(x)$ satisfies 
for small $x$ the bound 
\be
C_1 x ^{\beta+3}r_c^2<\rho^2(x)<C_2r_c^2 x ^{\beta+2}, \qquad 4>\beta>3\ee
then mass \eqref{M} behaves as $M=r_c x+o(x)$ and
temperatures \eqref{Tp} and \eqref{Tc} for $x\to 0$ behave  as $T_+\to 0$ and  $T_c\to const$.
\\

For example one can take in \eqref{Qrho}
\bea\label{betam}
 \rho&=& C_1 r_c^2x^{\beta+3},
 \eea
then for small $x$ we get
\bea
M&=&  r_+ -\frac{1}{2} C_1 \frac{r_+^{\beta +1}}{r_c^{\beta +1}}+...
\\
T_{+}&=&-\frac{3 r_+}{4 \pi 
   r_c^2}+ \frac{C_1 }{4 \pi  r_c^{1+\beta}}r_+^{\beta }+...,
  \eea
 i.e. we get that the temperature goes to zero for $\beta>0$. 

\subsection{RNAdS}
\begin{figure}[h!]
  \centering
\hspace{-25pt} \includegraphics[scale=0.28]
 {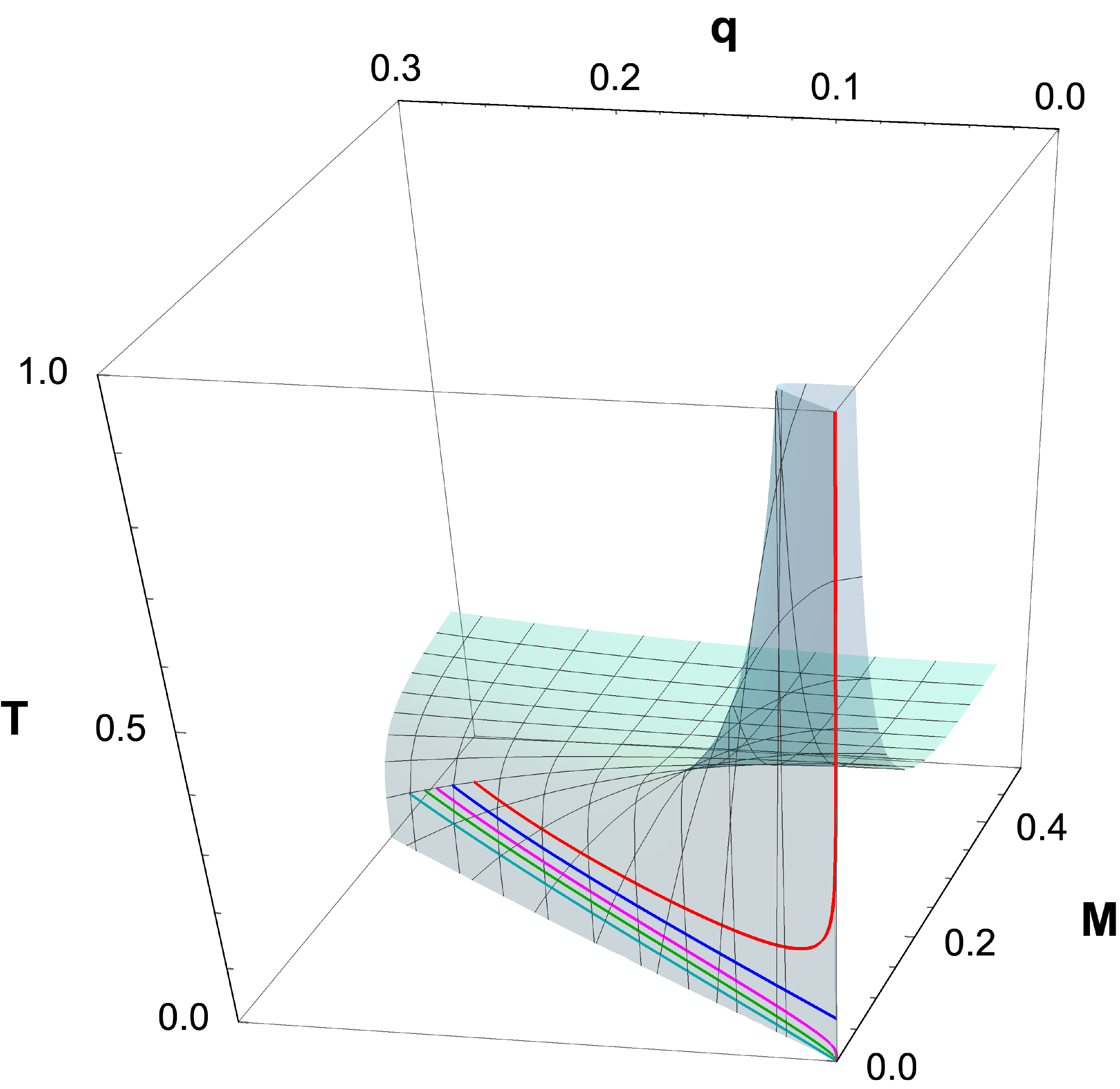}\qquad
 \includegraphics[scale=0.27]
 {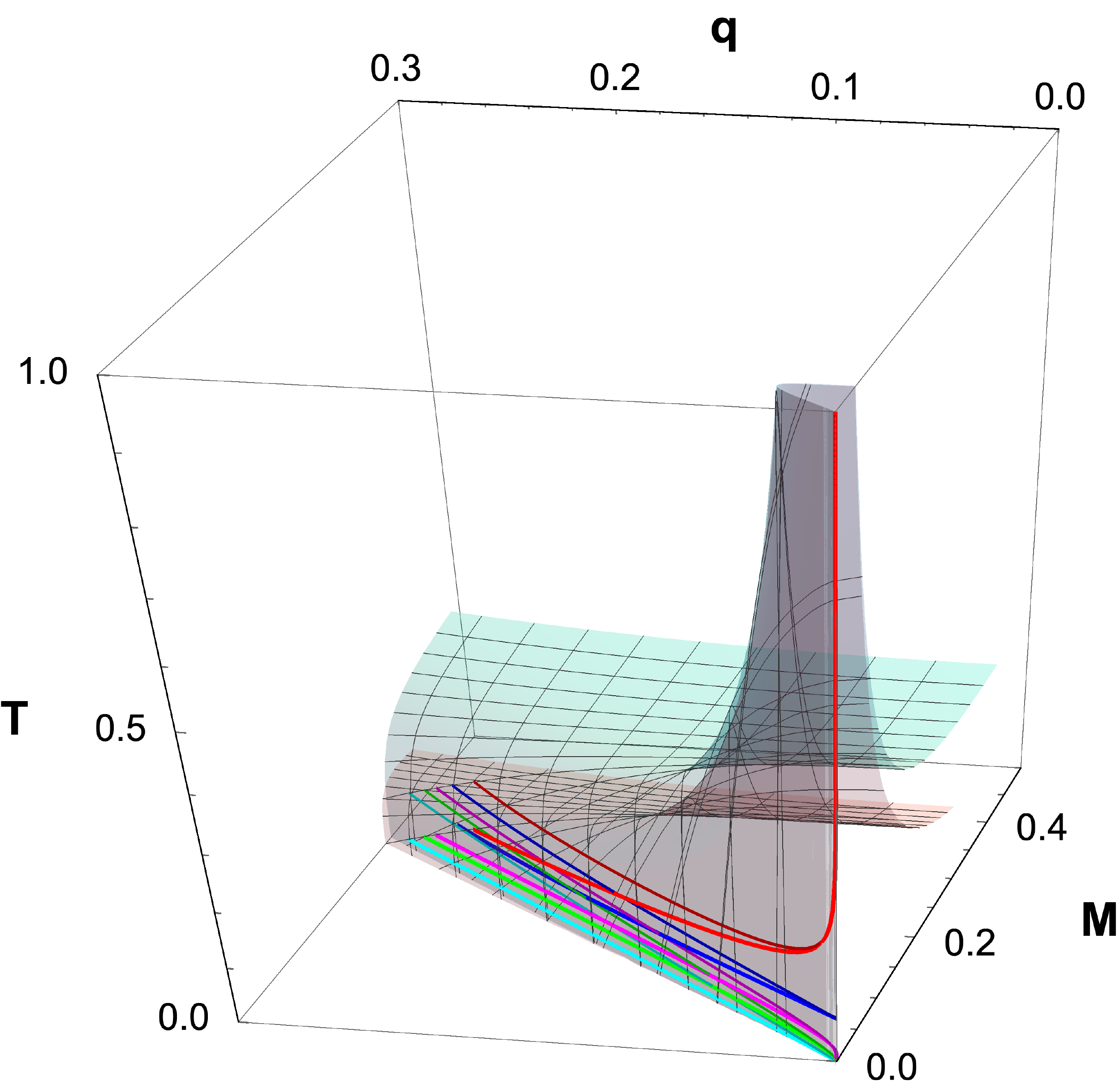}
 \\
\includegraphics[scale=0.17]{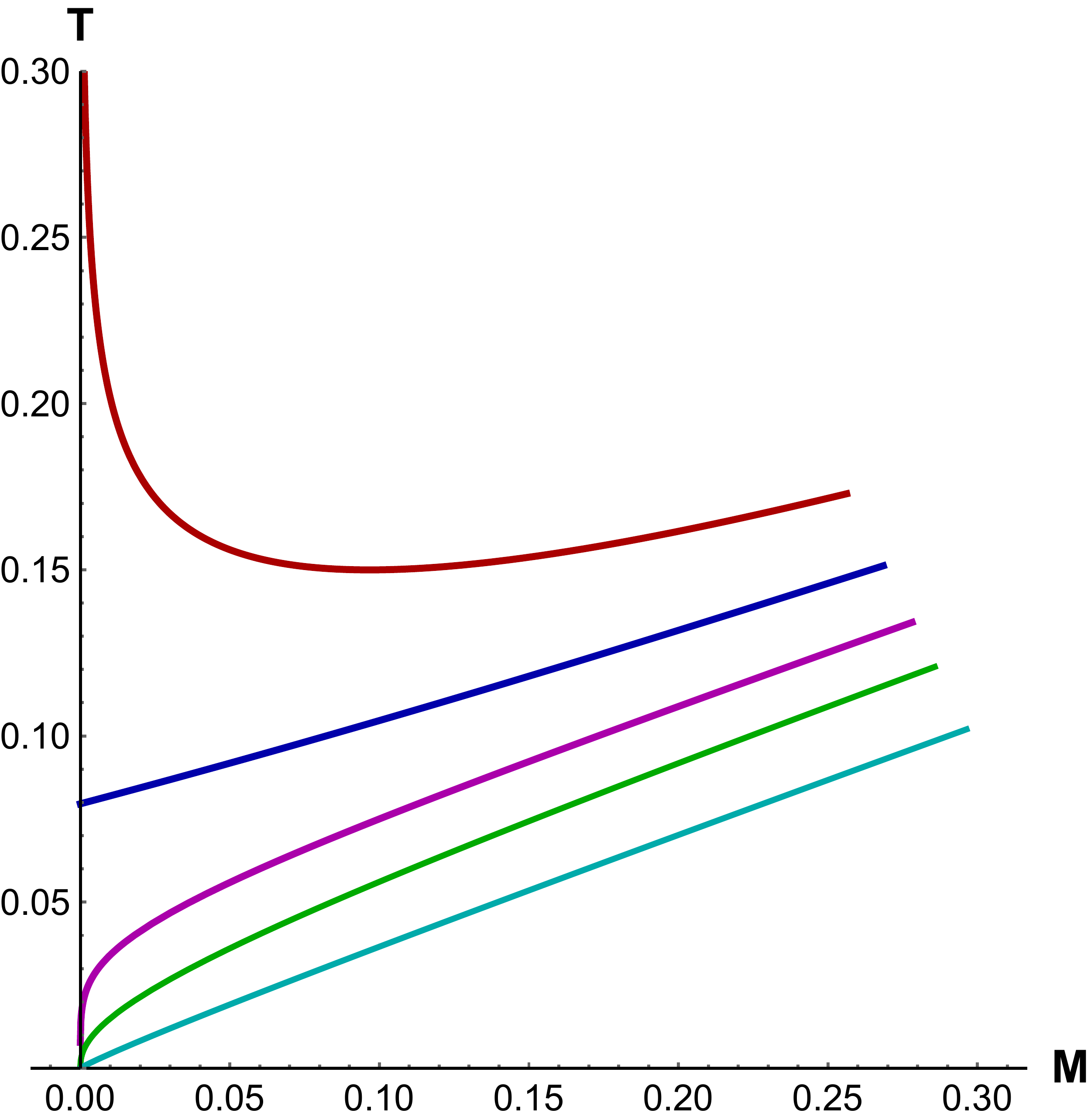}
 \includegraphics[scale=0.3]{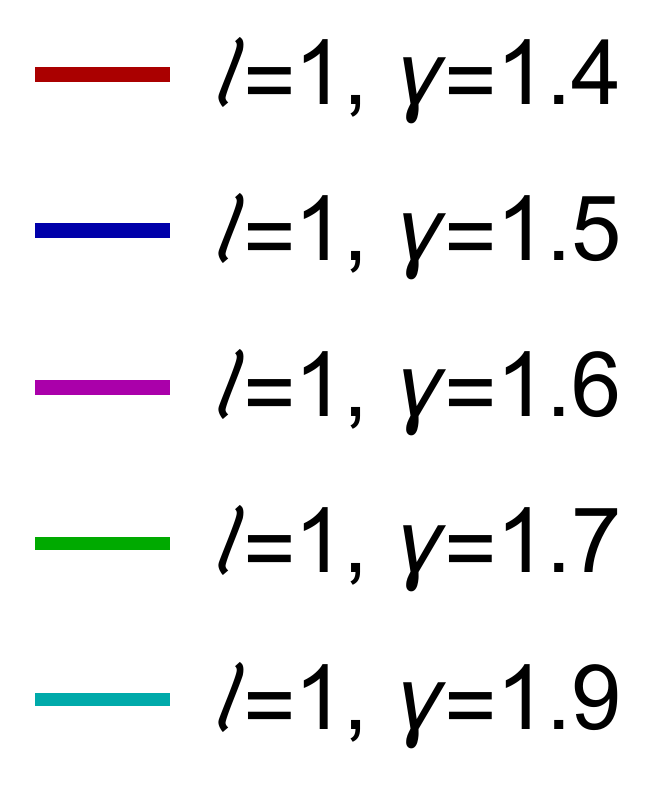}
 \qquad \qquad
\includegraphics[scale=0.17]{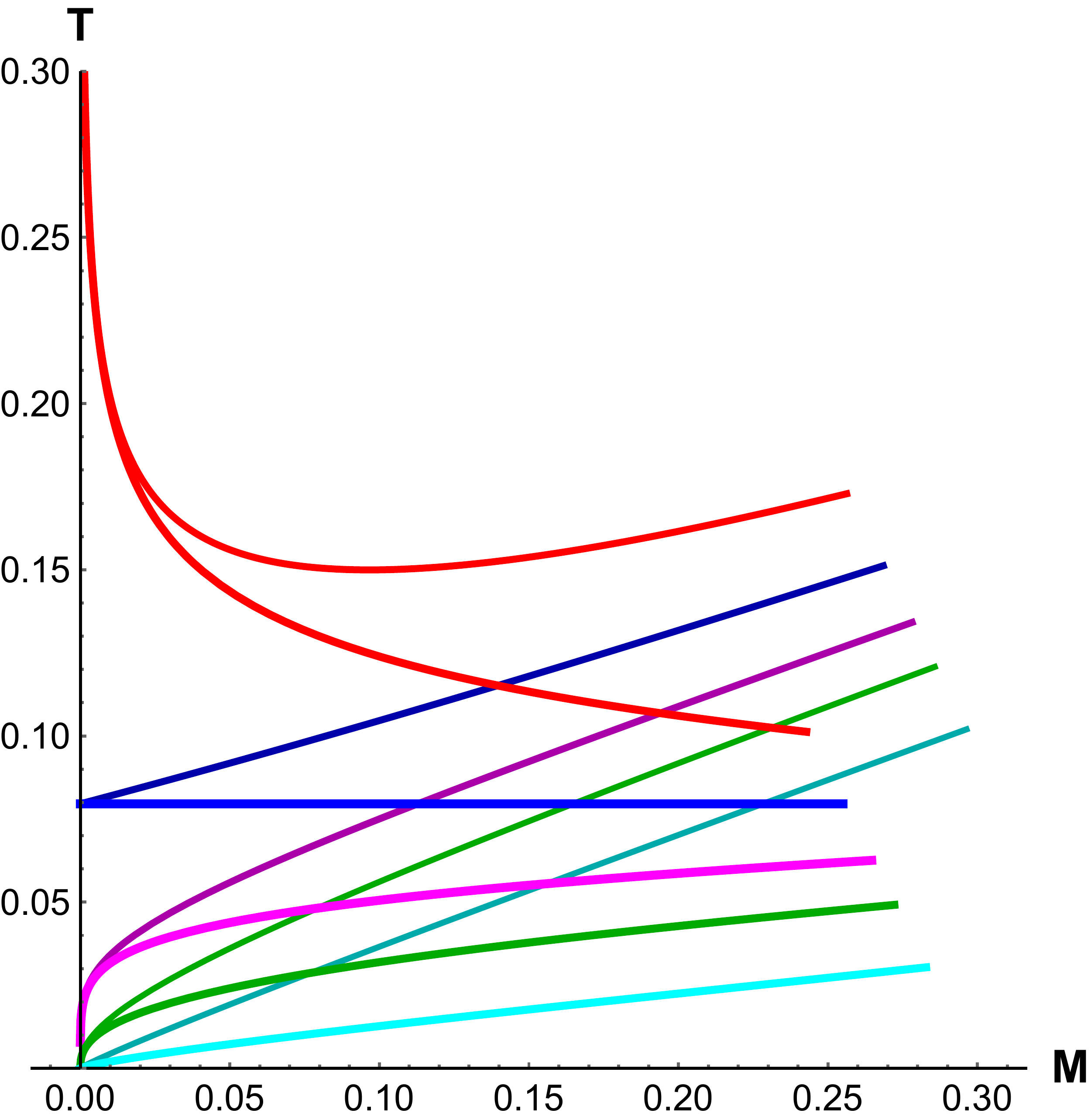}
\includegraphics[scale=0.3]
 {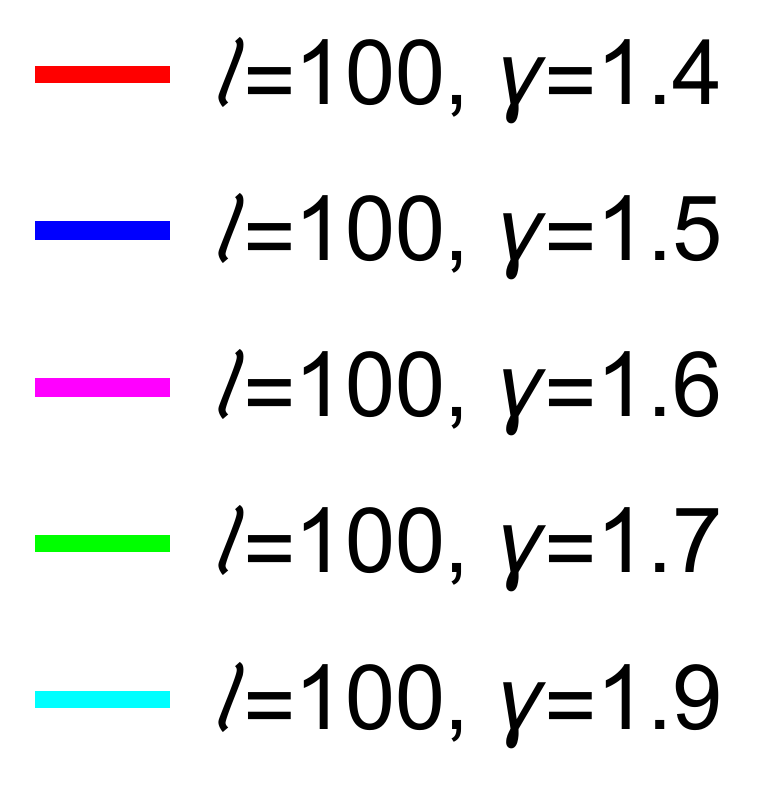}\\
  {\bf A}\hspace{150pt} {\bf B}
  \caption{{\bf A}. Top:  The cyan surface shows the dependence of temperature on mass $M$ and $q$. The color curves show dependencies of temperature along curves $q=q(x)$ given by equation \eqref{q-x} and \eqref{lambdacm} with different $\gamma$. Here $\ell=1$. Bottom: dependence of the temperature along the curves given by equations 
  \eqref{q-x} and  \eqref{lambdacm}  with  different $\gamma $ indicated on the legends to this plot.
 {\bf B}. Top: The pink surface  show  the dependence of temperature on mass $M$ and $q$
for large $\ell$ ($\ell=100$). 
 The light color curves show dependencies of temperature along curves $q=q(x)$ given by equation \eqref{q-x} and \eqref{lambdacm} with different $\gamma$ for large $\ell$. The cyan surface and darker curves are the same as on {\bf A}.
  }
  \label{fig:M-T-Ads}
\end{figure}

In this subsection we consider the AdS, and deal with formula  \eqref{M-RNAdSmm} and \eqref{T-RNAdSmm} for mass and temperature for $\varepsilon=1$.
As in the previous subsection we consider 
the same parametization as in \eqref{q-x} and
   take 
\be
\label{lambdacm}
\lambda(x)=C x^\gamma,
\qquad \frac{3}{2}<\gamma<2
\ee
(we take for simplicity $C=1$)
and consider the behaviour of the temperature near $x=0$.
The typical behaviour
is shown in Fig.\ref{fig:M-T-Ads}.
We see that for $\frac{3}{2}<\gamma<2$
the temperature decreases, for $\gamma=1.5$ goes to finit value and for $\gamma<3/2 $ goes to infinity.

\newpage

\section{Conclusions and discussions}\label{sect:SD}

As already mentioned in the Introduction,
the problem of complete evaporation of Schwarzschild black holes is that when the black hole mass $M$ tends to zero, an explosion of temperature $T=1/8\pi M$ occurs. Models of complete evaporation of black holes without blow-up of temperature are considered. 
In these models, the black holes metric depends not only on the mass $M$ but also on  additional parameters (thermodynamics variables)  such as the charge $Q$ and the angular momentum $a$ and special  relations between the mass and  these parameters are assumed.
\\

The Hawking temperature defines a
state equation surface $\Sigma$  in the space of thermodynamics variables. Curves on the surface $\Sigma$ such that \sout{the motion} \IA {evaporation} along them provides complete evaporation without blow-up of temperature are described. 
  In the models under consideration, there are two possible forms of projections of these curves on 
$(M,\cQ)$-plane (here $\cQ$ can be $Q$ or $a$), see Table 1.

\begin{table}[h]
\centering
\begin{tabular}{llll}
\hline
\multicolumn{1}{|c|}{}     & \multicolumn{1}{c|}{\textbf{$\cQ=\cQ(M)$}} & \multicolumn{1}{c|}{\textbf{$T=T(M)$}} & \multicolumn{1}{c|}{\textbf{$S_R(M)$}} \\ \hline
\multicolumn{1}{|c|}{\begin{picture}(0,0)
\put(-10,30){1-st}
\end{picture}} & \multicolumn{1}{c|}
{\begin{picture}(0,0)
\put(1,60){$\cQ$}
\end{picture}\includegraphics[scale=0.15]{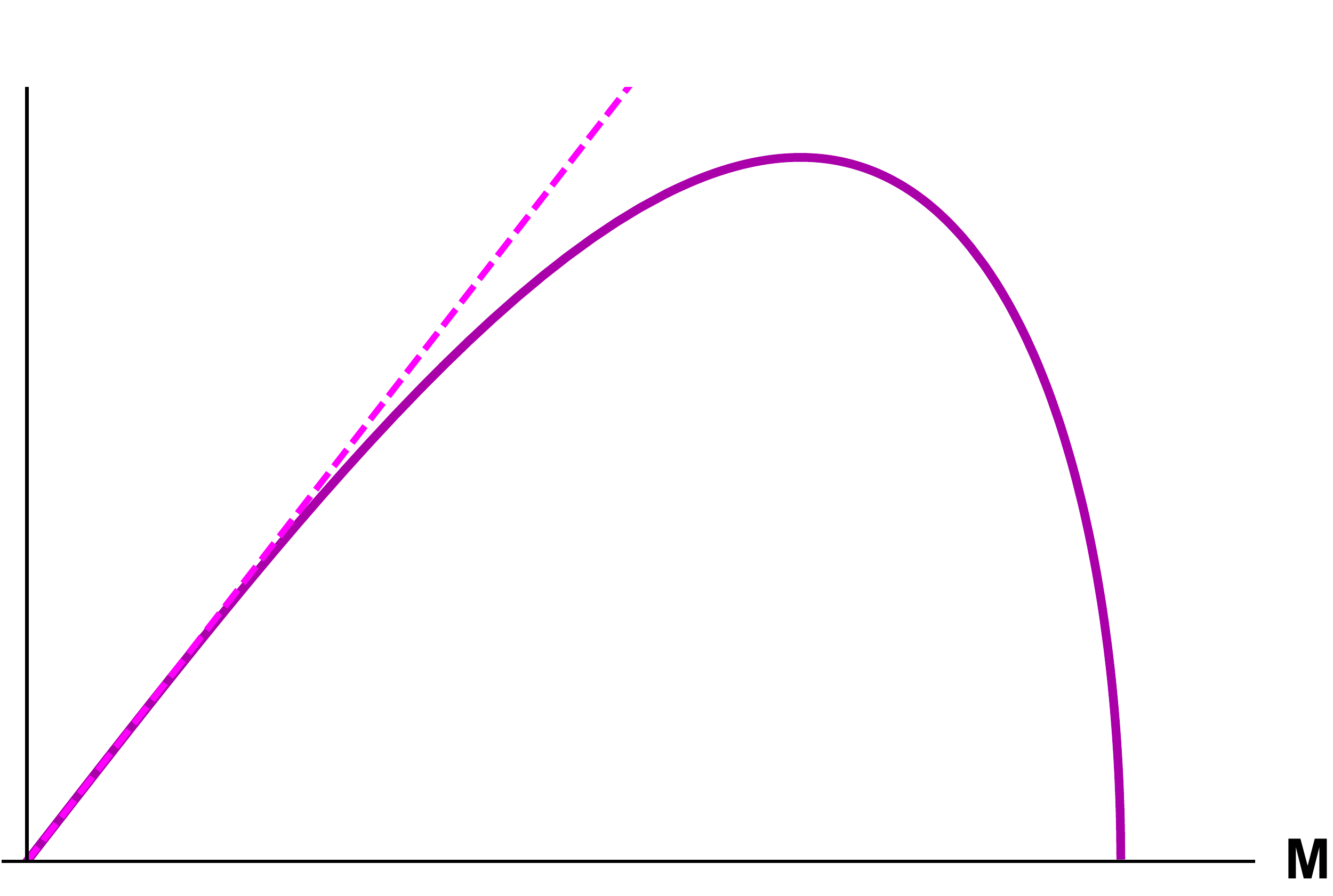}}                & 
\multicolumn{1}{c|}
{\begin{picture}(0,0)
\put(-120,-20){$\cQ^2$}
\end{picture}
\includegraphics[scale=0.15]{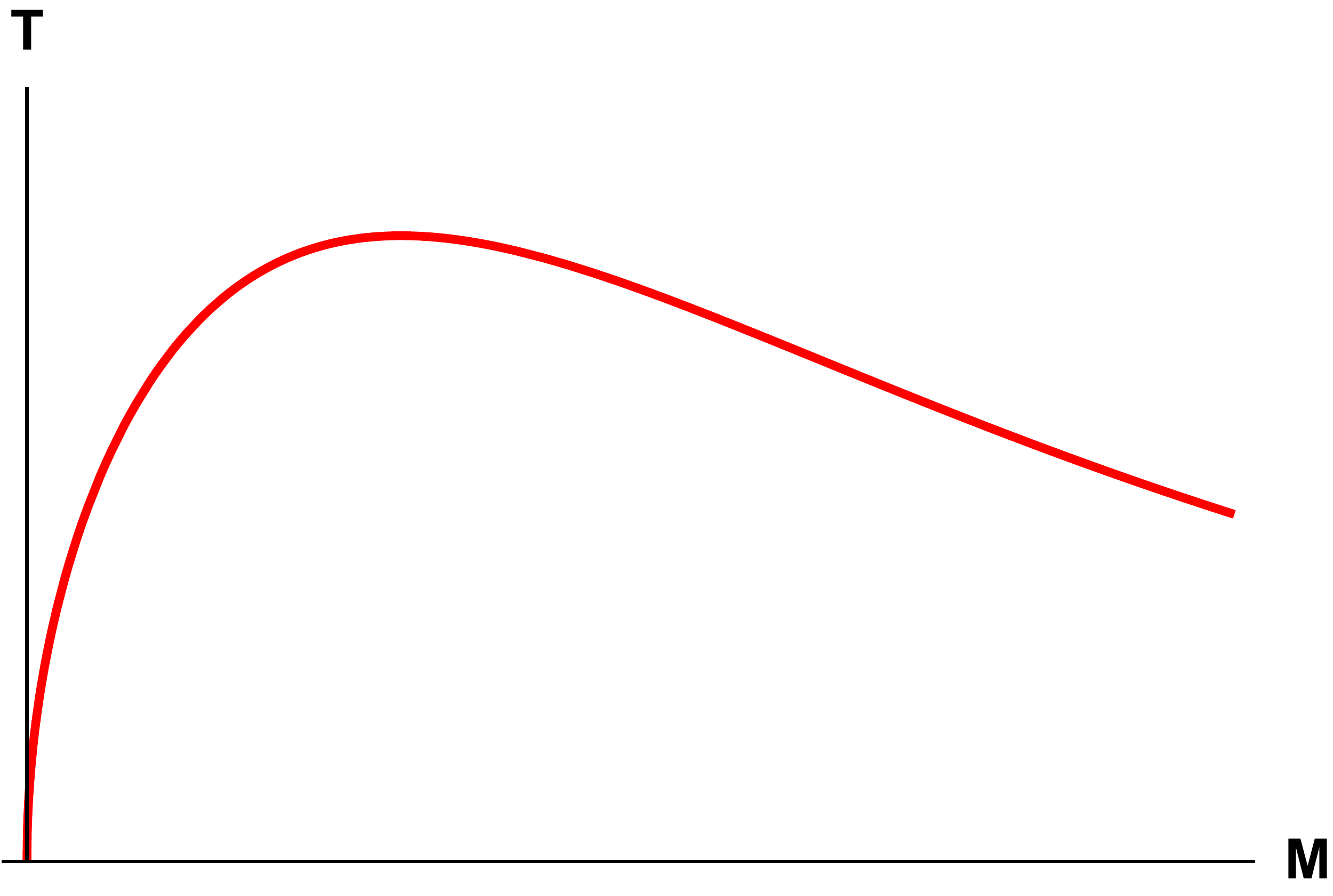}}                & \multicolumn{1}{c|}{\includegraphics[scale=0.15]{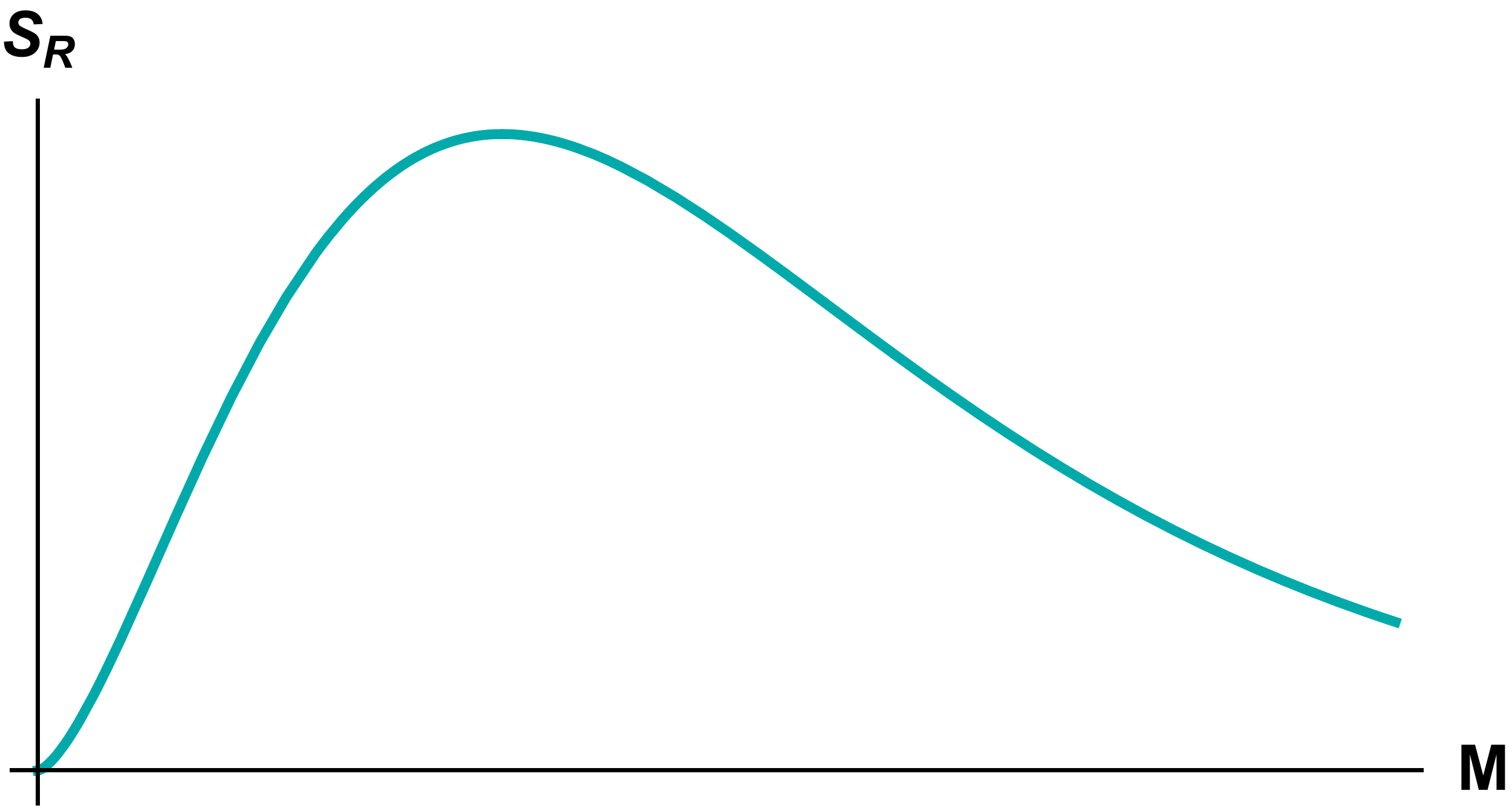}}                \\ \hline
\multicolumn{1}{|c|}{\begin{picture}(20,20)
\put(0,30){2-nd}
\end{picture}} & \multicolumn{1}{c|}{\includegraphics[scale=0.15]{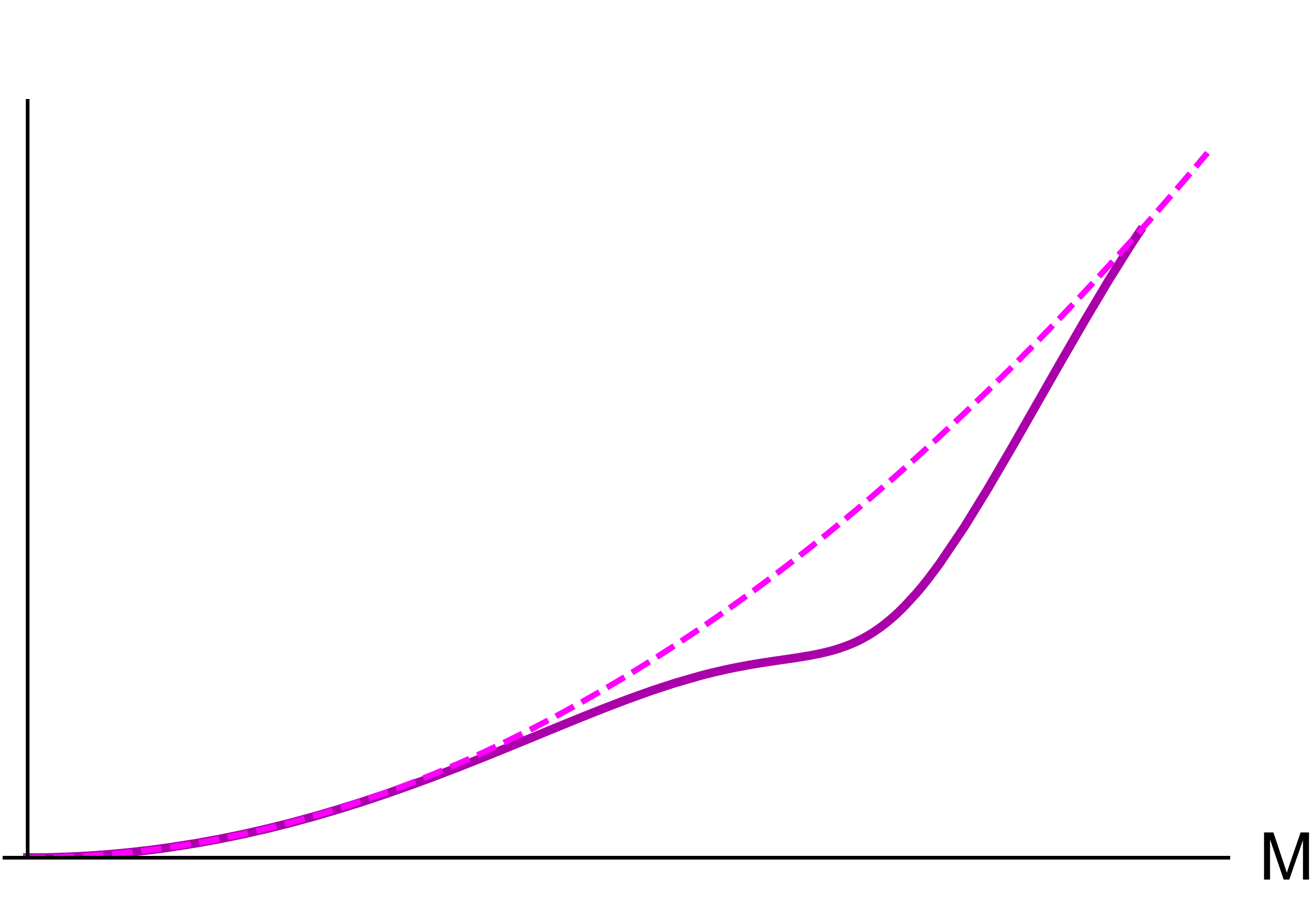}}                & \multicolumn{1}{c|}{\includegraphics[scale=0.15]{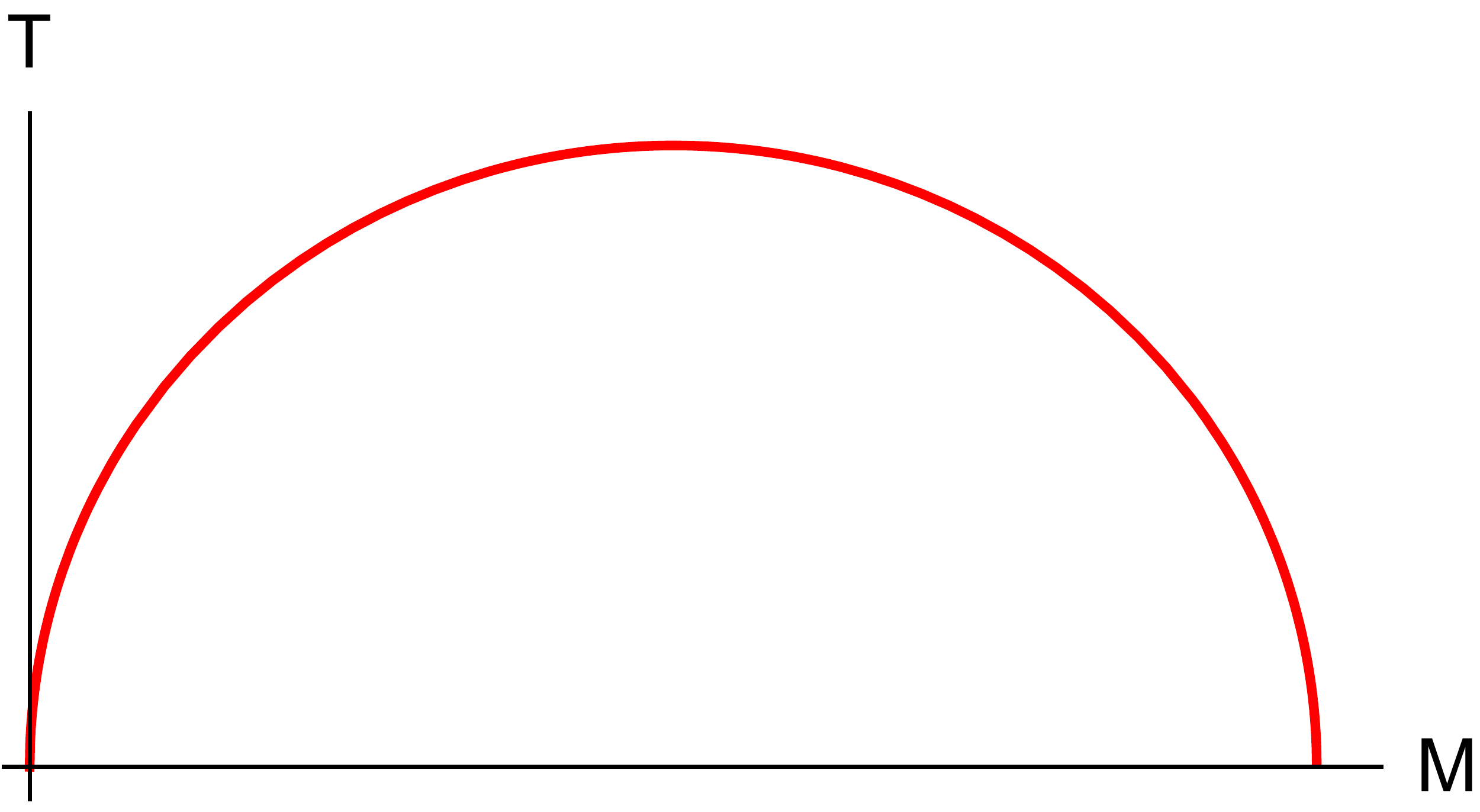}}                & \multicolumn{1}{c|}{\includegraphics[scale=0.15]{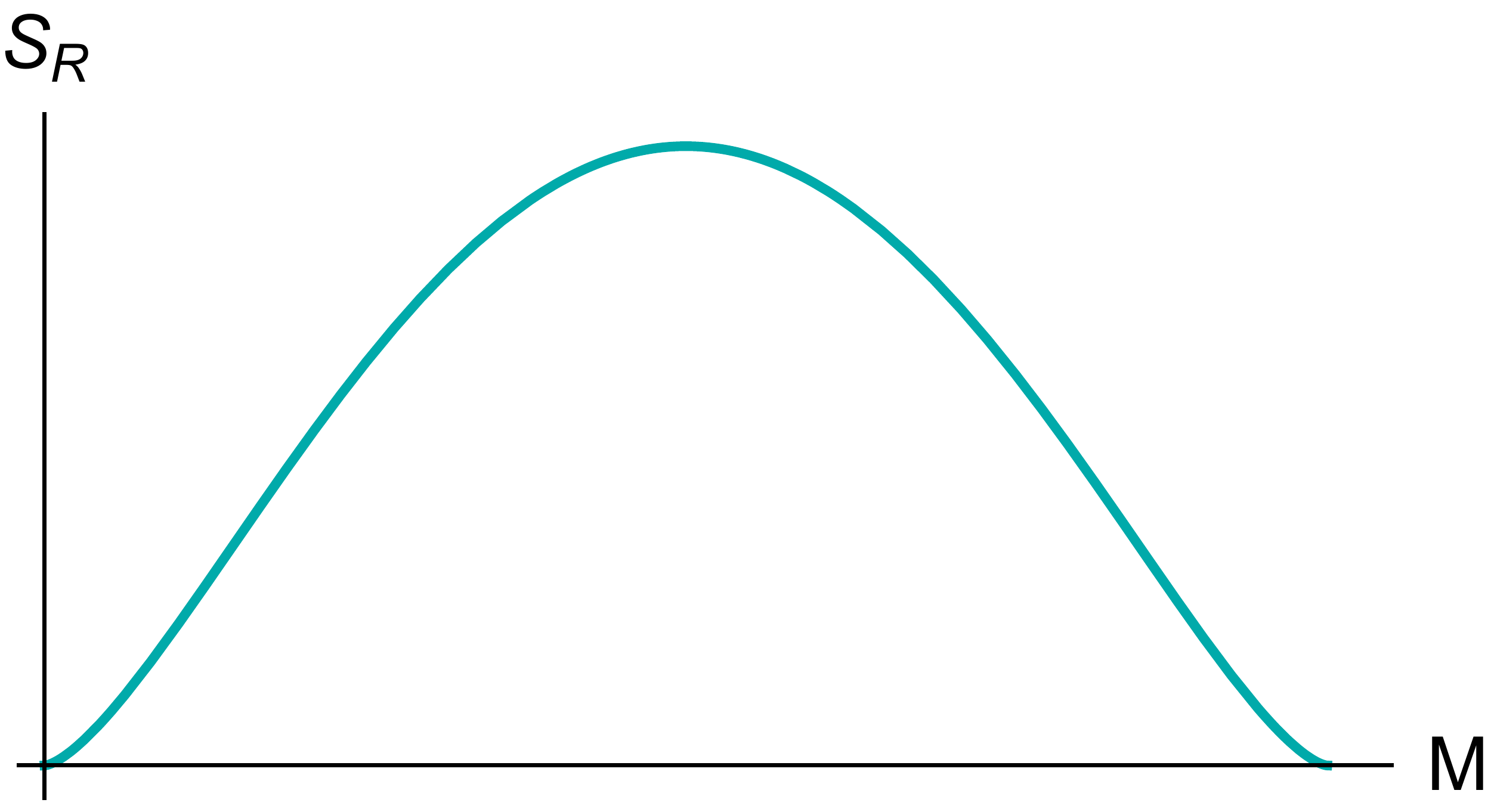}}                \\ \hline
                                  
\end{tabular}
\caption {$\,$}
\end{table}
\begin{table}[]
\centering
\begin{tabular}{llll}
\hline
\multicolumn{1}{|c|}{}     & \multicolumn{1}{c|}{\textbf{$S_{BH}=S_{BH}(M)$}} & \multicolumn{1}{c|}{\textbf{$G=G(M)$}} & \multicolumn{1}{c|}{\textbf{$S_R=S_R(T)$}} \\ \hline
\multicolumn{1}{|c|}{\begin{picture}(0,0)
\put(-10,30){1-st}
\end{picture}} & \multicolumn{1}{c|}{\includegraphics[scale=0.15]{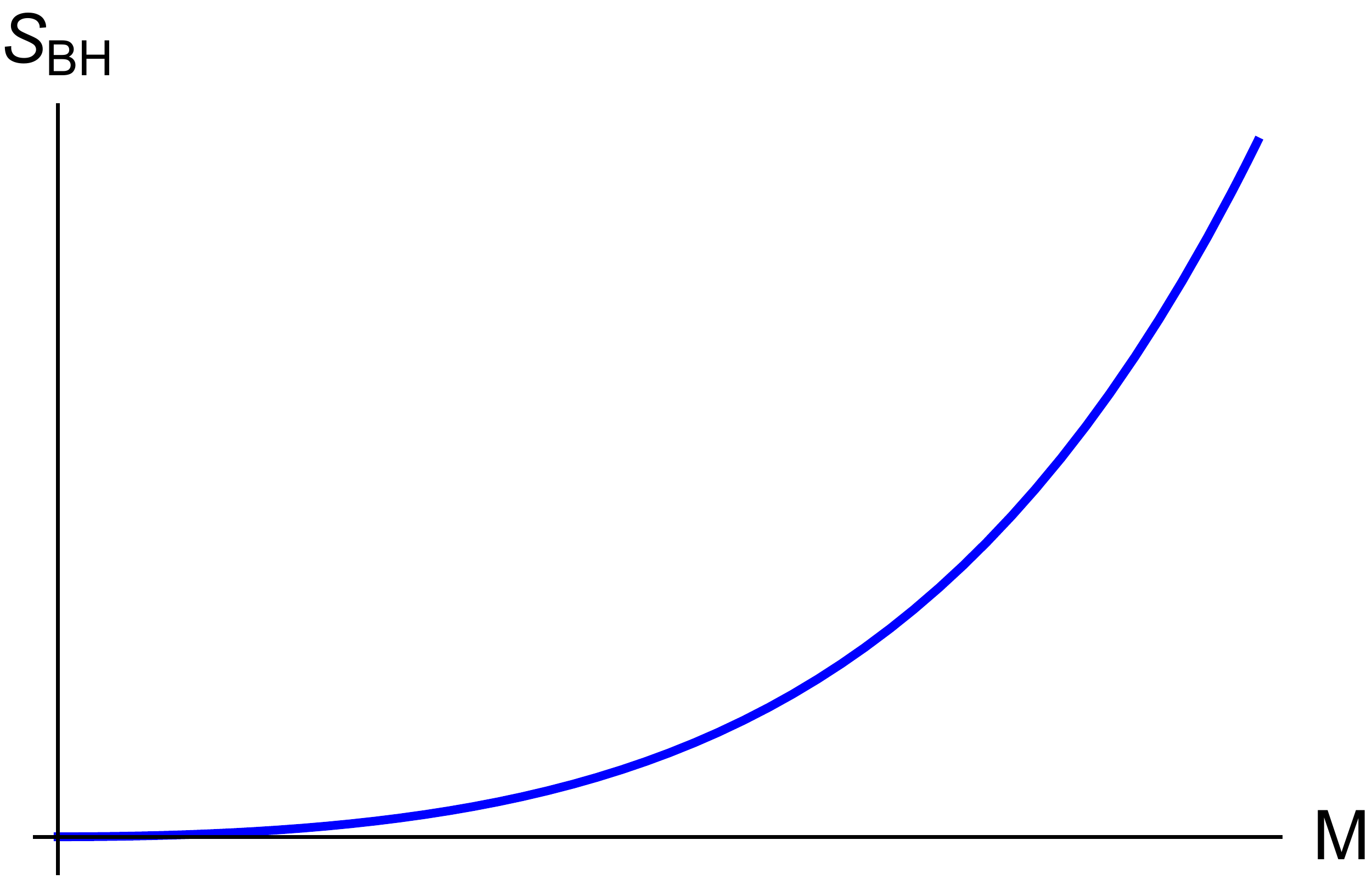}}                & \multicolumn{1}{c|}{\includegraphics[scale=0.15]{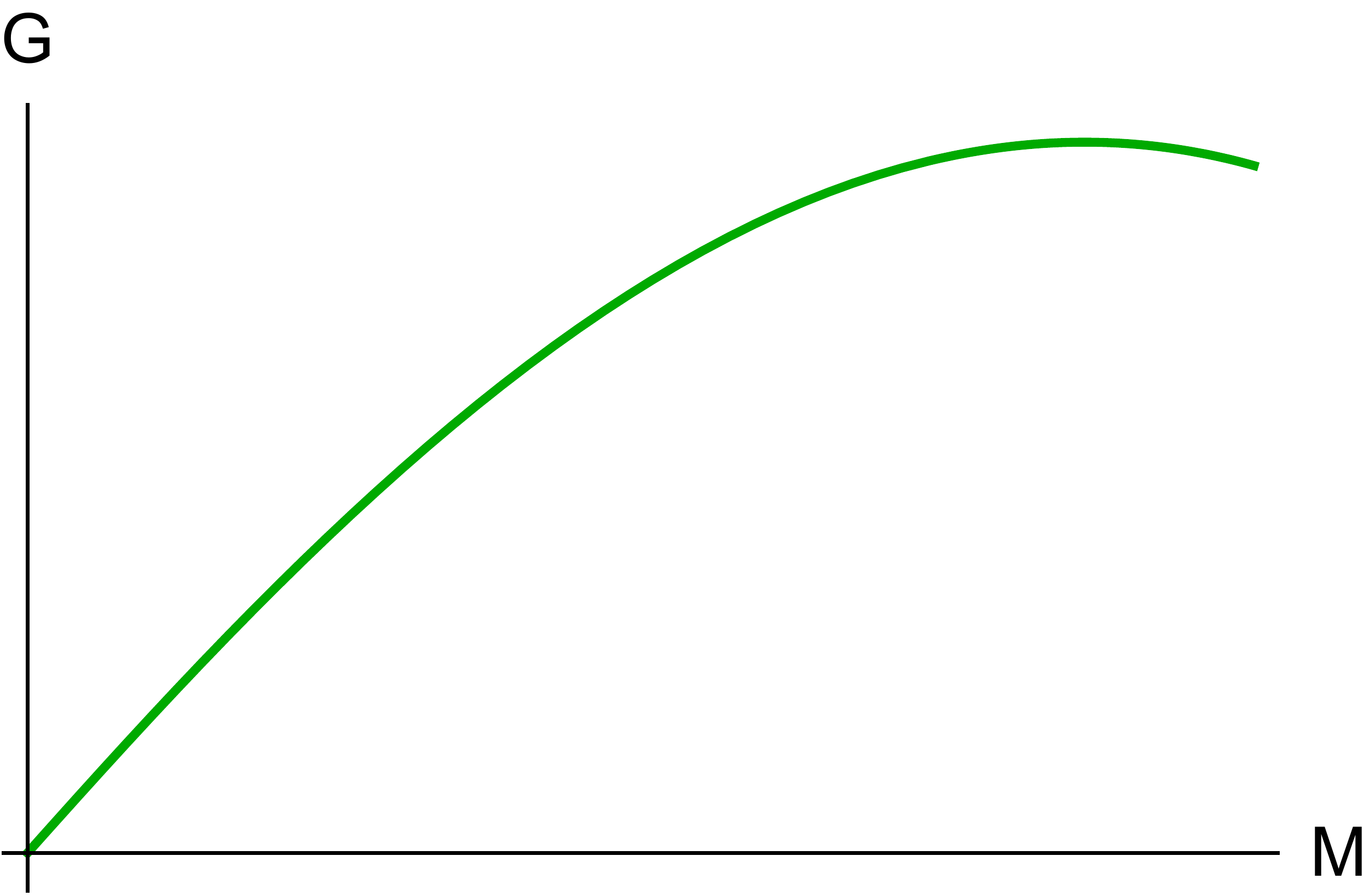}}                & \multicolumn{1}{c|}{\includegraphics[scale=0.15]{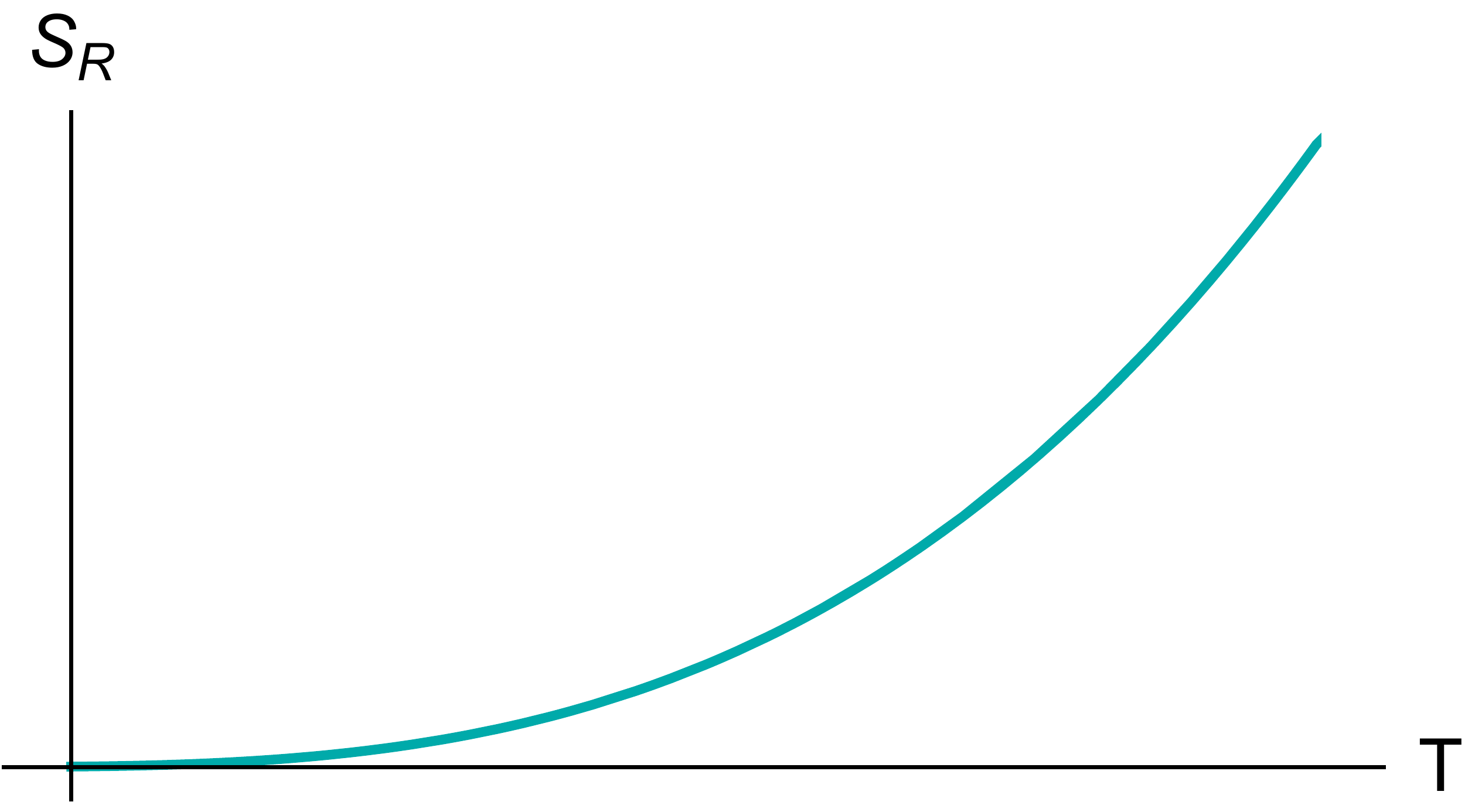}}                    \\ \hline
\multicolumn{1}{|c|}{\begin{picture}(20,20)
\put(0,30){2-nd}
\end{picture}} & \multicolumn{1}{c|}{\includegraphics[scale=0.15]{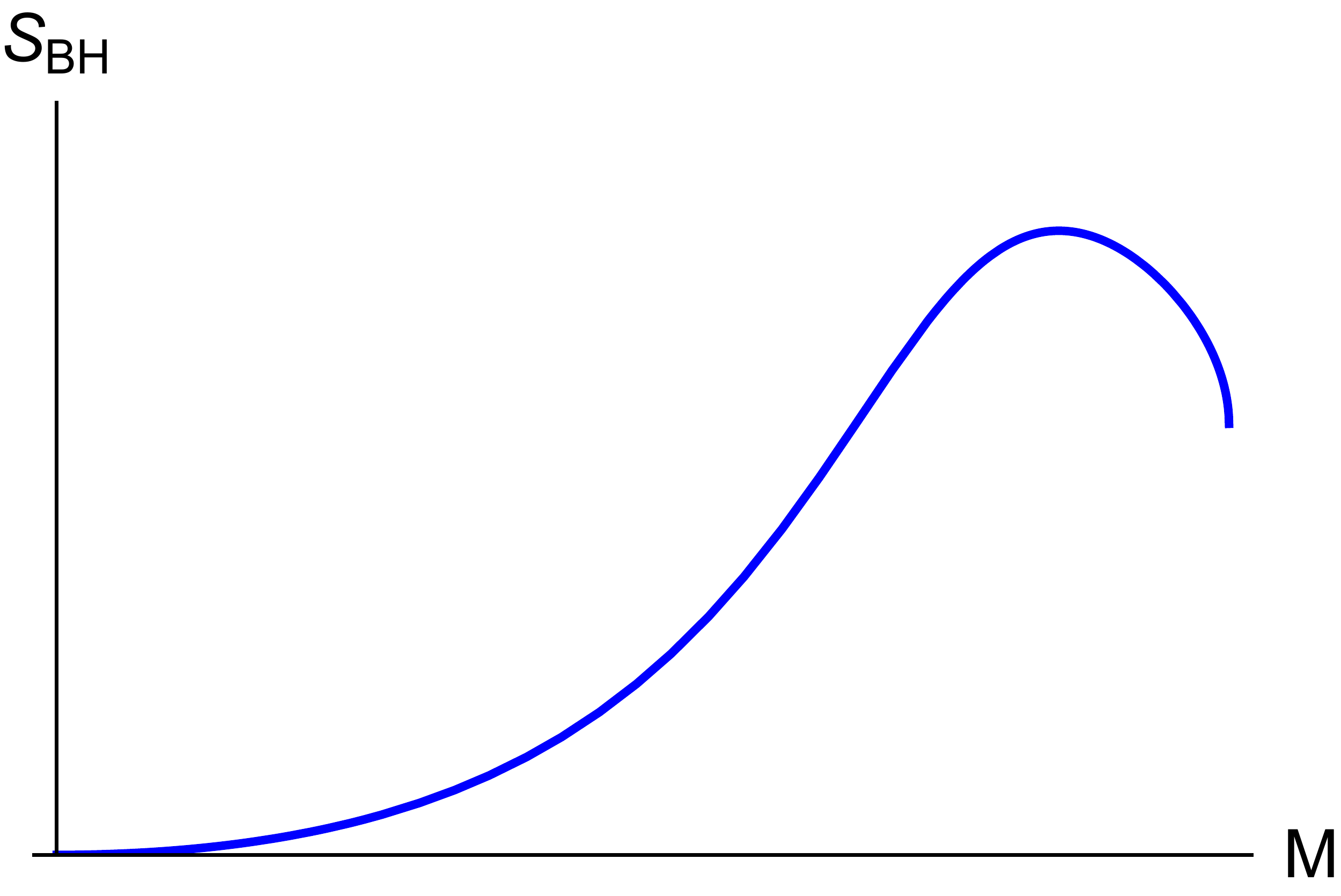}}                & \multicolumn{1}{c|}{\includegraphics[scale=0.15]{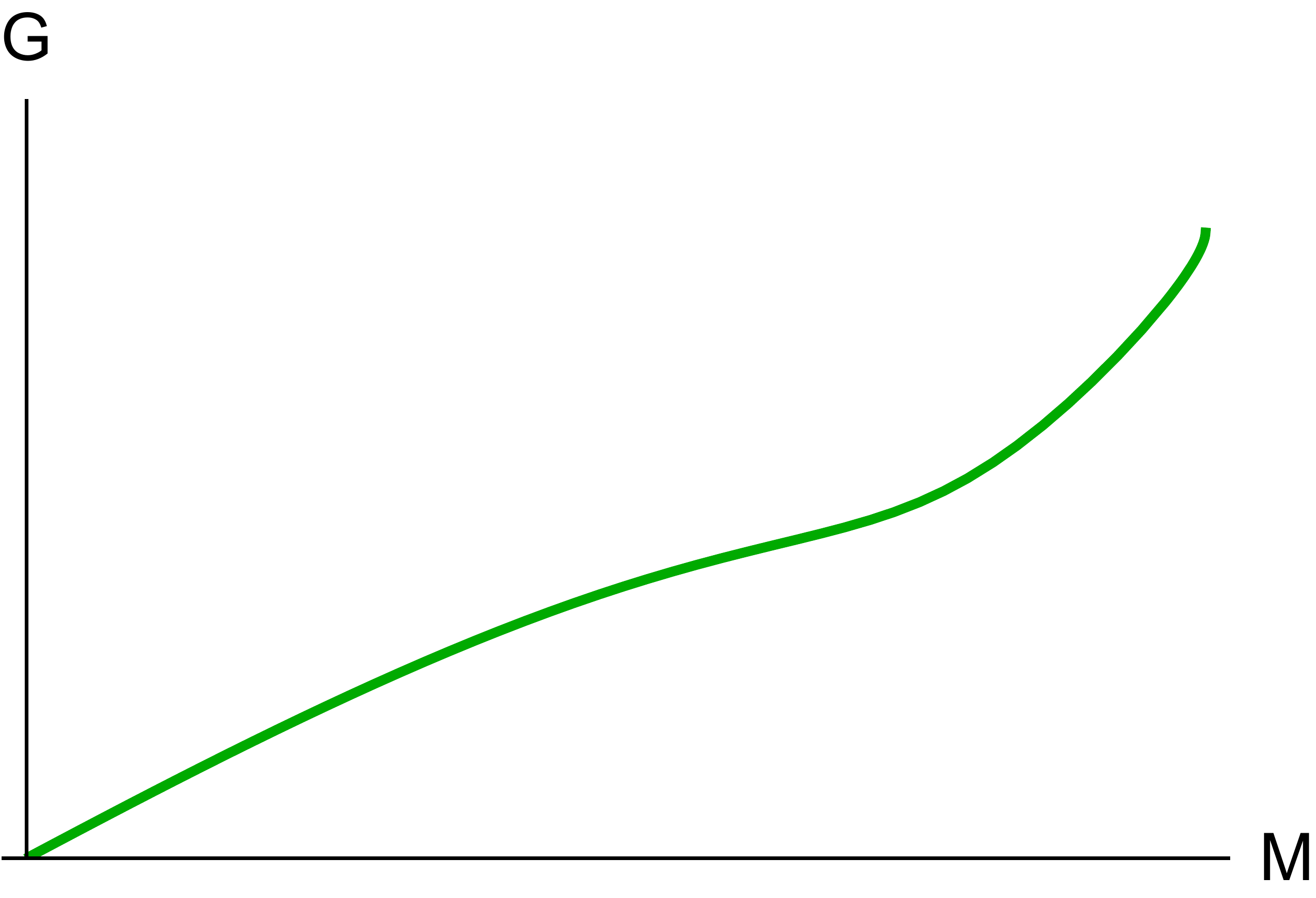}}                & \multicolumn{1}{c|}{\includegraphics[scale=0.15]{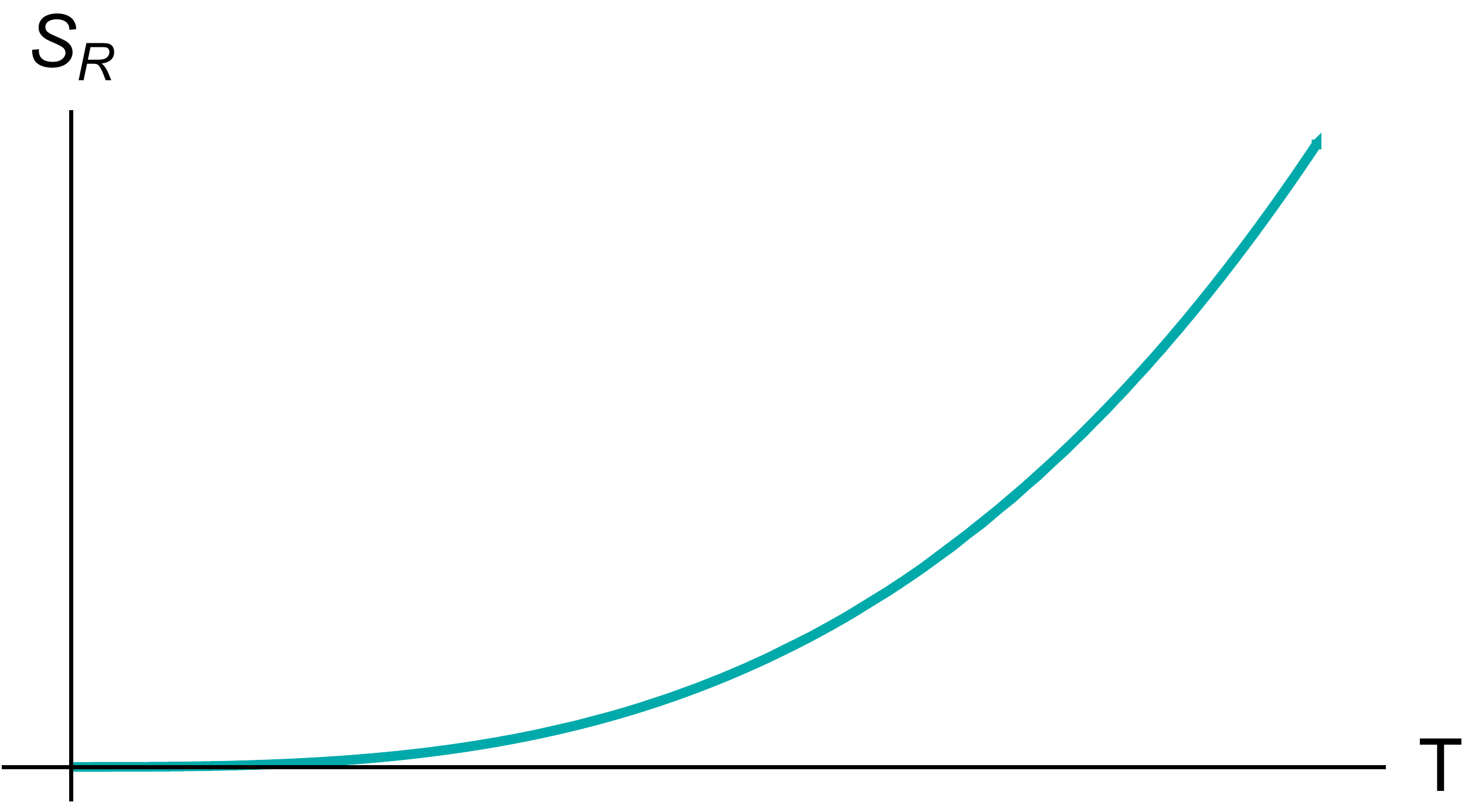}}                    \\ \hline 
\end{tabular}
\caption {$\,$}
\end{table}

\begin{itemize}
\item
In the first case, we are dealing with a deformed bell  form  of constraint (see a schematic plot in the first row, first coulomb in Table 1).
 Under additional restrictions on the parameter 
$\gamma$, specifying the form of   constraints \eqref{Q-M} with \eqref{lambda-gamma}, ($\gamma> 2$ in the text), 
we get complete evaporation of black holes with zero temperature at the end of evaporation. The Hawking temperature and the radiation entropy for this case
 first increase with decreasing of mass and get the maximal values, then they begin to decrease to zero values at zero mass (see a schematic plot in the first row, second and third  coulombs in Table 1). 
 Increasing of temperature with decreasing of mass corresponds to increasing of radiation entropy. Comparing the plot in  the first row, first coulomb, Table 1  and the plot in the first row, second coulomb, Table 2, we see that   recharging of the black hole  is accompanied by 
 increasing of free energy, that requires some extra forces.

\item In the second case,  constraints are given by curves in the $(M,\cQ)$-plane that correspond to  small deviations from corresponding extremal curves (see a schematic plot in the second row, first  coulombs in Table 1).  The mass dependence of temperature   has the form of the semi-circle
and the dependence of the radiation entropy  on mass  has the bell-shaped form (plots in the second row, second and third   coulombs in Table 1). The plots in second row of Table 2 show dependencies $S_{BH}(M)$, $G(M)$ and $S_R(T)$.

\end{itemize}
$$\,$$

Summing up the consideration in asymptotically flat cases we obtain the mass dependence of the entropy of radiation, which is schematically presented in Fig.\ref{fig:Page-curve-mass-time}.B, or in the third column of Table 1. 
Assuming a slow monotonic dependence of the mass on the time during the evaporation of the black hole, from such mass dependence of the radiation entropy we get the  Page curve for time dependence  of the radiation entropy.
We also generalized above consideration to  the case of  Schwarzschild-de Sitter and Reissner-Nordstrom-(Anti)-de-Sitter black holes.
\\

An important question  is why the black holes follow evaporation curves
avoided the blow-up of temperature?
 If one means realistic black holes, this question could be answered in the spirit of the anthropic principle, since otherwise an explosion of temperature occurs. Or in other words, we can say that we have to use a kind of black hole censorship. 
\\

The problem of complete evaporation is also discussed in \cite{Arefeva:2021byb}, where the singularity of temperature 
is attributed to the singularity of the Kruskal coordinates in the limit $M\to 0$ and alternative coordinates are proposed which describe a temperature distribution and are regular for vanishing mass.

\section*{Acknowledgments}
We would like to thank Victor Berezin, Valery Frolov and Michael Khramtsov  for useful discussions and remarks.

\end{document}